\documentclass[
    superscriptaddress,
    reprint,
    showkeys,
    showpacs,
    amsmath,
    amssymb,
    aps,
    prb,
    floatfix,
    tightenlines
]{revtex4-2}
\usepackage{upgreek}
\usepackage{graphicx}
\usepackage{dcolumn}
\usepackage{bm}
\usepackage{braket}
\usepackage[caption=false]{subfig}
\usepackage{float}
\usepackage{booktabs}
\usepackage{multirow}
\usepackage{tabularx}
\usepackage{microtype}
\usepackage{easyReview}
\usepackage{color}
\definecolor{link}{RGB}{57,106,177}
\definecolor{blue}{RGB}{0,0,0}
\usepackage{xr}
\newcommand{\ped}[1]{\ensuremath{_{\rm #1}}}
\newcommand{\apex}[1]{\ensuremath{^{\rm #1}}}

\usepackage{hyperref}
\hypersetup{
    colorlinks=true,
    linkcolor=link,
    citecolor=link,
    urlcolor=link
}
\AtBeginDocument{
\heavyrulewidth=.08em
\lightrulewidth=.05em
\cmidrulewidth=.03em
\belowrulesep=.65ex
\belowbottomsep=0pt
\aboverulesep=.4ex
\abovetopsep=0pt
\cmidrulesep=\doublerulesep
\cmidrulekern=.5em
\defaultaddspace=.5em
}
\begin{document}
\title{Reversible Tuning of Superconductivity in Ion-Gated NbN Ultrathin Films by Self-Encapsulation with a High-$\kappa$ Dielectric Layer}
\author{Erik Piatti}
\thanks{These authors contributed equally.}
\affiliation{
Department of Applied Science and Technology, Politecnico di Torino, I-10129 Torino, Italy
}
\author{Marco Colangelo}
\thanks{These authors contributed equally.}
\affiliation{
Department of Electrical Engineering and Computer Science, Massachusetts Institute of Technology, Cambridge, MA 02139, USA
}
\author{Mattia Bartoli}
\affiliation{
Center for Sustainable Future Technologies-CSFT@POLITO, I-10144 Torino, Italy
}
\affiliation{
Consorzio Interuniversitario Nazionale per la Scienza e Tecnologia dei Materiali (INSTM), I-850121 Firenze, Italy
}
\author{Owen Medeiros}
\affiliation{
Department of Electrical Engineering and Computer Science, Massachusetts Institute of Technology, Cambridge, MA 02139, USA
}
\author{Renato S. Gonnelli}
\affiliation{
Department of Applied Science and Technology, Politecnico di Torino, I-10129 Torino, Italy
}
\author{Karl K. Berggren}
\affiliation{
Department of Electrical Engineering and Computer Science, Massachusetts Institute of Technology, Cambridge, MA 02139, USA
}
\author{Dario Daghero}
\email{dario.daghero@polito.it}
\affiliation{
Department of Applied Science and Technology, Politecnico di Torino, I-10129 Torino, Italy
}

\begin{abstract}
Ionic gating is a powerful technique for tuning the physical properties of a material via electric-field-induced charge doping, but is prone to introduce extrinsic disorder and undesired electrochemical modifications in the gated material beyond pure electrostatics. Conversely, reversible, volatile, and electrostatic modulation is pivotal in the reliable design and operation of novel device concepts enabled by the ultrahigh induced charge densities attainable via ionic gating. Here we demonstrate a simple and effective method to achieve reversible and volatile gating of surface-sensitive ultrathin niobium nitride films via controlled oxidation of their surface. The resulting niobium oxide encapsulation layer exhibits a capacitance comparable to that of nonencapsulated ionic transistors, withstands gate voltages beyond the electrochemical stability window of the gate electrolyte, and enables a fully reversible tunability of both the normal-state resistivity and the superconducting transition temperature of the encapsulated films. Our approach should be transferable to other materials and device geometries where more standard encapsulation techniques are not readily applicable.\\\\
Cite this article as: E. Piatti, M. Colangelo, M. Bartoli, O. Medeiros, R. S. Gonnelli, K. K. Berggren, and D. Daghero. \textit{Phys. Rev. Applied} \href{https://doi.org/10.1103/PhysRevApplied.18.054023}{\textbf{18}, 054023 (2022)}.
\end{abstract}

\maketitle

\section{Introduction}

The ionic gating technique is a very powerful tool to tune the properties of a large variety of materials, including high-carrier density systems such as metals\,\cite{DagheroPRL2012, TortelloApSuSc2013, ShimizuPRL2013, DushenkoNatCommun2018, LiangSciAdv2018, LiangPRB2018}, BCS superconductors\,\cite{PiattiNE2021, ChoiAPL2014, PiattiJSNM2016, PiattiPRB2017, PaolucciNL2021}, thin flakes of metallic transition-metal dichalcogenides\,\cite{YoshidaAPL2016, LiNature2016, XiPRL2016} and iron-based superconductors\,\cite{ZhuPRB2017, ShiogaiNatPhys2016, LeiPRL2016, HanzawaPNAS2016, MiyakawaPRM2018, KounoSciRep2018, PiattiPRM2019} using a field-effect transistor (FET) configuration. In principle, the basic mechanism by which it operates is electrostatic and fully reversible: when the interface between an electrolyte and the material under study is polarized by a gate voltage, the mobile ions accumulate in the so-called electric double layer (EDL) and build up electric fields up to $\sim$\,100 times larger than those achievable in standard solid-dielectric FETs\,\cite{UenoJPSJ2014, FujimotoPCCP2013}. In practice, however, many processes beyond pure electrostatics can occur in an EDL-FET. These range from the introduction of extrinsic disorder in the form of charged-scattering centers\,\cite{PiattiPRM2019, GallagherNatCommun2015, PiattiApSuSc2017, SaitoACSNano2015, Gonnelli2dMater2017, PiattiNL2018, OvchinnikovNatCommun2016, PiattiEPJ2019, LuPNAS2018, PiattiApSuSc2020}, to field-induced distortions in the crystal lattice\,\cite{ZhangCPL2018, WangNJP2019, ZakidovACSNano2020}, to the intercalation of alkali ions\,\cite{YuNatNano2015, ShiSciRep2015, PiattiAPL2017, PiattiApSuSc2018mos2, LeiPRB2016, LeiPRB2017, WuAPL2018, KwabenaNature2018, CheAEM2019, ShangPRB2019, SongPRM2019} or protons\,\cite{LuNature2017, LengNPJQM2017, CuiSB2018, RafiqueNL2019, CuiCPL2019, LiarXiv, MengPRB2022, PiattiArXiv2022}, to the outright electrochemical modification of the gated material\,\cite{ShiogaiNatPhys2016, ZhangCPL2018, WangNJP2019, JeongScience2013, SchladtACSNano2013, PetachPRB2014, MaruyamaAPL2015, WalterACSNano2016, ZhangACSNAno2017, ZengPRL2018, SaleemAMI2019}. While these additional processes can be harnessed to provide additional degrees of freedom in modulating the properties of a material, it is often desirable to ensure that the modulation occurs only in the electrostatic regime. Indeed, reversible electrostatic switching is crucial for the realization of novel device concepts, such as chiral-light emitting transistors\,\cite{ZhangScience2014}, superconducting (SC) FETs\,\cite{ChenAdvMater2018, DeSimoniNatNano2018}, nano-constriction Josephson junctions\,\cite{PaolucciNL2018, PaolucciPRApp2019} and metallic SC quantum interference devices\,\cite{PaolucciNL2019}, as well as for reliable operation of stretchable and flexible devices\,\cite{ChoNatMater2008, LeeNanotech2014, PiattiNatElectron2021} and thermoelectric energy harvesters\,\cite{ShimizuNatCommun2019}.

The most straightforward way to ensure that the operation of an EDL-FET is dominated by reversible charge doping and that electrochemical interactions are suppressed, is to physically separate the active material from the electrolyte using an electrically insulating and electrochemically inert layer. This can be achieved by employing an electrolyte that partially decomposes when polarized, creating a passivation layer\,\cite{PiattiPRM2019}, but this strongly reduces the switching speed of the device\,\cite{PiattiPRM2019}. Another possibility is to employ encapsulation techniques widely used to protect unstable or reactive 2D materials in standard solid-state FETs\,\cite{Huang2021}. For example, one can cover the surface of the active materials, prior to the exposition to the electrolyte, with a high-quality ultrathin layered insulator obtained by micro-mechanical exfoliation of a bulk crystal\,\cite{GallagherNatCommun2015, XiPRL2016, LiNature2016}, or a protective dielectric layer\,\cite{JoSciRep2017, PiattiApSuSc2018mos2}. These alternatives, however, can present drawbacks when used in a ionic-gating setup. For instance, the first approach is not easily scalable to multiple integrated devices and large-area geometries\,\cite{GallagherNatCommun2015, XiPRL2016, LiNature2016}. In the second approach, the thickness of the protective layer is critical: thick passivation films strongly suppress the gate capacitance\,\cite{JoSciRep2017, PiattiApSuSc2018mos2}, while thin uniform films cannot be deposited on several materials of interest\,\cite{JoSciRep2017}. The development of an alternative, complementary encapsulation technique is therefore highly desirable. In this work, we demonstrate that growth of an ultrathin high-$\kappa$ dielectric layer on top of a surface-sensitive SC film by means of controlled in-situ oxidation ensures a fully reversible operation of the EDL-FET, a sizeable gate capacitance, a large induced charge-carrier density, and an enhanced tunability of the SC transition temperature with respect to the literature.

\section{Device Fabrication}

\begin{figure}
\centering
\includegraphics[width=\linewidth]{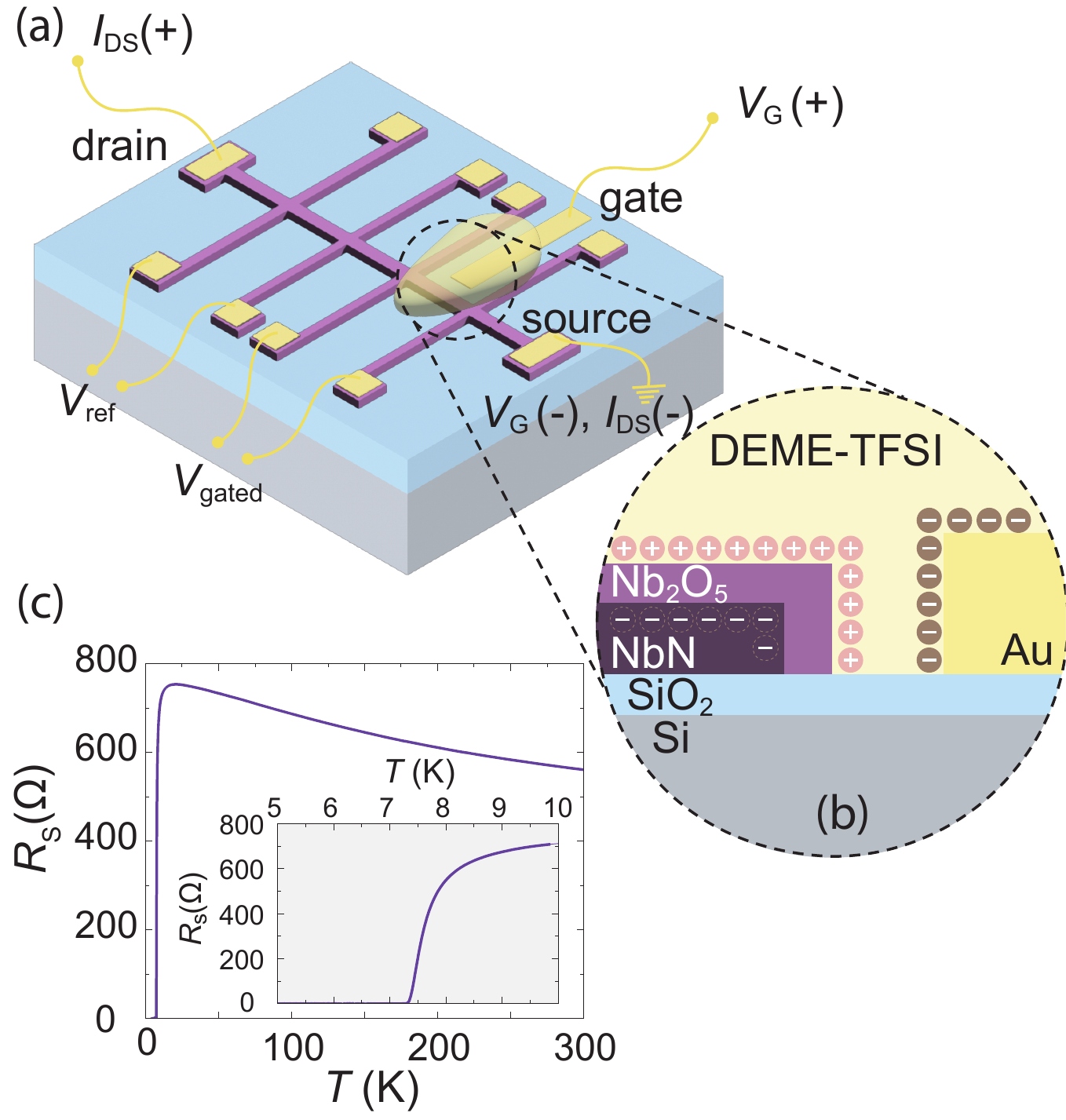}
\caption
{(a) Sketch of the multiple-Hall-bar structure and of the measurement configuration.  {\color{blue}Each channel is 1.15\,mm long and $100\,\upmu$m wide. The typical size of the Au side gate is $0.8\times1.2$\,mm$^2$. The ionic liquid droplet is drop-casted so as to cover the Au gate and the gated channel only.} (b) Sketch of the cross-section of the gated channel. (c) Sheet resistance $R\ped{s}$ as a function of temperature $T$ of the reference channel. Inset shows a magnification around the superconducting transition.}
\label{fig:device}
\end{figure}

Our device consists of a niobium nitride multiple-Hall-Bar structure. A $5\,\mathrm{nm}$-thick niobium nitride (NbN) layer was deposited\,\cite{dane2017bias} on a $300\,\mathrm{nm}$-thick thermal oxide layer on silicon. The NbN layer was patterned into a multiple-Hall-Bar geometry (see Fig.\,\ref{fig:device}) with direct writing photolithography followed by reactive ion etching. To facilitate the electrical contact with the measurements wires, gold pads were patterned and deposited on the outer lead regions of the Hall bar. The device was then annealed in oxygen atmosphere to grow a $\mathrm{Nb_2O_5}$ insulating barrier via direct oxidation of the superconducting layer. The thickness of the oxide barrier was $\approx 2.6\,\mathrm{nm}$, measured with ellipsometry\,\cite{medeiros2019measuring}. More details on the fabrication process are available in the Supplemental Section\,I\,\cite{SM}.

To characterize the device, we defined two measurement channels. The active (gated) channel is created by drop-casting the standard diethylmethyl\,(2-methoxyethyl)\,am- monium bis(trifluoromethylsulfonyl)imide (DEME-TFSI) ionic liquid on one section of the Hall bar and on the gate counterelectrode, made of a thin Au flake. The reference (ungated) channel is one of the other sections of the Hall bar, where no ionic liquid was casted. The droplet of liquid on the gated channel is covered with a thin ($10\,\upmu$m) kapton foil to tightly confine its coverage on the substrate and improve its thermo-mechanical stability.

\section{Gate-dependent electric transport}

Transport measurements were performed in the high-vacuum chamber of a Cryomech\textsuperscript{\textregistered} pulse-tube cryocooler by the four-wire method after the device was allowed to degas in vacuum at room temperature for at least $1$ day to minimize the water absorbed in the electrolyte. A small DC current ($I_{\mathrm{DS}}$) of $1\,\upmu$A was injected between the drain and source contacts with the first channel of an Agilent B2912 source-measure unit (SMU), and the voltage drops across the gated ($V_\mathrm{gated}$) and reference ($V_\mathrm{ref}$) channels were measured with two Agilent 34420 nanovoltmeters to determine the corresponding sheet resistances ($R\ped{s}$). Common-mode offsets such as thermoelectric voltages along the leads and contributions from the gate current were removed via the current-reversal method. The gate voltage ($V\ped{G}$) and current ($I\ped{G}$) were applied and measured between the gate and source contacts with the second channel of the same Agilent SMU. All the temperature ($T$)-dependent measurements were acquired during the slow, quasi-static warm-up of the devices to room $T$.
\begin{figure}[]
\centering
\includegraphics[width=\linewidth]{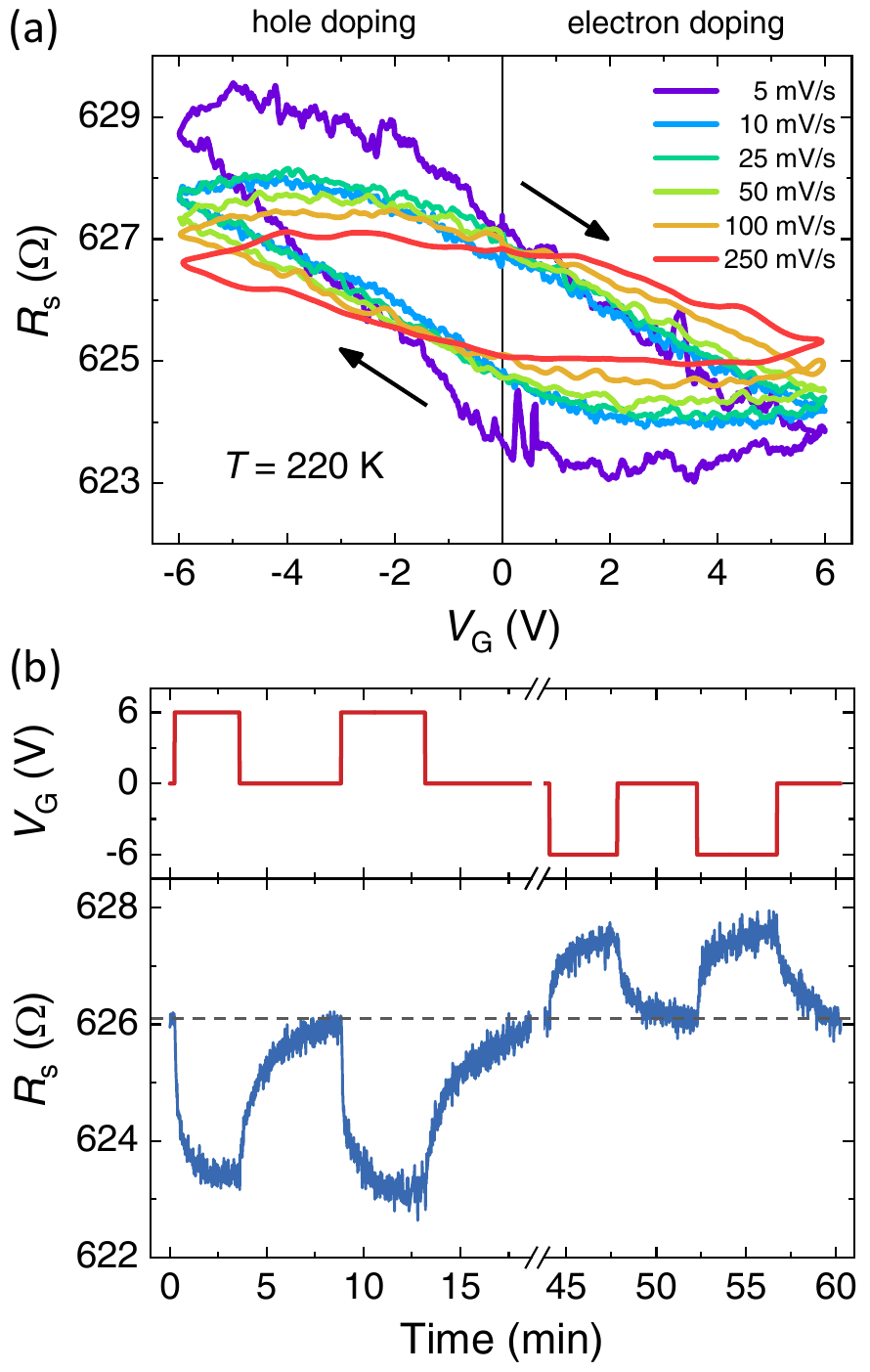}
\caption
{(a) Sheet resistance $R\ped{s}$ as a function of the gate voltage $V\ped{G}$ applied as a triangular wave at $T=220$\,K, for different values of the sweep rate.
(b) Typical response of $R\ped{s}$ (blue line, bottom panel) to the step-like application and removal of positive and negative values of $V\ped{G}$ (red line, top panel) at $T=220$\,K. Dashed line is a guide to the eye.}
\label{fig:gate_op}
\end{figure}

We first assessed the gate-dependent electric transport in our EDL-FETs through the Nb\ped{2}O\ped{5} encapsulation layer by sweeping $V\ped{G}$ in a triangular wave at $T=220$\,K and monitoring the modulation of the sheet resistance $R\ped{s}$ (Fig.\,\ref{fig:gate_op}a). Consistently with what we observed in thick, non-encapsulated films\,\cite{PiattiJSNM2016, PiattiPRB2017}, applying a positive $V\ped{G}$ (electron doping) reduces the value of $R\ped{s}$, while applying a negative $V\ped{G}$ (hole doping) increases it. After completing each sweep, $R\ped{s}$ returns to its original value (Fig.\,S1) irrespectively of the sweep rate within the uncertainty of the measurement\,\cite{SM}. The leakage current $I\ped{G}$ was always orders of magnitude smaller than $I\ped{DS}$. The tunability of $R\ped{s}$ decreases upon increasing the sweep rate, indicating that the relaxation time of the gate loop is dominated by the large resistance of the bulk ionic liquid due to the side-gate configuration\,\cite{PiattiPRM2019, ZhouJAP2012}. We thus investigated the tunability of $R\ped{s}$ over long time scales by applying and removing $V\ped{G}$ in a step-like fashion and waiting for the ion dynamics to settle (Fig.\,\ref{fig:gate_op}b). The total modulation of $R\ped{s}$ was found to be similar to that due to the triangular wave at the slowest sweep rate $5$\,mV/s. Most importantly, the modulation was completely reversible upon applying $V\ped{G}=0$ over a comparable time scale to that required for the saturation of $R\ped{s}$ upon application of a finite $V\ped{G}$. This complete reversibility was observed for both positive and negative applied $V\ped{G}$ in the Nb\ped{2}O\ped{5}-encapsulated devices, which is the typical feature of a modulation occurring via pure charge doping\,\cite{DagheroPRL2012, TortelloApSuSc2013, PiattiPRM2019, PiattiEPJ2019}. Note that these reversible modulations of R\ped{s} were retained when the devices were then cooled below the freezing point of the ionic liquid with a finite $V\ped{G}$ applied (Fig.\,S2\,\cite{SM}), further excluding the possibility that they might be an artifact due to a finite (even if small) gate leakage. Conversely, control measurements performed on non-encapsulated ultrathin NbN films resulted in modulations of $R\ped{s}$ which were largely irreversible upon $V\ped{G}$ removal (Fig.\,S3\,\cite{SM}){\color{blue}. This finding is in agreement with our earlier results on non-encapsulated films when their thickness was reduced below $\sim10$\,nm\,\cite{PiattiPRB2017}}. 
\begin{figure}[]
\centering
\includegraphics[width=\linewidth]{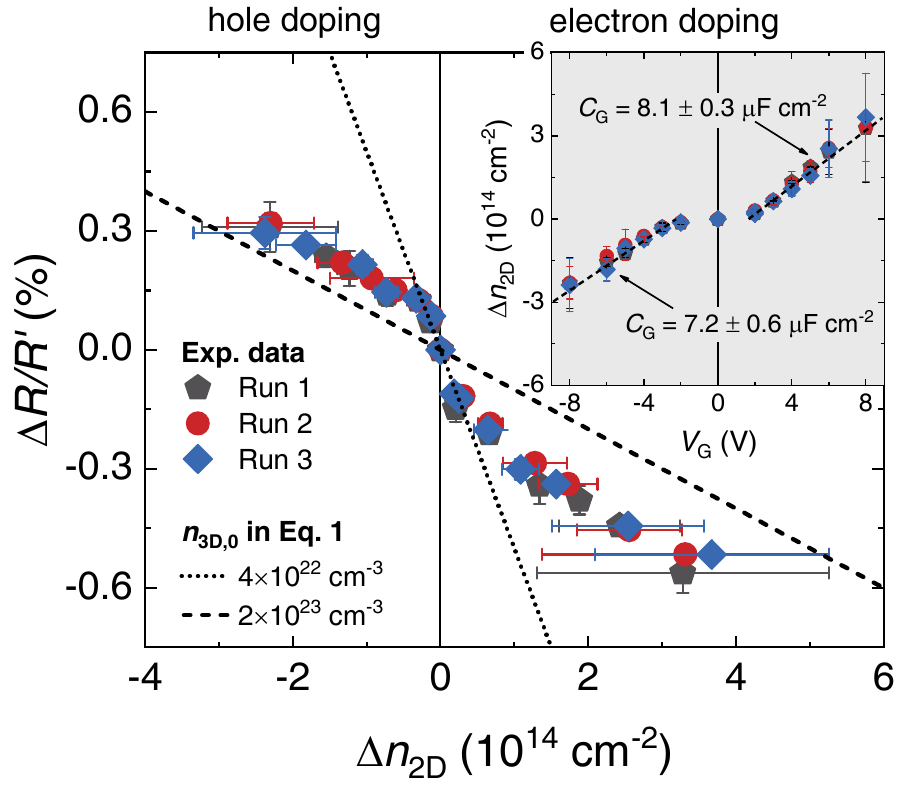}
\caption
{Normalized resistance variation, $\Delta R/R' = [R\ped{s}(\Delta n\ped{2D}) - R\ped{s}(0)] / R\ped{s}(\Delta n\ped{2D})$ as a function of the induced charge density $\Delta n\ped{2D}$ obtained upon the step-like application of $V\ped{G}$. Inset: $\Delta n\ped{2D}$ as a function of the applied gate voltage $V\ped{G}$ determined via double-step chronocoulometry in the same measurements. Dashed lines are linear fits to the data and allow to estimate the gate capacitance $C\ped{G}$.}
\label{fig:gate_sum}
\end{figure}

We gain further insight on the gate modulation process in our encapsulated NbN films by determining the surface density of induced charge, $\Delta n\ped{2D}$, as a function of $V\ped{G}$. This can be done by means of double-step chronocoulometry, a well-established electrochemical technique\,\cite{ScholtzBook} which allows determining the charge density stored in the EDL during the FET charging process\,\cite{DagheroPRL2012, TortelloApSuSc2013, PiattiJSNM2016, PiattiPRB2017, PiattiEPJ2019, PiattiPRM2019, PiattiApSuSc2020} through the analysis of the $I\ped{G}$ transients upon the step-like application and removal of a given value of $V\ped{G}$. As we show in the inset to Fig.\,\ref{fig:gate_sum}, the $V\ped{G}$-dependence of $\Delta n\ped{2D}$ further demonstrates that gate modulation occurs via charge doping: $\Delta n\ped{2D}$ linearly increases upon increasing $V\ped{G}$ for both electron and hole doping, as expected for the electrostatic charging of a capacitor. The corresponding gate capacitances ($C\ped{G} = 8.1\pm0.3$ and $7.2\pm0.6\,\upmu$F\,cm\apex{-2} as estimated from the linear fits in the electron and hole doping regimes respectively) are also in agreement with a simple estimation of the electrostatic capacitance of a Nb\ped{2}O\ped{5} layer with permittivity $\varepsilon_r \sim 30$\,\cite{PignoletTF1995} and thickness $d\ped{ox}\simeq 3$\,nm, $C\ped{ox} = \varepsilon_r \varepsilon_0 / d\ped{ox} \simeq 8.8\,\mu$F\,cm\apex{-2}. 
	
Let us now consider the normalized resistance modulation $\displaystyle \Delta R/R'=[R\ped{s}(\Delta n\ped{2D})-R\ped{s}(0)]/R\ped{s}(\Delta n\ped{2D})$ as a function of the induced charge density $\Delta n\ped{2D}$ (Fig.\,\ref{fig:gate_sum}). This quantity clearly follows two distinct linear trends (highlighted by the straight dashed and dotted lines) in the low- and high-$\Delta n\ped{2D}$ regimes. For the sake of comparison, in gated homogeneous films of elemental metals (Au, Ag, Cu) \,\cite{DagheroPRL2012, TortelloApSuSc2013}, $\displaystyle \Delta R/R'$ displays a simple linear trend, with the same slope in the whole range of $\Delta n\ped{2D}$, that can be described by a simple free-electron model with constant effective mass and relaxation time \,\cite{DagheroPRL2012, TortelloApSuSc2013, PiattiPRM2019}. The model predicts that $\Delta R/R'$ should depend on $\Delta n\ped{2D}$ according to the equation:
\begin{equation}
\frac{\Delta R}{R'} = \frac{R\ped{s}\left(\Delta n\ped{2D}\right)- R\ped{s}(0)}{R\ped{s}\left(\Delta n\ped{2D}\right)} = -\frac{\Delta n\ped{2D}}{n\ped{3D,0}t}
\label{eq:delta_R}
\end{equation}
where $n\ped{3D,0}$ is the intrinsic carrier density per unit volume and $t$ is the film thickness.  Deviations from this trend, with a reduction of the slope, were observed at high gate voltages in ultrathin metallic films ($t \simeq 5\, \mathrm{nm}$) \cite{DagheroPRL2012, TortelloApSuSc2013} and were ascribed to scattering phenomena at the film surface, which are not accounted for by Eq.\,\ref{eq:delta_R} but play a role when the thickness becomes comparable to the mean free path, as well predicted by quantum perturbative scattering models.

In our NbN films, the departure of the data from the initial linear trend cannot be interpreted in the same way, since: i) the mean free path of NbN is known to be very small (approximately 1/10 of the film thickness) so that these films are actually bulk-like \cite{Fuchs1938}; ii) if one uses the thickness of the films $t= 5\, \mathrm{nm}$, it turns out that the intrinsic carrier density of NbN $n\ped{3D,0}\simeq 2 \times 10^{23}\, \mathrm{cm^{-3}}$ \cite{ChockalingamPRB2008} would be compatible with the \textit{high-doping} slope $\Delta R/R'$ (see dashed line in Fig.\,\ref{fig:gate_sum}) rather than with the low-doping one, which would instead correspond to $n\ped{3D,0}\simeq 4 \times 10^{22}\, \mathrm{cm^{-3}}$ (dotted line). This indicates that, in the low-$V\ped{G}$ regime, the resistance modulation mainly stems from the charge doping of a layer which is less conducting than NbN. This conclusion is supported by the compositional analysis of the films (see X-ray photoelectron spectroscopy analyses below) which evidences the existence of an intermediate interfacial layer of the suboxide species NbO$_x$N$_{1-x}$ between the NbN film and the Nb$\ped{2}$O$\ped{5}$ oxide layer. At low gate voltages, this layer is less conductive and less capacitive than NbN\,\cite{TamerJPE2018,ChockalingamPRB2008}, absorbs most of the voltage drop through the device and is thus preferentially charge-doped. Eventually, on increasing the gate voltage, its charge density may become similar to that of NbN and the charge induction into the whole NbN film dominates. A more detailed analysis of the $\displaystyle \Delta R/R'$ trend, that takes into account the existence of \emph{two} layers of different materials, is reported in the Supplemental Section\,V\,\cite{SM}. 
	

Incidentally, Fig.\,\ref{fig:gate_sum} also shows that the Nb\ped{2}O\ped{5} encapsulation allows safely operating the EDL-FET beyond the electrochemical stability window of the ionic liquid ($|V\ped{G}| \leq 6$\,V at $T\sim220$\,K): All the resistance modulations induced at \mbox{$V\ped{G} = \pm 8$\,V} (that correspond to $|\Delta n\ped{2D}| \gtrsim 2\times 10^{14}\, \mathrm{cm^{-2}}$, see the inset) extrapolate nicely to the linear scaling observed at lower doping levels and the relevant resistance modulations remain reversible  -- even though a large uncertainty is introduced in the determination of $\Delta n\ped{2D}$ due to the large increase in $I\ped{G}$ caused by the decomposition of the ionic liquid.
\begin{figure}[]
\centering
\includegraphics[width=\linewidth]{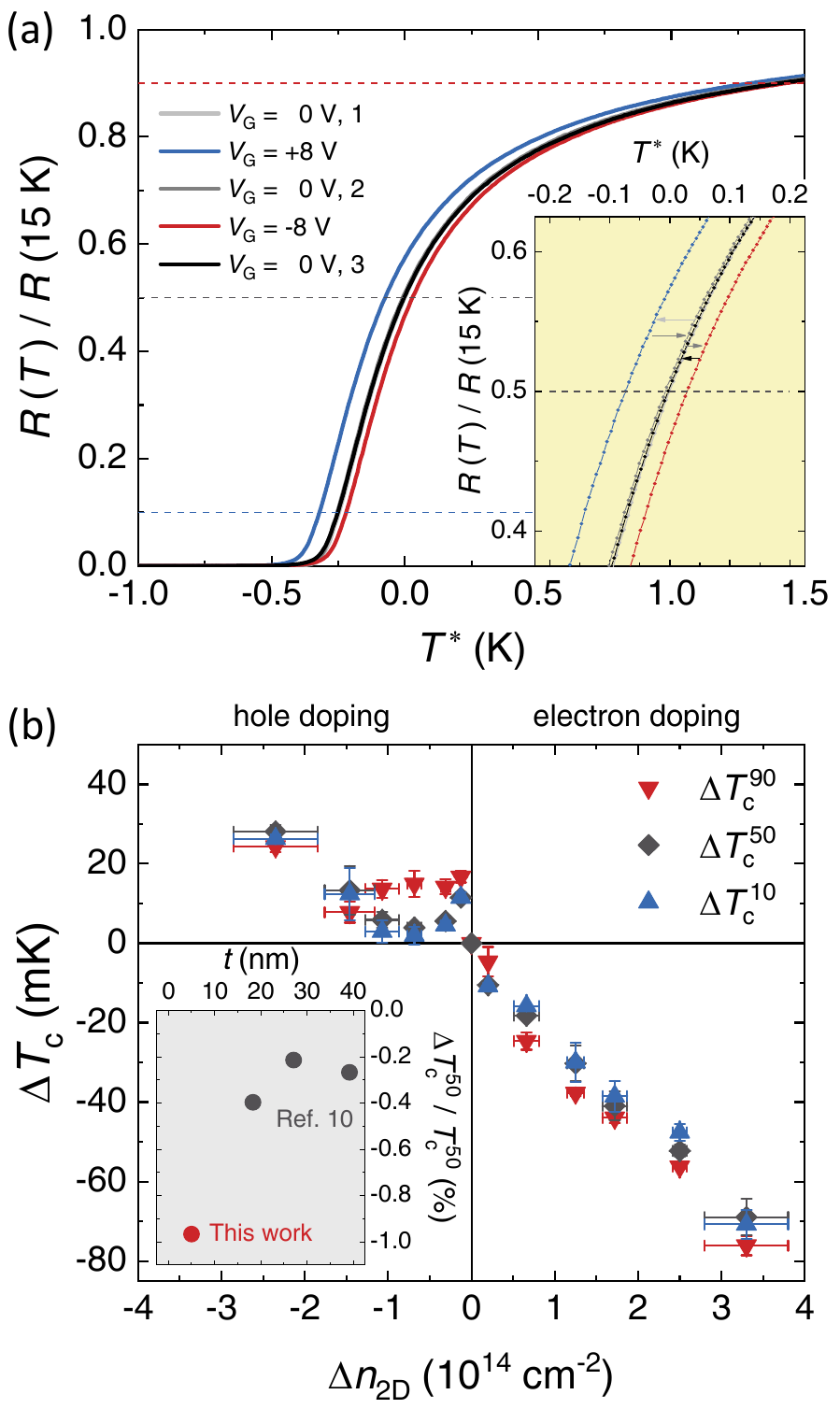}
\caption
{(a) Normalized resistance $R\ped{s} / R\ped{s}(15 \mathrm{K})$ as a function of the referenced temperature $T^* = \left[T - T\ped{c}\apex{ref}\right]_{V\ped{G}} - [T\ped{c}\apex{act} - T\ped{c}\apex{ref}]_{0}$ for different values of the applied gate voltage $V\ped{G}$. Dashed lines highlight the criteria used to obtain $T\ped{c}^{10}$, $T\ped{c}^{50}$, and $T\ped{c}^{90}$ from the resistive transitions. Inset: zoom of the same data close to the midpoint of the transition ($T\ped{c}^{50}$).
(b) $T\ped{c}$ shift $\Delta T\ped{c}$ as a function of the induced charge density $\Delta n\ped{2D}$ determined for $T\ped{c}^{10}$ (blue up triangles), $T\ped{c}^{50}$ (black diamonds), and $T\ped{c}^{90}$ (red down triangles). Inset: maximum $T\ped{c}$ tunability $\Delta T\ped{c}\apex{50} / T\ped{c}\apex{50}$ as a function of the NbN film thickness. Black dots are calculated from the data of Ref.\,\onlinecite{PiattiPRB2017}. The red dot is the maximum tunability achieved in this work.}
\label{fig:SC_transition}
\end{figure}

We now consider how the ionic gate modulates the SC properties of our encapsulated NbN ultrathin films, focusing on the dependence of the SC transition temperature $T\ped{c}$ on $\Delta n\ped{2D}$. Since the gate-induced $T\ped{c}$ shifts can be as small as few millikelvin, we adopt a differential technique allowed by the simultanous measurements of the resistive transition in the active ($T\ped{c}\apex{act}$) and reference ($T\ped{c}\apex{ref}$) channels\,\cite{PiattiJSNM2016, PiattiPRB2017, PiattiPRM2019}: For each threshold $\tau$=10, 50, 90 (i.e. 10\%, 50\% and 90\% of the resistive transition: see Fig.\,\ref{fig:SC_transition}a) the $T\ped{c}$ shift due to the application of a certain value of $V\ped{G}$ is determined as:
\begin{equation}
\Delta T\ped{c}\apex{\tau}(V\ped{G})= \left[T\ped{c}\apex{\tau,act} - T\ped{c}\apex{\tau,ref}\right]_{V\ped{G}} - [T\ped{c}\apex{\tau,act} - T\ped{c}\apex{\tau,ref}]_{0}.
\label{eq:DeltaTc1}
\end{equation}
We will also define a relative temperature scale $T^*$ whose zero falls on the midpoint (50\%) of the transition in the reference channel. {\color{blue} As a matter of fact, resistance vs. temperature curves at different values of $V\ped{G}$ are necessarily recorded in different runs, since the gate voltage can be changed only at high temperature ($> 200$ K), i.e. above the freezing point of the ionic liquid. Therefore, even the $R$ vs. $T$ curves of the ungated channel may not fall exactly on top of one another, due to a small thermal hysteresis. This does not affect in any way the determination of the $T_c$ shift due to charge accumulation, but may generate confusion when $R$ vs. $T$ curves measured at different $V\ped{G}$ are plotted in the same graph. To avoid this problem and improve the readability of the graphs, we will use $T^*$, defined as: \,\cite{PiattiJSNM2016, PiattiPRB2017, PiattiPRM2019}}
\begin{equation}
T^* = \left[T - T\ped{c}\apex{50,ref}\right]_{V\ped{G}} - [T\ped{c}\apex{50,act} - T\ped{c}\apex{50,ref}]_{0}.
\label{eq:DeltaTc2}
\end{equation}
The application of positive values of $V\ped{G}$ (electron doping) shifts the resistive transition to lower temperatures, while that of negative values of $V\ped{G}$ (hole doping) shifts it to higher temperatures (Fig.\,\ref{fig:SC_transition}a), consistently with what was reported on thick NbN films\,\cite{PiattiNE2021, PiattiJSNM2016, PiattiPRB2017}. Similarly to the $R\ped{s}$ modulations, also the shifts in the resistive transition are fully reversible by simply removing the applied $V\ped{G}$, as shown in the inset to Fig.\,\ref{fig:SC_transition}a). This reversible behavior must be compared with the control measurements performed on ultrathin non-encapsulated films, where the $T\ped{c}$ suppression upon electron doping was only partially reversible (Fig.\,S4a), and -- most importantly -- hole doping not only did not increase $T\ped{c}$ but suppressed it in a \textit{completely irreversible} fashion (Fig.\,S4b)\,\cite{SM}. Notably, the Nb\ped{2}O\ped{5} encapsulation allows for this fully reversible behavior to be maintained even for values of $V\ped{G}$ in excess of the electrochemical stability window of the ionic liquid (at least up to\mbox{$V\ped{G} = \pm 8$\,V}), while in non-encapsulated films much smaller values of $V\ped{G}$ were sufficient to trigger irreversible modifications {\color{blue} -- again, consistent with how irreversible $T\ped{c}$ shifts were induced in non-encapsulated NbN films when their thickness was reduced below $\sim 10$\,nm in our earlier report\,\cite{PiattiPRB2017}}.

In Fig.\,\ref{fig:SC_transition}b we summarize all the $T\ped{c}$ shifts measured as a function of $\Delta n\ped{2D}$ in our encapsulated films. In the \textit{electron doping} regime, $T\ped{c}$ is monotonically suppressed in a nearly linear fashion on increasing $\Delta n\ped{2D}$. Moreover, the $T\ped{c}$ shifts are nearly independent of the criterion used to define $T\ped{c}$, i.e. on the threshold $\tau$, which indicates that the resistive transition is rigidly shifted by the charge doping without any appreciable broadening. This is an expected feature for a gated SC film with a thickness smaller than the coherence length\,\cite{PiattiNE2021, PiattiPRB2017, UmmarinoPRB2017, UmmarinoPSSB2020}, since proximity effect ``spreads" the perturbation to the SC order parameter well beyond its electrostatic screening length\,\cite{PiattiNE2021, PiattiPRB2017, PiattiApSuSc2018nbn} and potentially up to the London penetration depth\,\cite{DeSimoniNatNano2018}. 

In the \textit{hole doping} regime, things are more complicated. The $T\ped{c}$ enhancement is found to be almost independent of $\tau$ only at large $\Delta n\ped{2D} \lesssim -2\times10^{14}$\,cm\apex{-2}). At smaller hole doping, $\Delta T\ped{c}\apex{90}$ turns out to be nearly doping-independent, but $\Delta T\ped{c}\apex{50}$ and $\Delta T\ped{c}\apex{10}$ vary in a non-monotonic fashion as a function of $\Delta n\ped{2D}$ and, although always positive, are smaller than $\Delta T\ped{c}\apex{90}$. This indicates a broadening of the SC transition which is instead typically observed in films where the SC order parameter is perturbed in a non-homogeneous way\,\cite{PiattiPRM2019, DeSimoniNatNano2018}. Overall, this asymmetric tuning of $T\ped{c}$ was already observed in thicker, non-encapsulated NbN films\,\cite{PiattiPRB2017, PiattiJSNM2016} and can be ascribed to the similarly asymmetric energy-dependence of the density of states above and below the undoped Fermi level.

Another figure of merit of our encapsulated ultrathin films is the maximum $T\ped{c}$ tunability, defined as the maximum value of $|\Delta T\ped{c}\apex{50}| / T\ped{c}\apex{50}$ observed in a given film. If compared to previous results obtained in thicker NbN films \,\cite{PiattiPRB2017}, the maximum tunability achieved in these ultrathin films is nearly three times larger and approaches 1\% (see the inset to Fig.\,\ref{fig:SC_transition}b). Notably, this strongly improved tunability is obtained at much lower values of charge doping{\color{blue}: $\Delta T\ped{c}\apex{50}\approx -70$\,mK is obtained at $\Delta n\ped{2D} \simeq 3\times 10^{14}$\,cm\apex{-2} in ultrathin encapsulated films, whereas the same $T\ped{c}$ shift required attaining $\Delta n\ped{2D} > 1\times 10^{15}$\,cm\apex{-2} in $\sim10$\,nm-thick non-encapsulated films in Ref.\,\onlinecite{PiattiPRB2017}}. Further large improvements can be expected by properly optimizing the growth process of the Nb\ped{2}O\ped{5} encapsulation layer and increasing its relative permittivity up to $\varepsilon_r\sim90$\,\cite{PignoletTF1995}.

\section{Spectroscopic characterization of the gate interface}

\begin{figure}[]
	\centering
	\includegraphics[width=\linewidth]{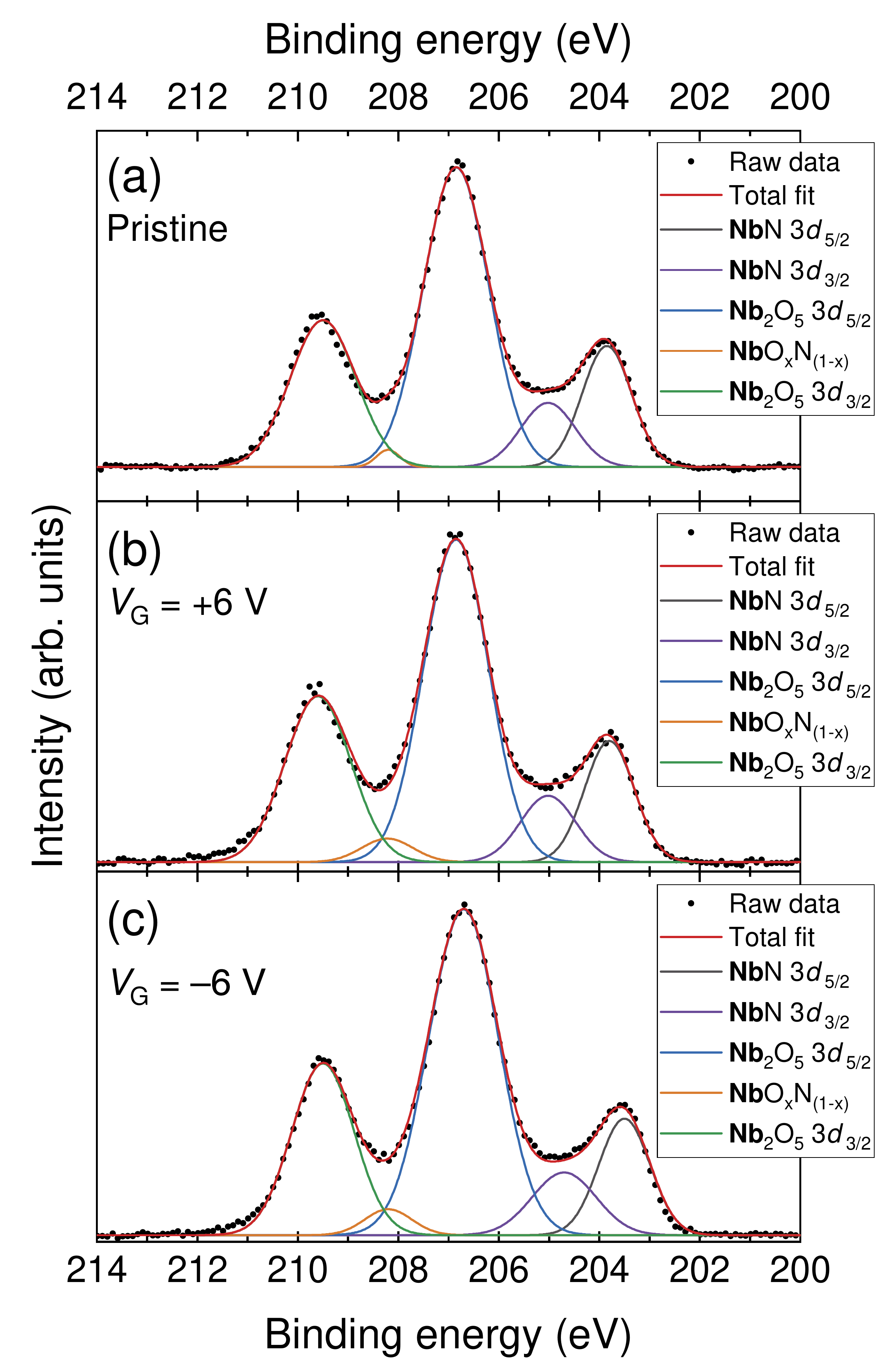}
	\caption
	{High-resolution X-ray photoelectron spectroscopy spectra of the Nb3\textit{d} region in an ungated NbN film (a), a NbN film gated at $V\ped{G}=+6$\,V (b), and a NbN film gated at $V\ped{G}=-6$\,V (c). Filled circles are the experimental data, solid lines are the fitted signals and relative components.}
	\label{fig:XPS_Nb}
\end{figure}
\begin{table}[]
	\begin{tabular}{ccccccc}
		& & & & & &\\
		\toprule
		\textbf{Species} & & \textbf{Pristine} & &\textbf{ Gated +6\,V} & & \textbf{Gated --6\,V} \\
		& & (atom \%) & & (atom \%) & & (atom \%) \\
		\midrule
		NbN & & $24.2 \pm 0.7$ & & $22.6 \pm 0.7$ & & $22.4 \pm 0.4$ \\
		\midrule
		Nb\ped{2}O\ped{5} & & $74.6 \pm 0.7$ & & $74.6 \pm 0.6$ & & $74.9 \pm 0.6$ \\
		\midrule
		NbO$_x$N$_{1-x}$ & & $1.0 \pm 0.2$ & & $2.8 \pm 0.4$ & & $2.7 \pm 0.6$ \\
		\bottomrule
	\end{tabular}
	\caption
	{XPS peak ratios of the Nb3$d$ region in a pristine NbN film, a NbN film gated at $V\ped{G}=+6$\,V, and a NbN film gated at $V\ped{G}=-6$\,V.}
	\label{tab:XPS_Nb}
\end{table}

As a further support of the effectiveness of the Nb\ped{2}O\ped{5} encapsulation layer in ensuring an electrostatic operation of the gated NbN devices, we carried out detailed analyses by means of X-ray photoelectron spectroscopy (XPS). Following a similar protocol as in our previous work\,\cite{PiattiPRM2019}, three unpatterned films were covered by DEME-TFSI ionic liquid and loaded in the cryocooler with the same procedure as the patterned devices. The first film was not electrically contacted and served as the pristine reference. The other two films were electrically contacted and subjected to $V\ped{G}=+6$\,V and $-6$\,V at $T=220$\,K, respectively, for $\sim30$\,min. All the films were then cooled down to the base temperature and warmed up, after which $V\ped{G}$ was released in the gated films. All films were then cleaned by subsequent sonications in soapy water, acetone and ethanol ($\sim 30$\,min each{\color{blue}; the procedure is safe against modifications of the physical and chemical state of the inorganic components\,\cite{CharaevJAP2017, NeylonACA2002}}) and blow-dried with a nitrogen gun, after which they were immediately transferred to the ultra-high vacuum chamber of a PHI 5000 Versaprobe scanning X-ray photoelectron spectrometer. XPS spectra were acquired using a monochromatic Al K-alpha X-ray source with 1486.6\,eV energy, 15\,kV voltage, and 1\,mA anode current.
Despite the cleaning process, the survey spectra of all samples (Fig.\,S6\,\cite{SM}) showed a massive presence of carbon contamination, which is unavoidable since ultra-thin samples cannot be subjected to in-situ Ar-ion milling before the acquisition of the XPS spectra\,\cite{PiattiPRM2019}. Such presence of organic species with unknown stoichiometry makes an unambiguous peak assignment of the N and O signals impossible, making their analysis highly speculative at best. We therefore focus on the high-resolution Nb3\textit{d} spectra shown in Fig.\,\ref{fig:XPS_Nb}, which are unaffected by impurities and ionic-liquid residues, and highly sensitive to the chemical environment in both the Nb\ped{2}O\ped{5} encapsulation layer and the underlying NbN film. The spectrum of the pristine sample (Fig.\,\ref{fig:XPS_Nb}a) comprises two peaks belonging to NbN (3\textit{d}\ped{5/2}, $\sim$\,203.8\,eV; 3\textit{d}\ped{3/2}, $\sim$\,205.1\,eV)\,\cite{ErmolieffApSuSc1985, LanCEJ2022}, two peaks due to the massive presence of Nb\ped{2}O\ped{5} (3\textit{d}\ped{5/2}, $\sim$\,206.9\,eV; 3\textit{d}\ped{3/2},  $\sim$\,209.5\,eV)\,\cite{ErmolieffApSuSc1985, LanCEJ2022}, and a fifth peak centered at $\sim$\,208.2\,eV which can be reasonably assigned to the intermediate suboxide species NbO$_x$N$_{1-x}$\,\cite{DarlinskiSIA1987}. This suggests that, even in the pristine sample, the Nb\ped{2}O\ped{5} and NbN layers are not separated by a sharp interface but, rather, by an intermediate transition region formed by sub-stoichiometric niobium oxynitride. The spectra of both the gated samples (Fig.\,\ref{fig:XPS_Nb}b, $V\ped{G}=+6$\,V; Fig.\,\ref{fig:XPS_Nb}c, $V\ped{G}=-6$\,V) do not show any appreciable difference with respect to the pristine sample in 
the Nb\ped{2}O\ped{5} peaks, as evidenced by the peak areas reported in Table\,\ref{tab:XPS_Nb}. Minute differences $\lesssim2\%$ are instead observed in the NbN peaks and in the NbO\ped{x}N\ped{1-x} peak, which are however extremely sensitive to both fitting procedure and baseline correction.
Overall, the XPS analysis indicates that the thickness of the oxynitride transition region might be significantly increased by the gating process, certainly at the expenses of both the NbN film and the Nb\ped{2}O\ped{5} encapsulation layer, even though the reduction in the thickness of the latter is, in percentage, very small and experimentally undetectable. Any change to the electronic properties of either the Nb\ped{2}O\ped{5} or the NbN layers is instead completely volatile upon removal of the gate voltage. 

\begin{figure}[ht!]
	\centering
	\includegraphics[width=\linewidth,angle=0]{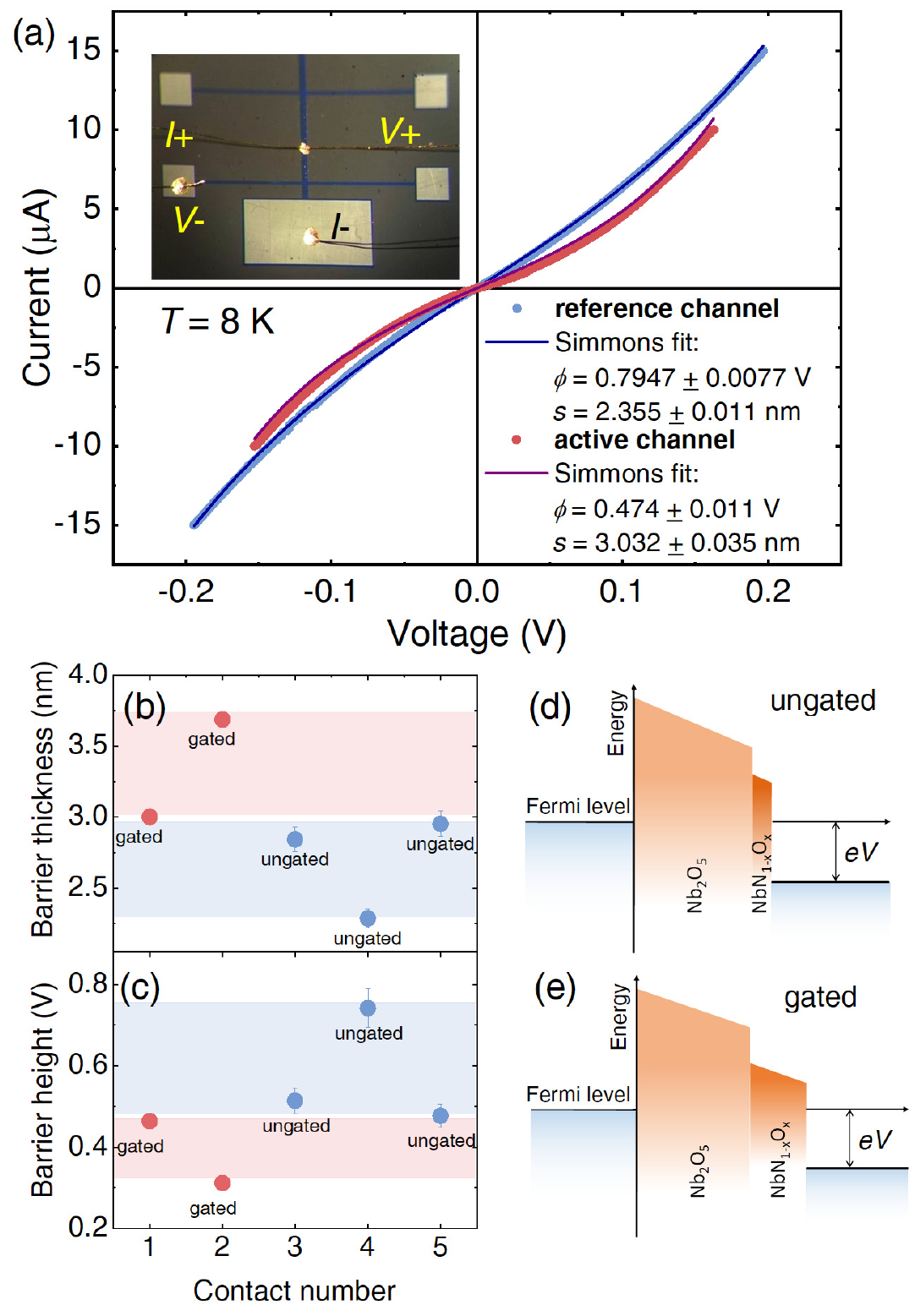}
	\caption
	{(a) Two examples of $I(V)$ characteristics of N/I/N junctions made through the Nb oxide layer, in a region of the reference channel (blue symbols) and in a region of the active channel (red symbols). The curves have already been corrected to eliminate the contribution of the spreading resistance. The blue and the red lines represent their fit with the Simmons model\,\cite{Simmons1963}. The values of $\phi$ (barrier height) and $s$ (barrier thickness) extracted from the fit are indicated. For these particular fits, the area of the junctions was fixed to $1.2 \times 10^{-8}\, \mathrm{m}$, based on the geometric estimation. The inset shows a picture of the setup. (b) Effective thickness $s$ and (c) effective height $\phi$ of the potential barrier, as extracted from the Simmon's fit of some I(V) curves in the ungated (blue symbols) and gated (red symbols) channels. (d) and (e) summarize in a schematic picture the evolution of the potential barrier upon gating, as it results from XPS and tunnel measurements.}
	\label{fig:tunnel}
\end{figure}

The increase of the oxynitride transition region and the consequent change in the mean potential barrier of the encapsulation layer is also confirmed by tunnel spectroscopy through the Nb\ped{2}O\ped{5} barrier. After a set of ionic gating measurements, we cleaned the surface of the devices and made point contacts (with a conductive Ag paste) on top of either the active (gated) channel or the reference (ungated) channel. A picture of the setup is shown in the inset to Fig.\,\ref{fig:tunnel}a. We then injected a current $I$ into the Ag/Nb\ped{2}O\ped{5}/NbN junctions and measured the voltage drop $V^+-V^- = V\ped{exp}$ both in the superconductive and in the normal state. This was done in order to determine and cancel the contribution of the spreading resistance r\ped{s} (i.e. the portion of NbN film between the point contact and the $V^-$ contact) to the measured $I(V\ped{exp})$ curve. Once the measured voltage is suitably corrected [$V(I) = V\ped{exp}(I) - r\ped{s}I$], we determine the $I(V)$ curve of each junction in the normal state. Fig.\,\ref{fig:tunnel}a reports two examples of such curves, measured on the reference channel (blue symbols) and on the active one (red symbols). Clearly, the latter shows a greater degree of non-linearity that, once the curves are fitted to the simple Simmons' model\,\cite{Simmons1963}, can be rationalized as being due to a higher thickness and smaller height of the potential barrier that separates the normal electrodes (see Supplemental Section\,VII for more details\,\cite{SM}). As shown in Fig.\,\ref{fig:tunnel}b and c, this is a general trend; the mean height of the potential barrier is $\langle \phi\rangle \ped{u}=0.60\pm0.13$\,V before gating, and decreases to $\langle \phi \rangle \ped{g}=0.39\pm0.07$\,V after gating; at the same time, the thickness of the potential barrier increases from $\langle s \rangle \ped{u}=2.62\pm0.33$\,nm to $\langle s \rangle \ped{g}=3.34\pm0.34$\,nm. Both these results are compatible with the expansion (by a factor 3: see last line of Table \ref{tab:XPS_Nb}) of the intermediate interfacial layer of sub-stoichiometric oxide, as observed via XPS, at the expenses of the NbN film and possibly of the Nb\ped{2}O\ped{5} encapsulation layer -- provided that one admits, as it seems reasonable, that the relevant potential barrier is lower than that of the insulating oxide. A rough estimation based on the XPS and tunnel data would indicate that the interfacial layer has a thickness of the order of 0.3\,nm in the ungated devices, and expands to about 1\,nm in the gated ones. A schematic picture of the junction is shown in Figs.\,\ref{fig:tunnel}d and \ref{fig:tunnel}e in the ungated and gated devices respectively.

\section{Conclusions}
In summary, we have demonstrated a simple and effective method to ensure the volatile and reversible operation of ion-gated superconducting films by means of encapsulation in an ultrathin high-$\kappa$ dielectric niobium oxide layer. Our gate-dependent electric transport measurements show that encapsulated devices exhibit fully-reversible tunability of both the normal-state resistivity and the superconducting transition temperature, a gate capacitance comparable to that found in non-encapsulated ionic transistors, and stability even beyond the electrochemical stability window of the electrolyte. X-ray photoelectron and tunnel spectroscopy characterizations confirm the effectiveness of the encapsulation layer in suppressing undesired electrochemical interactions between the superconducting film and the electrolyte, and reveal how the only non-volatile alteration to the devices is an increase in the thickness of the substoichiometric interfacial region between the superconducting film and the encapsulation layer. Our approach should be readily transferable to other materials and devices where ensuring a reversible and volatile ionic gate operation without major losses in gate capacitance is required for successful device operation.

\begin{acknowledgments}
We thank S. Guastella and F. Galanti for assistance in the XPS and transport measurements, respectively. We thanks J. Daley and M. K. Mondol of the MIT Nanostructures Laboratory Facility for
technical support. We thank E. Batson for assistance in editing the final manuscript.
E.P., D.D. and R.S.G. acknowledge funding from the Italian Ministry of Education, University and Research (Project PRIN “Quantum2D”, Grant No. \textsc{2017Z8TS5B}).
M.C., O.M., and K.K.B. acknowledge funding from the National Science Foundation grant ECCS-2000743. M.C. acknowledges support from the Claude E. Shannon Award.
\end{acknowledgments}

\bibliography{Bibliography}

\begin{thebibliography}{92}%
\makeatletter
\providecommand \@ifxundefined [1]{%
 \@ifx{#1\undefined}
}%
\providecommand \@ifnum [1]{%
 \ifnum #1\expandafter \@firstoftwo
 \else \expandafter \@secondoftwo
 \fi
}%
\providecommand \@ifx [1]{%
 \ifx #1\expandafter \@firstoftwo
 \else \expandafter \@secondoftwo
 \fi
}%
\providecommand \natexlab [1]{#1}%
\providecommand \enquote  [1]{``#1''}%
\providecommand \bibnamefont  [1]{#1}%
\providecommand \bibfnamefont [1]{#1}%
\providecommand \citenamefont [1]{#1}%
\providecommand \href@noop [0]{\@secondoftwo}%
\providecommand \href [0]{\begingroup \@sanitize@url \@href}%
\providecommand \@href[1]{\@@startlink{#1}\@@href}%
\providecommand \@@href[1]{\endgroup#1\@@endlink}%
\providecommand \@sanitize@url [0]{\catcode `\\12\catcode `\$12\catcode
  `\&12\catcode `\#12\catcode `\^12\catcode `\_12\catcode `\%12\relax}%
\providecommand \@@startlink[1]{}%
\providecommand \@@endlink[0]{}%
\providecommand \url  [0]{\begingroup\@sanitize@url \@url }%
\providecommand \@url [1]{\endgroup\@href {#1}{\urlprefix }}%
\providecommand \urlprefix  [0]{URL }%
\providecommand \Eprint [0]{\href }%
\providecommand \doibase [0]{https://doi.org/}%
\providecommand \selectlanguage [0]{\@gobble}%
\providecommand \bibinfo  [0]{\@secondoftwo}%
\providecommand \bibfield  [0]{\@secondoftwo}%
\providecommand \translation [1]{[#1]}%
\providecommand \BibitemOpen [0]{}%
\providecommand \bibitemStop [0]{}%
\providecommand \bibitemNoStop [0]{.\EOS\space}%
\providecommand \EOS [0]{\spacefactor3000\relax}%
\providecommand \BibitemShut  [1]{\csname bibitem#1\endcsname}%
\let\auto@bib@innerbib\@empty
\bibitem [{\citenamefont {Daghero}\ \emph {et~al.}(2012)\citenamefont
  {Daghero}, \citenamefont {Paolucci}, \citenamefont {Sola}, \citenamefont
  {Tortello}, \citenamefont {Ummarino}, \citenamefont {Agosto}, \citenamefont
  {Gonnelli}, \citenamefont {Nair},\ and\ \citenamefont
  {Gerbaldi}}]{DagheroPRL2012}%
  \BibitemOpen
  \bibfield  {author} {\bibinfo {author} {\bibfnamefont {D.}~\bibnamefont
  {Daghero}}, \bibinfo {author} {\bibfnamefont {F.}~\bibnamefont {Paolucci}},
  \bibinfo {author} {\bibfnamefont {A.}~\bibnamefont {Sola}}, \bibinfo {author}
  {\bibfnamefont {M.}~\bibnamefont {Tortello}}, \bibinfo {author}
  {\bibfnamefont {G.~A.}\ \bibnamefont {Ummarino}}, \bibinfo {author}
  {\bibfnamefont {M.}~\bibnamefont {Agosto}}, \bibinfo {author} {\bibfnamefont
  {R.~S.}\ \bibnamefont {Gonnelli}}, \bibinfo {author} {\bibfnamefont {J.~R.}\
  \bibnamefont {Nair}},\ and\ \bibinfo {author} {\bibfnamefont
  {C.}~\bibnamefont {Gerbaldi}},\ }\bibfield  {title} {\bibinfo {title} {Large
  conductance modulation of gold thin films by huge charge injection via
  electrochemical gating},\ }\href
  {https://doi.org/https://doi.org/10.1103/PhysRevLett.108.066807} {\bibfield
  {journal} {\bibinfo  {journal} {Phys. Rev. Lett.}\ }\textbf {\bibinfo
  {volume} {108}},\ \bibinfo {pages} {066807} (\bibinfo {year}
  {2012})}\BibitemShut {NoStop}%
\bibitem [{\citenamefont {Tortello}\ \emph {et~al.}(2013)\citenamefont
  {Tortello}, \citenamefont {Sola}, \citenamefont {Sharda}, \citenamefont
  {Paolucci}, \citenamefont {Nair}, \citenamefont {Gerbaldi}, \citenamefont
  {Daghero},\ and\ \citenamefont {Gonnelli}}]{TortelloApSuSc2013}%
  \BibitemOpen
  \bibfield  {author} {\bibinfo {author} {\bibfnamefont {M.}~\bibnamefont
  {Tortello}}, \bibinfo {author} {\bibfnamefont {A.}~\bibnamefont {Sola}},
  \bibinfo {author} {\bibfnamefont {K.}~\bibnamefont {Sharda}}, \bibinfo
  {author} {\bibfnamefont {F.}~\bibnamefont {Paolucci}}, \bibinfo {author}
  {\bibfnamefont {J.~R.}\ \bibnamefont {Nair}}, \bibinfo {author}
  {\bibfnamefont {C.}~\bibnamefont {Gerbaldi}}, \bibinfo {author}
  {\bibfnamefont {D.}~\bibnamefont {Daghero}},\ and\ \bibinfo {author}
  {\bibfnamefont {R.~S.}\ \bibnamefont {Gonnelli}},\ }\bibfield  {title}
  {\bibinfo {title} {Huge field-effect surface charge injection and conductance
  modulation in metallic thin films by electrochemical gating},\ }\href
  {https://doi.org/https://doi.org/10.1016/j.apsusc.2012.09.157} {\bibfield
  {journal} {\bibinfo  {journal} {Appl. Surf. Sci.}\ }\textbf {\bibinfo
  {volume} {269}},\ \bibinfo {pages} {17} (\bibinfo {year} {2013})}\BibitemShut
  {NoStop}%
\bibitem [{\citenamefont {Shimizu}\ \emph {et~al.}(2013)\citenamefont
  {Shimizu}, \citenamefont {Takahashi}, \citenamefont {Hatano}, \citenamefont
  {Kawasaki}, \citenamefont {Tokura},\ and\ \citenamefont
  {Iwasa}}]{ShimizuPRL2013}%
  \BibitemOpen
  \bibfield  {author} {\bibinfo {author} {\bibfnamefont {S.}~\bibnamefont
  {Shimizu}}, \bibinfo {author} {\bibfnamefont {K.~S.}\ \bibnamefont
  {Takahashi}}, \bibinfo {author} {\bibfnamefont {T.}~\bibnamefont {Hatano}},
  \bibinfo {author} {\bibfnamefont {M.}~\bibnamefont {Kawasaki}}, \bibinfo
  {author} {\bibfnamefont {Y.}~\bibnamefont {Tokura}},\ and\ \bibinfo {author}
  {\bibfnamefont {Y.}~\bibnamefont {Iwasa}},\ }\bibfield  {title} {\bibinfo
  {title} {Electrically tunable anomalous {Hall} effect in {Pt} thin films},\
  }\href {https://doi.org/https://doi.org/10.1103/PhysRevLett.111.216803}
  {\bibfield  {journal} {\bibinfo  {journal} {Phys. Rev. Lett.}\ }\textbf
  {\bibinfo {volume} {111}},\ \bibinfo {pages} {216803} (\bibinfo {year}
  {2013})}\BibitemShut {NoStop}%
\bibitem [{\citenamefont {Dushenko}\ \emph {et~al.}(2018)\citenamefont
  {Dushenko}, \citenamefont {Hokazono}, \citenamefont {Nakamura}, \citenamefont
  {Ando}, \citenamefont {Shinjo},\ and\ \citenamefont
  {Shiraishi}}]{DushenkoNatCommun2018}%
  \BibitemOpen
  \bibfield  {author} {\bibinfo {author} {\bibfnamefont {S.}~\bibnamefont
  {Dushenko}}, \bibinfo {author} {\bibfnamefont {M.}~\bibnamefont {Hokazono}},
  \bibinfo {author} {\bibfnamefont {K.}~\bibnamefont {Nakamura}}, \bibinfo
  {author} {\bibfnamefont {Y.}~\bibnamefont {Ando}}, \bibinfo {author}
  {\bibfnamefont {T.}~\bibnamefont {Shinjo}},\ and\ \bibinfo {author}
  {\bibfnamefont {M.}~\bibnamefont {Shiraishi}},\ }\bibfield  {title} {\bibinfo
  {title} {Tunable inverse spin {Hall} effect in nanometer-thick platinum films
  by ionic gating},\ }\href
  {https://doi.org/https://doi.org/10.1038/s41467-018-05611-9} {\bibfield
  {journal} {\bibinfo  {journal} {Nat. Commun.}\ }\textbf {\bibinfo {volume}
  {9}},\ \bibinfo {pages} {3118} (\bibinfo {year} {2018})}\BibitemShut
  {NoStop}%
\bibitem [{\citenamefont {Liang}\ \emph
  {et~al.}(2018{\natexlab{a}})\citenamefont {Liang}, \citenamefont {Chen},
  \citenamefont {Lu}, \citenamefont {Talsma}, \citenamefont {Shan},
  \citenamefont {Blake}, \citenamefont {Palstra},\ and\ \citenamefont
  {Ye}}]{LiangSciAdv2018}%
  \BibitemOpen
  \bibfield  {author} {\bibinfo {author} {\bibfnamefont {L.}~\bibnamefont
  {Liang}}, \bibinfo {author} {\bibfnamefont {Q.}~\bibnamefont {Chen}},
  \bibinfo {author} {\bibfnamefont {J.}~\bibnamefont {Lu}}, \bibinfo {author}
  {\bibfnamefont {W.}~\bibnamefont {Talsma}}, \bibinfo {author} {\bibfnamefont
  {J.}~\bibnamefont {Shan}}, \bibinfo {author} {\bibfnamefont {G.~R.}\
  \bibnamefont {Blake}}, \bibinfo {author} {\bibfnamefont {T.~T.}\ \bibnamefont
  {Palstra}},\ and\ \bibinfo {author} {\bibfnamefont {J.}~\bibnamefont {Ye}},\
  }\bibfield  {title} {\bibinfo {title} {Inducing ferromagnetism and kondo
  effect in platinum by paramagnetic ionic gating},\ }\href
  {https://doi.org/https://doi.org/10.1126/sciadv.aar2030} {\bibfield
  {journal} {\bibinfo  {journal} {Sci. Adv.}\ }\textbf {\bibinfo {volume}
  {4}},\ \bibinfo {pages} {eaar2030} (\bibinfo {year}
  {2018}{\natexlab{a}})}\BibitemShut {NoStop}%
\bibitem [{\citenamefont {Liang}\ \emph
  {et~al.}(2018{\natexlab{b}})\citenamefont {Liang}, \citenamefont {Shan},
  \citenamefont {Chen}, \citenamefont {Lu}, \citenamefont {Blake},
  \citenamefont {Palstra}, \citenamefont {Bauer}, \citenamefont {Van~Wees},\
  and\ \citenamefont {Ye}}]{LiangPRB2018}%
  \BibitemOpen
  \bibfield  {author} {\bibinfo {author} {\bibfnamefont {L.}~\bibnamefont
  {Liang}}, \bibinfo {author} {\bibfnamefont {J.}~\bibnamefont {Shan}},
  \bibinfo {author} {\bibfnamefont {Q.}~\bibnamefont {Chen}}, \bibinfo {author}
  {\bibfnamefont {J.}~\bibnamefont {Lu}}, \bibinfo {author} {\bibfnamefont
  {G.~R.}\ \bibnamefont {Blake}}, \bibinfo {author} {\bibfnamefont {T.~T.}\
  \bibnamefont {Palstra}}, \bibinfo {author} {\bibfnamefont {G.~E.}\
  \bibnamefont {Bauer}}, \bibinfo {author} {\bibfnamefont {B.}~\bibnamefont
  {Van~Wees}},\ and\ \bibinfo {author} {\bibfnamefont {J.}~\bibnamefont {Ye}},\
  }\bibfield  {title} {\bibinfo {title} {Gate-controlled magnetoresistance of a
  paramagnetic-insulator|platinum interface},\ }\href
  {https://doi.org/https://doi.org/10.1103/PhysRevB.98.134402} {\bibfield
  {journal} {\bibinfo  {journal} {Phys. Rev. B}\ }\textbf {\bibinfo {volume}
  {98}},\ \bibinfo {pages} {134402} (\bibinfo {year}
  {2018}{\natexlab{b}})}\BibitemShut {NoStop}%
\bibitem [{\citenamefont {Piatti}(2021)}]{PiattiNE2021}%
  \BibitemOpen
  \bibfield  {author} {\bibinfo {author} {\bibfnamefont {E.}~\bibnamefont
  {Piatti}},\ }\bibfield  {title} {\bibinfo {title} {Ionic gating in metallic
  superconductors: {A} brief review},\ }\href
  {https://doi.org/https://doi.org/10.1088/2632-959X/ac011d} {\bibfield
  {journal} {\bibinfo  {journal} {Nano Ex.}\ }\textbf {\bibinfo {volume} {2}},\
  \bibinfo {pages} {024003} (\bibinfo {year} {2021})}\BibitemShut {NoStop}%
\bibitem [{\citenamefont {Choi}\ \emph {et~al.}(2014)\citenamefont {Choi},
  \citenamefont {Pradheesh}, \citenamefont {Kim}, \citenamefont {Im},
  \citenamefont {Chong},\ and\ \citenamefont {Chae}}]{ChoiAPL2014}%
  \BibitemOpen
  \bibfield  {author} {\bibinfo {author} {\bibfnamefont {J.}~\bibnamefont
  {Choi}}, \bibinfo {author} {\bibfnamefont {R.}~\bibnamefont {Pradheesh}},
  \bibinfo {author} {\bibfnamefont {H.}~\bibnamefont {Kim}}, \bibinfo {author}
  {\bibfnamefont {H.}~\bibnamefont {Im}}, \bibinfo {author} {\bibfnamefont
  {Y.}~\bibnamefont {Chong}},\ and\ \bibinfo {author} {\bibfnamefont {D.-H.}\
  \bibnamefont {Chae}},\ }\bibfield  {title} {\bibinfo {title} {Electrical
  modulation of superconducting critical temperature in liquid-gated thin
  niobium films},\ }\href {https://doi.org/https://doi.org/10.1063/1.4890085}
  {\bibfield  {journal} {\bibinfo  {journal} {Appl. Phys. Lett.}\ }\textbf
  {\bibinfo {volume} {105}},\ \bibinfo {pages} {012601} (\bibinfo {year}
  {2014})}\BibitemShut {NoStop}%
\bibitem [{\citenamefont {Piatti}\ \emph {et~al.}(2016)\citenamefont {Piatti},
  \citenamefont {Sola}, \citenamefont {Daghero}, \citenamefont {Ummarino},
  \citenamefont {Laviano}, \citenamefont {Nair}, \citenamefont {Gerbaldi},
  \citenamefont {Cristiano}, \citenamefont {Casaburi},\ and\ \citenamefont
  {Gonnelli}}]{PiattiJSNM2016}%
  \BibitemOpen
  \bibfield  {author} {\bibinfo {author} {\bibfnamefont {E.}~\bibnamefont
  {Piatti}}, \bibinfo {author} {\bibfnamefont {A.}~\bibnamefont {Sola}},
  \bibinfo {author} {\bibfnamefont {D.}~\bibnamefont {Daghero}}, \bibinfo
  {author} {\bibfnamefont {G.~A.}\ \bibnamefont {Ummarino}}, \bibinfo {author}
  {\bibfnamefont {F.}~\bibnamefont {Laviano}}, \bibinfo {author} {\bibfnamefont
  {J.~R.}\ \bibnamefont {Nair}}, \bibinfo {author} {\bibfnamefont
  {C.}~\bibnamefont {Gerbaldi}}, \bibinfo {author} {\bibfnamefont
  {R.}~\bibnamefont {Cristiano}}, \bibinfo {author} {\bibfnamefont
  {A.}~\bibnamefont {Casaburi}},\ and\ \bibinfo {author} {\bibfnamefont
  {R.~S.}\ \bibnamefont {Gonnelli}},\ }\bibfield  {title} {\bibinfo {title}
  {Superconducting transition temperature modulation in {NbN} via {EDL}
  gating},\ }\href {https://doi.org/https://doi.org/10.1007/s10948-015-3306-0}
  {\bibfield  {journal} {\bibinfo  {journal} {J. Supercond. Novel Magn.}\
  }\textbf {\bibinfo {volume} {29}},\ \bibinfo {pages} {587} (\bibinfo {year}
  {2016})}\BibitemShut {NoStop}%
\bibitem [{\citenamefont {Piatti}\ \emph
  {et~al.}(2017{\natexlab{a}})\citenamefont {Piatti}, \citenamefont {Daghero},
  \citenamefont {Ummarino}, \citenamefont {Laviano}, \citenamefont {Nair},
  \citenamefont {Cristiano}, \citenamefont {Casaburi}, \citenamefont {Portesi},
  \citenamefont {Sola},\ and\ \citenamefont {Gonnelli}}]{PiattiPRB2017}%
  \BibitemOpen
  \bibfield  {author} {\bibinfo {author} {\bibfnamefont {E.}~\bibnamefont
  {Piatti}}, \bibinfo {author} {\bibfnamefont {D.}~\bibnamefont {Daghero}},
  \bibinfo {author} {\bibfnamefont {G.~A.}\ \bibnamefont {Ummarino}}, \bibinfo
  {author} {\bibfnamefont {F.}~\bibnamefont {Laviano}}, \bibinfo {author}
  {\bibfnamefont {J.~R.}\ \bibnamefont {Nair}}, \bibinfo {author}
  {\bibfnamefont {R.}~\bibnamefont {Cristiano}}, \bibinfo {author}
  {\bibfnamefont {A.}~\bibnamefont {Casaburi}}, \bibinfo {author}
  {\bibfnamefont {C.}~\bibnamefont {Portesi}}, \bibinfo {author} {\bibfnamefont
  {A.}~\bibnamefont {Sola}},\ and\ \bibinfo {author} {\bibfnamefont {R.~S.}\
  \bibnamefont {Gonnelli}},\ }\bibfield  {title} {\bibinfo {title} {Control of
  bulk superconductivity in a {BCS} superconductor by surface charge doping via
  electrochemical gating},\ }\href
  {https://doi.org/https://doi.org/10.1103/PhysRevB.95.140501} {\bibfield
  {journal} {\bibinfo  {journal} {Phys. Rev. B}\ }\textbf {\bibinfo {volume}
  {95}},\ \bibinfo {pages} {140501(R)} (\bibinfo {year}
  {2017}{\natexlab{a}})}\BibitemShut {NoStop}%
\bibitem [{\citenamefont {Paolucci}\ \emph {et~al.}(2021)\citenamefont
  {Paolucci}, \citenamefont {Cris{\'a}}, \citenamefont {De~Simoni},
  \citenamefont {Bours}, \citenamefont {Puglia}, \citenamefont {Strambini},
  \citenamefont {Roddaro},\ and\ \citenamefont {Giazotto}}]{PaolucciNL2021}%
  \BibitemOpen
  \bibfield  {author} {\bibinfo {author} {\bibfnamefont {F.}~\bibnamefont
  {Paolucci}}, \bibinfo {author} {\bibfnamefont {F.}~\bibnamefont {Cris{\'a}}},
  \bibinfo {author} {\bibfnamefont {G.}~\bibnamefont {De~Simoni}}, \bibinfo
  {author} {\bibfnamefont {L.}~\bibnamefont {Bours}}, \bibinfo {author}
  {\bibfnamefont {C.}~\bibnamefont {Puglia}}, \bibinfo {author} {\bibfnamefont
  {E.}~\bibnamefont {Strambini}}, \bibinfo {author} {\bibfnamefont
  {S.}~\bibnamefont {Roddaro}},\ and\ \bibinfo {author} {\bibfnamefont
  {F.}~\bibnamefont {Giazotto}},\ }\bibfield  {title} {\bibinfo {title}
  {Electrostatic field-driven supercurrent suppression in ionic-gated metallic
  superconducting nanotransistors},\ }\href
  {https://doi.org/https://doi.org/10.1021/acs.nanolett.1c03481} {\bibfield
  {journal} {\bibinfo  {journal} {Nano Lett.}\ }\textbf {\bibinfo {volume}
  {21}},\ \bibinfo {pages} {10309} (\bibinfo {year} {2021})}\BibitemShut
  {NoStop}%
\bibitem [{\citenamefont {Yoshida}\ \emph {et~al.}(2016)\citenamefont
  {Yoshida}, \citenamefont {Ye}, \citenamefont {Nishizaki}, \citenamefont
  {Kobayashi},\ and\ \citenamefont {Iwasa}}]{YoshidaAPL2016}%
  \BibitemOpen
  \bibfield  {author} {\bibinfo {author} {\bibfnamefont {M.}~\bibnamefont
  {Yoshida}}, \bibinfo {author} {\bibfnamefont {J.}~\bibnamefont {Ye}},
  \bibinfo {author} {\bibfnamefont {T.}~\bibnamefont {Nishizaki}}, \bibinfo
  {author} {\bibfnamefont {N.}~\bibnamefont {Kobayashi}},\ and\ \bibinfo
  {author} {\bibfnamefont {Y.}~\bibnamefont {Iwasa}},\ }\bibfield  {title}
  {\bibinfo {title} {Electrostatic and electrochemical tuning of
  superconductivity in two-dimensional {NbSe\ped{2}} crystals},\ }\href
  {https://doi.org/https://doi.org/10.1063/1.4950804} {\bibfield  {journal}
  {\bibinfo  {journal} {Appl. Phys. Lett.}\ }\textbf {\bibinfo {volume}
  {108}},\ \bibinfo {pages} {202602} (\bibinfo {year} {2016})}\BibitemShut
  {NoStop}%
\bibitem [{\citenamefont {Li}\ \emph {et~al.}(2016)\citenamefont {Li},
  \citenamefont {O'Farrell}, \citenamefont {Loh}, \citenamefont {Eda},
  \citenamefont {{\"O}zyilmaz},\ and\ \citenamefont
  {Castro~Neto}}]{LiNature2016}%
  \BibitemOpen
  \bibfield  {author} {\bibinfo {author} {\bibfnamefont {L.~J.}\ \bibnamefont
  {Li}}, \bibinfo {author} {\bibfnamefont {E.~C.~T.}\ \bibnamefont
  {O'Farrell}}, \bibinfo {author} {\bibfnamefont {K.~P.}\ \bibnamefont {Loh}},
  \bibinfo {author} {\bibfnamefont {G.}~\bibnamefont {Eda}}, \bibinfo {author}
  {\bibfnamefont {B.}~\bibnamefont {{\"O}zyilmaz}},\ and\ \bibinfo {author}
  {\bibfnamefont {A.~H.}\ \bibnamefont {Castro~Neto}},\ }\bibfield  {title}
  {\bibinfo {title} {Controlling many-body states by the electric-field effect
  in a two-dimensional material},\ }\href
  {https://doi.org/https://doi.org/10.1038/nature16175} {\bibfield  {journal}
  {\bibinfo  {journal} {Nature}\ }\textbf {\bibinfo {volume} {529}},\ \bibinfo
  {pages} {185} (\bibinfo {year} {2016})}\BibitemShut {NoStop}%
\bibitem [{\citenamefont {Xi}\ \emph {et~al.}(2016)\citenamefont {Xi},
  \citenamefont {Berger}, \citenamefont {Forr{\'o}}, \citenamefont {Shan},\
  and\ \citenamefont {Mak}}]{XiPRL2016}%
  \BibitemOpen
  \bibfield  {author} {\bibinfo {author} {\bibfnamefont {X.}~\bibnamefont
  {Xi}}, \bibinfo {author} {\bibfnamefont {H.}~\bibnamefont {Berger}}, \bibinfo
  {author} {\bibfnamefont {L.}~\bibnamefont {Forr{\'o}}}, \bibinfo {author}
  {\bibfnamefont {J.}~\bibnamefont {Shan}},\ and\ \bibinfo {author}
  {\bibfnamefont {K.~F.}\ \bibnamefont {Mak}},\ }\bibfield  {title} {\bibinfo
  {title} {Gate tuning of electronic phase transitions in two-dimensional
  {NbSe\ped{2}}},\ }\href
  {https://doi.org/https://doi.org/10.1103/PhysRevLett.117.106801} {\bibfield
  {journal} {\bibinfo  {journal} {Phys. Rev. Lett.}\ }\textbf {\bibinfo
  {volume} {117}},\ \bibinfo {pages} {106801} (\bibinfo {year}
  {2016})}\BibitemShut {NoStop}%
\bibitem [{\citenamefont {Zhu}\ \emph {et~al.}(2017)\citenamefont {Zhu},
  \citenamefont {Cui}, \citenamefont {Lei}, \citenamefont {Wang}, \citenamefont
  {Shang}, \citenamefont {Meng}, \citenamefont {Ma}, \citenamefont {Luo},
  \citenamefont {Wu}, \citenamefont {Sun} \emph {et~al.}}]{ZhuPRB2017}%
  \BibitemOpen
  \bibfield  {author} {\bibinfo {author} {\bibfnamefont {C.}~\bibnamefont
  {Zhu}}, \bibinfo {author} {\bibfnamefont {J.}~\bibnamefont {Cui}}, \bibinfo
  {author} {\bibfnamefont {B.}~\bibnamefont {Lei}}, \bibinfo {author}
  {\bibfnamefont {N.}~\bibnamefont {Wang}}, \bibinfo {author} {\bibfnamefont
  {C.}~\bibnamefont {Shang}}, \bibinfo {author} {\bibfnamefont
  {F.}~\bibnamefont {Meng}}, \bibinfo {author} {\bibfnamefont {L.}~\bibnamefont
  {Ma}}, \bibinfo {author} {\bibfnamefont {X.}~\bibnamefont {Luo}}, \bibinfo
  {author} {\bibfnamefont {T.}~\bibnamefont {Wu}}, \bibinfo {author}
  {\bibfnamefont {Z.}~\bibnamefont {Sun}}, \emph {et~al.},\ }\bibfield  {title}
  {\bibinfo {title} {Tuning electronic properties of {FeSe\ped{0.5}Te\ped{0.5}}
  thin flakes using a solid ion conductor field-effect transistor},\ }\href
  {https://doi.org/https://doi.org/10.1103/PhysRevB.95.174513} {\bibfield
  {journal} {\bibinfo  {journal} {Phys. Rev. B}\ }\textbf {\bibinfo {volume}
  {95}},\ \bibinfo {pages} {174513} (\bibinfo {year} {2017})}\BibitemShut
  {NoStop}%
\bibitem [{\citenamefont {Shiogai}\ \emph {et~al.}(2016)\citenamefont
  {Shiogai}, \citenamefont {Ito}, \citenamefont {Mitsuhashi}, \citenamefont
  {Nojima},\ and\ \citenamefont {Tsukazaki}}]{ShiogaiNatPhys2016}%
  \BibitemOpen
  \bibfield  {author} {\bibinfo {author} {\bibfnamefont {J.}~\bibnamefont
  {Shiogai}}, \bibinfo {author} {\bibfnamefont {Y.}~\bibnamefont {Ito}},
  \bibinfo {author} {\bibfnamefont {T.}~\bibnamefont {Mitsuhashi}}, \bibinfo
  {author} {\bibfnamefont {T.}~\bibnamefont {Nojima}},\ and\ \bibinfo {author}
  {\bibfnamefont {A.}~\bibnamefont {Tsukazaki}},\ }\bibfield  {title} {\bibinfo
  {title} {Electric-field-induced superconductivity in electrochemically etched
  ultrathin {FeSe} films on {SrTiO\ped{3}} and {MgO}},\ }\href
  {https://doi.org/https://doi.org/10.1038/nphys3530} {\bibfield  {journal}
  {\bibinfo  {journal} {Nat. Phys.}\ }\textbf {\bibinfo {volume} {12}},\
  \bibinfo {pages} {42} (\bibinfo {year} {2016})}\BibitemShut {NoStop}%
\bibitem [{\citenamefont {Lei}\ \emph {et~al.}(2016{\natexlab{a}})\citenamefont
  {Lei}, \citenamefont {Cui}, \citenamefont {Xiang}, \citenamefont {Shang},
  \citenamefont {Wang}, \citenamefont {Ye}, \citenamefont {Luo}, \citenamefont
  {Wu}, \citenamefont {Sun},\ and\ \citenamefont {Chen}}]{LeiPRL2016}%
  \BibitemOpen
  \bibfield  {author} {\bibinfo {author} {\bibfnamefont {B.}~\bibnamefont
  {Lei}}, \bibinfo {author} {\bibfnamefont {J.}~\bibnamefont {Cui}}, \bibinfo
  {author} {\bibfnamefont {Z.}~\bibnamefont {Xiang}}, \bibinfo {author}
  {\bibfnamefont {C.}~\bibnamefont {Shang}}, \bibinfo {author} {\bibfnamefont
  {N.}~\bibnamefont {Wang}}, \bibinfo {author} {\bibfnamefont {G.}~\bibnamefont
  {Ye}}, \bibinfo {author} {\bibfnamefont {X.}~\bibnamefont {Luo}}, \bibinfo
  {author} {\bibfnamefont {T.}~\bibnamefont {Wu}}, \bibinfo {author}
  {\bibfnamefont {Z.}~\bibnamefont {Sun}},\ and\ \bibinfo {author}
  {\bibfnamefont {X.}~\bibnamefont {Chen}},\ }\bibfield  {title} {\bibinfo
  {title} {Evolution of high-temperature superconductivity from a
  low-{T\ped{c}} phase tuned by carrier concentration in {FeSe} thin flakes},\
  }\href {https://doi.org/https://doi.org/10.1103/PhysRevLett.116.077002}
  {\bibfield  {journal} {\bibinfo  {journal} {Phys. Rev. Lett.}\ }\textbf
  {\bibinfo {volume} {116}},\ \bibinfo {pages} {077002} (\bibinfo {year}
  {2016}{\natexlab{a}})}\BibitemShut {NoStop}%
\bibitem [{\citenamefont {Hanzawa}\ \emph {et~al.}(2016)\citenamefont
  {Hanzawa}, \citenamefont {Sato}, \citenamefont {Hiramatsu}, \citenamefont
  {Kamiya},\ and\ \citenamefont {Hosono}}]{HanzawaPNAS2016}%
  \BibitemOpen
  \bibfield  {author} {\bibinfo {author} {\bibfnamefont {K.}~\bibnamefont
  {Hanzawa}}, \bibinfo {author} {\bibfnamefont {H.}~\bibnamefont {Sato}},
  \bibinfo {author} {\bibfnamefont {H.}~\bibnamefont {Hiramatsu}}, \bibinfo
  {author} {\bibfnamefont {T.}~\bibnamefont {Kamiya}},\ and\ \bibinfo {author}
  {\bibfnamefont {H.}~\bibnamefont {Hosono}},\ }\bibfield  {title} {\bibinfo
  {title} {Electric field-induced superconducting transition of insulating
  {FeSe} thin film at {35~K}},\ }\href
  {https://doi.org/https://doi.org/10.1073/pnas.1520810113} {\bibfield
  {journal} {\bibinfo  {journal} {Proc. Natl. Acad. Sci. USA}\ }\textbf
  {\bibinfo {volume} {113}},\ \bibinfo {pages} {3986} (\bibinfo {year}
  {2016})}\BibitemShut {NoStop}%
\bibitem [{\citenamefont {Miyakawa}\ \emph {et~al.}(2018)\citenamefont
  {Miyakawa}, \citenamefont {Shiogai}, \citenamefont {Shimizu}, \citenamefont
  {Matsumoto}, \citenamefont {Ito}, \citenamefont {Harada}, \citenamefont
  {Fujiwara}, \citenamefont {Nojima}, \citenamefont {Itoh}, \citenamefont
  {Aida} \emph {et~al.}}]{MiyakawaPRM2018}%
  \BibitemOpen
  \bibfield  {author} {\bibinfo {author} {\bibfnamefont {T.}~\bibnamefont
  {Miyakawa}}, \bibinfo {author} {\bibfnamefont {J.}~\bibnamefont {Shiogai}},
  \bibinfo {author} {\bibfnamefont {S.}~\bibnamefont {Shimizu}}, \bibinfo
  {author} {\bibfnamefont {M.}~\bibnamefont {Matsumoto}}, \bibinfo {author}
  {\bibfnamefont {Y.}~\bibnamefont {Ito}}, \bibinfo {author} {\bibfnamefont
  {T.}~\bibnamefont {Harada}}, \bibinfo {author} {\bibfnamefont
  {K.}~\bibnamefont {Fujiwara}}, \bibinfo {author} {\bibfnamefont
  {T.}~\bibnamefont {Nojima}}, \bibinfo {author} {\bibfnamefont
  {Y.}~\bibnamefont {Itoh}}, \bibinfo {author} {\bibfnamefont {T.}~\bibnamefont
  {Aida}}, \emph {et~al.},\ }\bibfield  {title} {\bibinfo {title} {Enhancement
  of superconducting transition temperature in {FeSe} electric-double-layer
  transistor with multivalent ionic liquids},\ }\href
  {https://doi.org/https://doi.org/10.1103/PhysRevMaterials.2.031801}
  {\bibfield  {journal} {\bibinfo  {journal} {Phys. Rev. Materials}\ }\textbf
  {\bibinfo {volume} {2}},\ \bibinfo {pages} {031801(R)} (\bibinfo {year}
  {2018})}\BibitemShut {NoStop}%
\bibitem [{\citenamefont {Kouno}\ \emph {et~al.}(2018)\citenamefont {Kouno},
  \citenamefont {Sato}, \citenamefont {Katayama}, \citenamefont {Ichinose},
  \citenamefont {Asami}, \citenamefont {Nabeshima}, \citenamefont {Imai},
  \citenamefont {Maeda},\ and\ \citenamefont {Ueno}}]{KounoSciRep2018}%
  \BibitemOpen
  \bibfield  {author} {\bibinfo {author} {\bibfnamefont {S.}~\bibnamefont
  {Kouno}}, \bibinfo {author} {\bibfnamefont {Y.}~\bibnamefont {Sato}},
  \bibinfo {author} {\bibfnamefont {Y.}~\bibnamefont {Katayama}}, \bibinfo
  {author} {\bibfnamefont {A.}~\bibnamefont {Ichinose}}, \bibinfo {author}
  {\bibfnamefont {D.}~\bibnamefont {Asami}}, \bibinfo {author} {\bibfnamefont
  {F.}~\bibnamefont {Nabeshima}}, \bibinfo {author} {\bibfnamefont
  {Y.}~\bibnamefont {Imai}}, \bibinfo {author} {\bibfnamefont {A.}~\bibnamefont
  {Maeda}},\ and\ \bibinfo {author} {\bibfnamefont {K.}~\bibnamefont {Ueno}},\
  }\bibfield  {title} {\bibinfo {title} {Superconductivity at {38~K} at an
  electrochemical interface between an ionic liquid and
  {FeSe\ped{0.8}Te\ped{0.2}} on various substrates},\ }\href
  {https://doi.org/https://doi.org/10.1038/s41598-018-33121-7} {\bibfield
  {journal} {\bibinfo  {journal} {Sci. Rep.}\ }\textbf {\bibinfo {volume}
  {8}},\ \bibinfo {pages} {14731} (\bibinfo {year} {2018})}\BibitemShut
  {NoStop}%
\bibitem [{\citenamefont {Piatti}\ \emph
  {et~al.}(2019{\natexlab{a}})\citenamefont {Piatti}, \citenamefont {Hatano},
  \citenamefont {Daghero}, \citenamefont {Galanti}, \citenamefont {Gerbaldi},
  \citenamefont {Guastella}, \citenamefont {Portesi}, \citenamefont {Nakamura},
  \citenamefont {Fujimoto}, \citenamefont {Iida} \emph
  {et~al.}}]{PiattiPRM2019}%
  \BibitemOpen
  \bibfield  {author} {\bibinfo {author} {\bibfnamefont {E.}~\bibnamefont
  {Piatti}}, \bibinfo {author} {\bibfnamefont {T.}~\bibnamefont {Hatano}},
  \bibinfo {author} {\bibfnamefont {D.}~\bibnamefont {Daghero}}, \bibinfo
  {author} {\bibfnamefont {F.}~\bibnamefont {Galanti}}, \bibinfo {author}
  {\bibfnamefont {C.}~\bibnamefont {Gerbaldi}}, \bibinfo {author}
  {\bibfnamefont {S.}~\bibnamefont {Guastella}}, \bibinfo {author}
  {\bibfnamefont {C.}~\bibnamefont {Portesi}}, \bibinfo {author} {\bibfnamefont
  {I.}~\bibnamefont {Nakamura}}, \bibinfo {author} {\bibfnamefont
  {R.}~\bibnamefont {Fujimoto}}, \bibinfo {author} {\bibfnamefont
  {K.}~\bibnamefont {Iida}}, \emph {et~al.},\ }\bibfield  {title} {\bibinfo
  {title} {Ambipolar suppression of superconductivity by ionic gating in
  optimally doped {BaFe\ped{2}(As,P)\ped{2}} ultrathin films},\ }\href
  {https://doi.org/https://doi.org/10.1103/PhysRevMaterials.3.044801}
  {\bibfield  {journal} {\bibinfo  {journal} {Phys. Rev. Materials}\ }\textbf
  {\bibinfo {volume} {3}},\ \bibinfo {pages} {044801} (\bibinfo {year}
  {2019}{\natexlab{a}})}\BibitemShut {NoStop}%
\bibitem [{\citenamefont {Ueno}\ \emph {et~al.}(2014)\citenamefont {Ueno},
  \citenamefont {Shimotani}, \citenamefont {Yuan}, \citenamefont {Ye},
  \citenamefont {Kawasaki},\ and\ \citenamefont {Iwasa}}]{UenoJPSJ2014}%
  \BibitemOpen
  \bibfield  {author} {\bibinfo {author} {\bibfnamefont {K.}~\bibnamefont
  {Ueno}}, \bibinfo {author} {\bibfnamefont {H.}~\bibnamefont {Shimotani}},
  \bibinfo {author} {\bibfnamefont {H.}~\bibnamefont {Yuan}}, \bibinfo {author}
  {\bibfnamefont {J.}~\bibnamefont {Ye}}, \bibinfo {author} {\bibfnamefont
  {M.}~\bibnamefont {Kawasaki}},\ and\ \bibinfo {author} {\bibfnamefont
  {Y.}~\bibnamefont {Iwasa}},\ }\bibfield  {title} {\bibinfo {title}
  {Field-induced superconductivity in electric double layer transistors},\
  }\href {https://doi.org/https://doi.org/10.7566/JPSJ.83.032001} {\bibfield
  {journal} {\bibinfo  {journal} {J. Phys. Soc. Jpn.}\ }\textbf {\bibinfo
  {volume} {83}},\ \bibinfo {pages} {032001} (\bibinfo {year}
  {2014})}\BibitemShut {NoStop}%
\bibitem [{\citenamefont {Fujimoto}\ and\ \citenamefont
  {Awaga}(2013)}]{FujimotoPCCP2013}%
  \BibitemOpen
  \bibfield  {author} {\bibinfo {author} {\bibfnamefont {T.}~\bibnamefont
  {Fujimoto}}\ and\ \bibinfo {author} {\bibfnamefont {K.}~\bibnamefont
  {Awaga}},\ }\bibfield  {title} {\bibinfo {title} {Electric-double-layer
  field-effect transistors with ionic liquids},\ }\href
  {https://doi.org/https://doi.org/10.1039/C3CP50755F} {\bibfield  {journal}
  {\bibinfo  {journal} {Phys. Chem. Chem. Phys.}\ }\textbf {\bibinfo {volume}
  {15}},\ \bibinfo {pages} {8983} (\bibinfo {year} {2013})}\BibitemShut
  {NoStop}%
\bibitem [{\citenamefont {Gallagher}\ \emph {et~al.}(2015)\citenamefont
  {Gallagher}, \citenamefont {Lee}, \citenamefont {Petach}, \citenamefont
  {Stanwyck}, \citenamefont {Williams}, \citenamefont {Watanabe}, \citenamefont
  {Taniguchi},\ and\ \citenamefont
  {Goldhaber-Gordon}}]{GallagherNatCommun2015}%
  \BibitemOpen
  \bibfield  {author} {\bibinfo {author} {\bibfnamefont {P.}~\bibnamefont
  {Gallagher}}, \bibinfo {author} {\bibfnamefont {M.}~\bibnamefont {Lee}},
  \bibinfo {author} {\bibfnamefont {T.~A.}\ \bibnamefont {Petach}}, \bibinfo
  {author} {\bibfnamefont {S.~W.}\ \bibnamefont {Stanwyck}}, \bibinfo {author}
  {\bibfnamefont {J.~R.}\ \bibnamefont {Williams}}, \bibinfo {author}
  {\bibfnamefont {K.}~\bibnamefont {Watanabe}}, \bibinfo {author}
  {\bibfnamefont {T.}~\bibnamefont {Taniguchi}},\ and\ \bibinfo {author}
  {\bibfnamefont {D.}~\bibnamefont {Goldhaber-Gordon}},\ }\bibfield  {title}
  {\bibinfo {title} {A high-mobility electronic system at an electrolyte-gated
  oxide surface},\ }\href {https://doi.org/https://doi.org/10.1038/ncomms7437}
  {\bibfield  {journal} {\bibinfo  {journal} {Nat. Commun.}\ }\textbf {\bibinfo
  {volume} {6}},\ \bibinfo {pages} {6437} (\bibinfo {year} {2015})}\BibitemShut
  {NoStop}%
\bibitem [{\citenamefont {Piatti}\ \emph
  {et~al.}(2017{\natexlab{b}})\citenamefont {Piatti}, \citenamefont {Galasso},
  \citenamefont {Tortello}, \citenamefont {Nair}, \citenamefont {Gerbaldi},
  \citenamefont {Bruna}, \citenamefont {Borini}, \citenamefont {Daghero},\ and\
  \citenamefont {Gonnelli}}]{PiattiApSuSc2017}%
  \BibitemOpen
  \bibfield  {author} {\bibinfo {author} {\bibfnamefont {E.}~\bibnamefont
  {Piatti}}, \bibinfo {author} {\bibfnamefont {S.}~\bibnamefont {Galasso}},
  \bibinfo {author} {\bibfnamefont {M.}~\bibnamefont {Tortello}}, \bibinfo
  {author} {\bibfnamefont {J.~R.}\ \bibnamefont {Nair}}, \bibinfo {author}
  {\bibfnamefont {C.}~\bibnamefont {Gerbaldi}}, \bibinfo {author}
  {\bibfnamefont {M.}~\bibnamefont {Bruna}}, \bibinfo {author} {\bibfnamefont
  {S.}~\bibnamefont {Borini}}, \bibinfo {author} {\bibfnamefont
  {D.}~\bibnamefont {Daghero}},\ and\ \bibinfo {author} {\bibfnamefont {R.~S.}\
  \bibnamefont {Gonnelli}},\ }\bibfield  {title} {\bibinfo {title} {Carrier
  mobility and scattering lifetime in electric double-layer gated few-layer
  graphene},\ }\href
  {https://doi.org/https://doi.org/10.1016/j.apsusc.2016.06.192} {\bibfield
  {journal} {\bibinfo  {journal} {Appl. Surf. Sci.}\ }\textbf {\bibinfo
  {volume} {395}},\ \bibinfo {pages} {37} (\bibinfo {year}
  {2017}{\natexlab{b}})}\BibitemShut {NoStop}%
\bibitem [{\citenamefont {Saito}\ and\ \citenamefont
  {Iwasa}(2015)}]{SaitoACSNano2015}%
  \BibitemOpen
  \bibfield  {author} {\bibinfo {author} {\bibfnamefont {Y.}~\bibnamefont
  {Saito}}\ and\ \bibinfo {author} {\bibfnamefont {Y.}~\bibnamefont {Iwasa}},\
  }\bibfield  {title} {\bibinfo {title} {Ambipolar insulator-to-metal
  transition in black phosphorus by ionic-liquid gating},\ }\href
  {https://doi.org/https://doi.org/10.1021/acsnano.5b00497} {\bibfield
  {journal} {\bibinfo  {journal} {ACS Nano}\ }\textbf {\bibinfo {volume} {9}},\
  \bibinfo {pages} {3192} (\bibinfo {year} {2015})}\BibitemShut {NoStop}%
\bibitem [{\citenamefont {Gonnelli}\ \emph {et~al.}(2017)\citenamefont
  {Gonnelli}, \citenamefont {Piatti}, \citenamefont {Sola}, \citenamefont
  {Tortello}, \citenamefont {Dolcini}, \citenamefont {Galasso}, \citenamefont
  {Nair}, \citenamefont {Gerbaldi}, \citenamefont {Cappelluti}, \citenamefont
  {Bruna} \emph {et~al.}}]{Gonnelli2dMater2017}%
  \BibitemOpen
  \bibfield  {author} {\bibinfo {author} {\bibfnamefont {R.~S.}\ \bibnamefont
  {Gonnelli}}, \bibinfo {author} {\bibfnamefont {E.}~\bibnamefont {Piatti}},
  \bibinfo {author} {\bibfnamefont {A.}~\bibnamefont {Sola}}, \bibinfo {author}
  {\bibfnamefont {M.}~\bibnamefont {Tortello}}, \bibinfo {author}
  {\bibfnamefont {F.}~\bibnamefont {Dolcini}}, \bibinfo {author} {\bibfnamefont
  {S.}~\bibnamefont {Galasso}}, \bibinfo {author} {\bibfnamefont {J.~R.}\
  \bibnamefont {Nair}}, \bibinfo {author} {\bibfnamefont {C.}~\bibnamefont
  {Gerbaldi}}, \bibinfo {author} {\bibfnamefont {E.}~\bibnamefont
  {Cappelluti}}, \bibinfo {author} {\bibfnamefont {M.}~\bibnamefont {Bruna}},
  \emph {et~al.},\ }\bibfield  {title} {\bibinfo {title} {Weak localization in
  electric-double-layer gated few-layer graphene},\ }\href
  {https://doi.org/https://doi.org/10.1088/2053-1583/aa5afe} {\bibfield
  {journal} {\bibinfo  {journal} {2D Mater.}\ }\textbf {\bibinfo {volume}
  {4}},\ \bibinfo {pages} {035006} (\bibinfo {year} {2017})}\BibitemShut
  {NoStop}%
\bibitem [{\citenamefont {Piatti}\ \emph
  {et~al.}(2018{\natexlab{a}})\citenamefont {Piatti}, \citenamefont {De~Fazio},
  \citenamefont {Daghero}, \citenamefont {Tamalampudi}, \citenamefont {Yoon},
  \citenamefont {Ferrari},\ and\ \citenamefont {Gonnelli}}]{PiattiNL2018}%
  \BibitemOpen
  \bibfield  {author} {\bibinfo {author} {\bibfnamefont {E.}~\bibnamefont
  {Piatti}}, \bibinfo {author} {\bibfnamefont {D.}~\bibnamefont {De~Fazio}},
  \bibinfo {author} {\bibfnamefont {D.}~\bibnamefont {Daghero}}, \bibinfo
  {author} {\bibfnamefont {S.~R.}\ \bibnamefont {Tamalampudi}}, \bibinfo
  {author} {\bibfnamefont {D.}~\bibnamefont {Yoon}}, \bibinfo {author}
  {\bibfnamefont {A.~C.}\ \bibnamefont {Ferrari}},\ and\ \bibinfo {author}
  {\bibfnamefont {R.~S.}\ \bibnamefont {Gonnelli}},\ }\bibfield  {title}
  {\bibinfo {title} {Multi-valley superconductivity in ion-gated {MoS\ped{2}}
  layers},\ }\href
  {https://doi.org/https://doi.org/10.1021/acs.nanolett.8b01390} {\bibfield
  {journal} {\bibinfo  {journal} {Nano Lett.}\ }\textbf {\bibinfo {volume}
  {18}},\ \bibinfo {pages} {4821} (\bibinfo {year}
  {2018}{\natexlab{a}})}\BibitemShut {NoStop}%
\bibitem [{\citenamefont {Ovchinnikov}\ \emph {et~al.}(2016)\citenamefont
  {Ovchinnikov}, \citenamefont {Gargiulo}, \citenamefont {Allain},
  \citenamefont {Pasquier}, \citenamefont {Dumcenco}, \citenamefont {Ho},
  \citenamefont {Yazyev},\ and\ \citenamefont
  {Kis}}]{OvchinnikovNatCommun2016}%
  \BibitemOpen
  \bibfield  {author} {\bibinfo {author} {\bibfnamefont {D.}~\bibnamefont
  {Ovchinnikov}}, \bibinfo {author} {\bibfnamefont {F.}~\bibnamefont
  {Gargiulo}}, \bibinfo {author} {\bibfnamefont {A.}~\bibnamefont {Allain}},
  \bibinfo {author} {\bibfnamefont {D.~J.}\ \bibnamefont {Pasquier}}, \bibinfo
  {author} {\bibfnamefont {D.}~\bibnamefont {Dumcenco}}, \bibinfo {author}
  {\bibfnamefont {C.-H.}\ \bibnamefont {Ho}}, \bibinfo {author} {\bibfnamefont
  {O.~V.}\ \bibnamefont {Yazyev}},\ and\ \bibinfo {author} {\bibfnamefont
  {A.}~\bibnamefont {Kis}},\ }\bibfield  {title} {\bibinfo {title} {Disorder
  engineering and conductivity dome in {ReS\ped{2}} with electrolyte gating},\
  }\href {https://doi.org/https://doi.org/10.1038/ncomms12391} {\bibfield
  {journal} {\bibinfo  {journal} {Nat. Commun.}\ }\textbf {\bibinfo {volume}
  {7}},\ \bibinfo {pages} {12391} (\bibinfo {year} {2016})}\BibitemShut
  {NoStop}%
\bibitem [{\citenamefont {Piatti}\ \emph
  {et~al.}(2019{\natexlab{b}})\citenamefont {Piatti}, \citenamefont {Galanti},
  \citenamefont {Pippione}, \citenamefont {Pasquarelli},\ and\ \citenamefont
  {Gonnelli}}]{PiattiEPJ2019}%
  \BibitemOpen
  \bibfield  {author} {\bibinfo {author} {\bibfnamefont {E.}~\bibnamefont
  {Piatti}}, \bibinfo {author} {\bibfnamefont {F.}~\bibnamefont {Galanti}},
  \bibinfo {author} {\bibfnamefont {G.}~\bibnamefont {Pippione}}, \bibinfo
  {author} {\bibfnamefont {A.}~\bibnamefont {Pasquarelli}},\ and\ \bibinfo
  {author} {\bibfnamefont {R.~S.}\ \bibnamefont {Gonnelli}},\ }\bibfield
  {title} {\bibinfo {title} {Towards the insulator-to-metal transition at the
  surface of ion-gated nanocrystalline diamond films},\ }\href
  {https://doi.org/https://doi.org/10.1140/epjst/e2019-800188-9} {\bibfield
  {journal} {\bibinfo  {journal} {Eur. Phys. J. Spec. Top.}\ }\textbf {\bibinfo
  {volume} {228}},\ \bibinfo {pages} {689} (\bibinfo {year}
  {2019}{\natexlab{b}})}\BibitemShut {NoStop}%
\bibitem [{\citenamefont {Lu}\ \emph {et~al.}(2017{\natexlab{a}})\citenamefont
  {Lu}, \citenamefont {Zheliuk}, \citenamefont {Chen}, \citenamefont
  {Leermakers}, \citenamefont {Hussey}, \citenamefont {Zeitler},\ and\
  \citenamefont {Ye}}]{LuPNAS2018}%
  \BibitemOpen
  \bibfield  {author} {\bibinfo {author} {\bibfnamefont {J.}~\bibnamefont
  {Lu}}, \bibinfo {author} {\bibfnamefont {O.}~\bibnamefont {Zheliuk}},
  \bibinfo {author} {\bibfnamefont {Q.}~\bibnamefont {Chen}}, \bibinfo {author}
  {\bibfnamefont {I.}~\bibnamefont {Leermakers}}, \bibinfo {author}
  {\bibfnamefont {N.~E.}\ \bibnamefont {Hussey}}, \bibinfo {author}
  {\bibfnamefont {U.}~\bibnamefont {Zeitler}},\ and\ \bibinfo {author}
  {\bibfnamefont {J.}~\bibnamefont {Ye}},\ }\bibfield  {title} {\bibinfo
  {title} {Full superconducting dome of strong ising protection in gated
  monolayer {WS\ped{2}}},\ }\href
  {https://doi.org/https://doi.org/10.1073/pnas.1716781115} {\bibfield
  {journal} {\bibinfo  {journal} {Proc. Natl. Acad. Sci. USA}\ }\textbf
  {\bibinfo {volume} {115}},\ \bibinfo {pages} {3551} (\bibinfo {year}
  {2017}{\natexlab{a}})}\BibitemShut {NoStop}%
\bibitem [{\citenamefont {Piatti}\ \emph {et~al.}(2020)\citenamefont {Piatti},
  \citenamefont {Pasquarelli},\ and\ \citenamefont
  {Gonnelli}}]{PiattiApSuSc2020}%
  \BibitemOpen
  \bibfield  {author} {\bibinfo {author} {\bibfnamefont {E.}~\bibnamefont
  {Piatti}}, \bibinfo {author} {\bibfnamefont {A.}~\bibnamefont
  {Pasquarelli}},\ and\ \bibinfo {author} {\bibfnamefont {R.~S.}\ \bibnamefont
  {Gonnelli}},\ }\bibfield  {title} {\bibinfo {title} {Orientation-dependent
  electric transport and band filling in hole co-doped epitaxial diamond
  films},\ }\href
  {https://doi.org/https://doi.org/10.1016/j.apsusc.2020.146795} {\bibfield
  {journal} {\bibinfo  {journal} {Appl. Surf. Sci.}\ }\textbf {\bibinfo
  {volume} {528}},\ \bibinfo {pages} {146795} (\bibinfo {year}
  {2020})}\BibitemShut {NoStop}%
\bibitem [{\citenamefont {Zhang}\ \emph {et~al.}(2018)\citenamefont {Zhang},
  \citenamefont {Gao}, \citenamefont {Fu}, \citenamefont {Wang}, \citenamefont
  {Ren},\ and\ \citenamefont {Chen}}]{ZhangCPL2018}%
  \BibitemOpen
  \bibfield  {author} {\bibinfo {author} {\bibfnamefont {S.}~\bibnamefont
  {Zhang}}, \bibinfo {author} {\bibfnamefont {M.-R.}\ \bibnamefont {Gao}},
  \bibinfo {author} {\bibfnamefont {H.-Y.}\ \bibnamefont {Fu}}, \bibinfo
  {author} {\bibfnamefont {X.-M.}\ \bibnamefont {Wang}}, \bibinfo {author}
  {\bibfnamefont {Z.-A.}\ \bibnamefont {Ren}},\ and\ \bibinfo {author}
  {\bibfnamefont {G.-F.}\ \bibnamefont {Chen}},\ }\bibfield  {title} {\bibinfo
  {title} {Electric field induced permanent superconductivity in layered metal
  nitride chlorides {HfNCl} and {ZrNCl}},\ }\href
  {https://doi.org/https://doi.org/10.1088/0256-307X/35/9/097401} {\bibfield
  {journal} {\bibinfo  {journal} {Chin. Phys. Lett.}\ }\textbf {\bibinfo
  {volume} {35}},\ \bibinfo {pages} {097401} (\bibinfo {year}
  {2018})}\BibitemShut {NoStop}%
\bibitem [{\citenamefont {Wang}\ \emph {et~al.}(2019)\citenamefont {Wang},
  \citenamefont {Zhang}, \citenamefont {Fu}, \citenamefont {Gao}, \citenamefont
  {Ren},\ and\ \citenamefont {Chen}}]{WangNJP2019}%
  \BibitemOpen
  \bibfield  {author} {\bibinfo {author} {\bibfnamefont {X.}~\bibnamefont
  {Wang}}, \bibinfo {author} {\bibfnamefont {S.}~\bibnamefont {Zhang}},
  \bibinfo {author} {\bibfnamefont {H.}~\bibnamefont {Fu}}, \bibinfo {author}
  {\bibfnamefont {M.}~\bibnamefont {Gao}}, \bibinfo {author} {\bibfnamefont
  {Z.}~\bibnamefont {Ren}},\ and\ \bibinfo {author} {\bibfnamefont
  {G.}~\bibnamefont {Chen}},\ }\bibfield  {title} {\bibinfo {title} {Dominant
  role of processing temperature in electric field induced superconductivity in
  layered {ZrNBr}},\ }\href
  {https://doi.org/https://doi.org/10.1088/1367-2630/ab00c1} {\bibfield
  {journal} {\bibinfo  {journal} {New J. Phys.}\ }\textbf {\bibinfo {volume}
  {21}},\ \bibinfo {pages} {023002} (\bibinfo {year} {2019})}\BibitemShut
  {NoStop}%
\bibitem [{\citenamefont {Zakhidov}\ \emph {et~al.}(2020)\citenamefont
  {Zakhidov}, \citenamefont {Rehn}, \citenamefont {Reed},\ and\ \citenamefont
  {Salleo}}]{ZakidovACSNano2020}%
  \BibitemOpen
  \bibfield  {author} {\bibinfo {author} {\bibfnamefont {D.}~\bibnamefont
  {Zakhidov}}, \bibinfo {author} {\bibfnamefont {D.~A.}\ \bibnamefont {Rehn}},
  \bibinfo {author} {\bibfnamefont {E.~J.}\ \bibnamefont {Reed}},\ and\
  \bibinfo {author} {\bibfnamefont {A.}~\bibnamefont {Salleo}},\ }\bibfield
  {title} {\bibinfo {title} {Reversible electrochemical phase change in
  monolayer to bulk-like {MoTe\ped{2}} by ionic liquid gating},\ }\href
  {https://doi.org/https://doi.org/10.1021/acsnano.9b07095} {\bibfield
  {journal} {\bibinfo  {journal} {ACS Nano}\ }\textbf {\bibinfo {volume}
  {14}},\ \bibinfo {pages} {2894} (\bibinfo {year} {2020})}\BibitemShut
  {NoStop}%
\bibitem [{\citenamefont {Yu}\ \emph {et~al.}(2015)\citenamefont {Yu},
  \citenamefont {Yang}, \citenamefont {Lu}, \citenamefont {Yan}, \citenamefont
  {Cho}, \citenamefont {Ma}, \citenamefont {Niu}, \citenamefont {Kim},
  \citenamefont {Son}, \citenamefont {Feng} \emph {et~al.}}]{YuNatNano2015}%
  \BibitemOpen
  \bibfield  {author} {\bibinfo {author} {\bibfnamefont {Y.}~\bibnamefont
  {Yu}}, \bibinfo {author} {\bibfnamefont {F.}~\bibnamefont {Yang}}, \bibinfo
  {author} {\bibfnamefont {X.~F.}\ \bibnamefont {Lu}}, \bibinfo {author}
  {\bibfnamefont {Y.~J.}\ \bibnamefont {Yan}}, \bibinfo {author} {\bibfnamefont
  {Y.-H.}\ \bibnamefont {Cho}}, \bibinfo {author} {\bibfnamefont
  {L.}~\bibnamefont {Ma}}, \bibinfo {author} {\bibfnamefont {X.}~\bibnamefont
  {Niu}}, \bibinfo {author} {\bibfnamefont {S.}~\bibnamefont {Kim}}, \bibinfo
  {author} {\bibfnamefont {Y.-W.}\ \bibnamefont {Son}}, \bibinfo {author}
  {\bibfnamefont {D.}~\bibnamefont {Feng}}, \emph {et~al.},\ }\bibfield
  {title} {\bibinfo {title} {Gate-tunable phase transitions in thin flakes of
  {1T-TaS\ped{2}}},\ }\href
  {https://doi.org/https://doi.org/10.1038/nnano.2014.323} {\bibfield
  {journal} {\bibinfo  {journal} {Nat. Nanotechnol.}\ }\textbf {\bibinfo
  {volume} {10}},\ \bibinfo {pages} {270} (\bibinfo {year} {2015})}\BibitemShut
  {NoStop}%
\bibitem [{\citenamefont {Shi}\ \emph {et~al.}(2015)\citenamefont {Shi},
  \citenamefont {Ye}, \citenamefont {Zhang}, \citenamefont {Suzuki},
  \citenamefont {Yoshida}, \citenamefont {Miyazaki}, \citenamefont {Inoue},
  \citenamefont {Saito},\ and\ \citenamefont {Iwasa}}]{ShiSciRep2015}%
  \BibitemOpen
  \bibfield  {author} {\bibinfo {author} {\bibfnamefont {W.}~\bibnamefont
  {Shi}}, \bibinfo {author} {\bibfnamefont {J.}~\bibnamefont {Ye}}, \bibinfo
  {author} {\bibfnamefont {Y.}~\bibnamefont {Zhang}}, \bibinfo {author}
  {\bibfnamefont {R.}~\bibnamefont {Suzuki}}, \bibinfo {author} {\bibfnamefont
  {M.}~\bibnamefont {Yoshida}}, \bibinfo {author} {\bibfnamefont
  {J.}~\bibnamefont {Miyazaki}}, \bibinfo {author} {\bibfnamefont
  {N.}~\bibnamefont {Inoue}}, \bibinfo {author} {\bibfnamefont
  {Y.}~\bibnamefont {Saito}},\ and\ \bibinfo {author} {\bibfnamefont
  {Y.}~\bibnamefont {Iwasa}},\ }\bibfield  {title} {\bibinfo {title}
  {Superconductivity series in transition metal dichalcogenides by ionic
  gating},\ }\href {https://doi.org/https://doi.org/10.1038/srep12534}
  {\bibfield  {journal} {\bibinfo  {journal} {Sci. Rep.}\ }\textbf {\bibinfo
  {volume} {5}},\ \bibinfo {pages} {12534} (\bibinfo {year}
  {2015})}\BibitemShut {NoStop}%
\bibitem [{\citenamefont {Piatti}\ \emph
  {et~al.}(2017{\natexlab{c}})\citenamefont {Piatti}, \citenamefont {Chen},\
  and\ \citenamefont {Ye}}]{PiattiAPL2017}%
  \BibitemOpen
  \bibfield  {author} {\bibinfo {author} {\bibfnamefont {E.}~\bibnamefont
  {Piatti}}, \bibinfo {author} {\bibfnamefont {Q.}~\bibnamefont {Chen}},\ and\
  \bibinfo {author} {\bibfnamefont {J.}~\bibnamefont {Ye}},\ }\bibfield
  {title} {\bibinfo {title} {Strong dopant dependence of electric transport in
  ion-gated {MoS\ped{2}}},\ }\href
  {https://doi.org/https://doi.org/10.1063/1.4992477} {\bibfield  {journal}
  {\bibinfo  {journal} {Appl. Phys. Lett.}\ }\textbf {\bibinfo {volume}
  {111}},\ \bibinfo {pages} {013106} (\bibinfo {year}
  {2017}{\natexlab{c}})}\BibitemShut {NoStop}%
\bibitem [{\citenamefont {Piatti}\ \emph
  {et~al.}(2018{\natexlab{b}})\citenamefont {Piatti}, \citenamefont {Chen},
  \citenamefont {Tortello}, \citenamefont {Ye},\ and\ \citenamefont
  {Gonnelli}}]{PiattiApSuSc2018mos2}%
  \BibitemOpen
  \bibfield  {author} {\bibinfo {author} {\bibfnamefont {E.}~\bibnamefont
  {Piatti}}, \bibinfo {author} {\bibfnamefont {Q.}~\bibnamefont {Chen}},
  \bibinfo {author} {\bibfnamefont {M.}~\bibnamefont {Tortello}}, \bibinfo
  {author} {\bibfnamefont {J.}~\bibnamefont {Ye}},\ and\ \bibinfo {author}
  {\bibfnamefont {R.~S.}\ \bibnamefont {Gonnelli}},\ }\bibfield  {title}
  {\bibinfo {title} {Possible charge-density-wave signatures in the anomalous
  resistivity of li-intercalated multilayer {MoS\ped{2}}},\ }\href
  {https://doi.org/https://doi.org/10.1016/j.apsusc.2018.05.232} {\bibfield
  {journal} {\bibinfo  {journal} {Appl. Surf. Sci.}\ }\textbf {\bibinfo
  {volume} {461}},\ \bibinfo {pages} {269} (\bibinfo {year}
  {2018}{\natexlab{b}})}\BibitemShut {NoStop}%
\bibitem [{\citenamefont {Lei}\ \emph {et~al.}(2016{\natexlab{b}})\citenamefont
  {Lei}, \citenamefont {Xiang}, \citenamefont {Lu}, \citenamefont {Wang},
  \citenamefont {Chang}, \citenamefont {Shang}, \citenamefont {Zhang},
  \citenamefont {Zhang}, \citenamefont {Luo}, \citenamefont {Wu} \emph
  {et~al.}}]{LeiPRB2016}%
  \BibitemOpen
  \bibfield  {author} {\bibinfo {author} {\bibfnamefont {B.}~\bibnamefont
  {Lei}}, \bibinfo {author} {\bibfnamefont {Z.}~\bibnamefont {Xiang}}, \bibinfo
  {author} {\bibfnamefont {X.}~\bibnamefont {Lu}}, \bibinfo {author}
  {\bibfnamefont {N.}~\bibnamefont {Wang}}, \bibinfo {author} {\bibfnamefont
  {J.}~\bibnamefont {Chang}}, \bibinfo {author} {\bibfnamefont
  {C.}~\bibnamefont {Shang}}, \bibinfo {author} {\bibfnamefont
  {A.}~\bibnamefont {Zhang}}, \bibinfo {author} {\bibfnamefont
  {Q.}~\bibnamefont {Zhang}}, \bibinfo {author} {\bibfnamefont
  {X.}~\bibnamefont {Luo}}, \bibinfo {author} {\bibfnamefont {T.}~\bibnamefont
  {Wu}}, \emph {et~al.},\ }\bibfield  {title} {\bibinfo {title} {Gate-tuned
  superconductor-insulator transition in {(Li,Fe)OHFeSe}},\ }\href
  {https://doi.org/https://doi.org/10.1103/PhysRevB.93.060501} {\bibfield
  {journal} {\bibinfo  {journal} {Phys. Rev. B}\ }\textbf {\bibinfo {volume}
  {93}},\ \bibinfo {pages} {060501(R)} (\bibinfo {year}
  {2016}{\natexlab{b}})}\BibitemShut {NoStop}%
\bibitem [{\citenamefont {Lei}\ \emph {et~al.}(2017)\citenamefont {Lei},
  \citenamefont {Wang}, \citenamefont {Shang}, \citenamefont {Meng},
  \citenamefont {Ma}, \citenamefont {Luo}, \citenamefont {Wu}, \citenamefont
  {Sun}, \citenamefont {Wang}, \citenamefont {Jiang} \emph
  {et~al.}}]{LeiPRB2017}%
  \BibitemOpen
  \bibfield  {author} {\bibinfo {author} {\bibfnamefont {B.}~\bibnamefont
  {Lei}}, \bibinfo {author} {\bibfnamefont {N.}~\bibnamefont {Wang}}, \bibinfo
  {author} {\bibfnamefont {C.}~\bibnamefont {Shang}}, \bibinfo {author}
  {\bibfnamefont {F.}~\bibnamefont {Meng}}, \bibinfo {author} {\bibfnamefont
  {L.}~\bibnamefont {Ma}}, \bibinfo {author} {\bibfnamefont {X.}~\bibnamefont
  {Luo}}, \bibinfo {author} {\bibfnamefont {T.}~\bibnamefont {Wu}}, \bibinfo
  {author} {\bibfnamefont {Z.}~\bibnamefont {Sun}}, \bibinfo {author}
  {\bibfnamefont {Y.}~\bibnamefont {Wang}}, \bibinfo {author} {\bibfnamefont
  {Z.}~\bibnamefont {Jiang}}, \emph {et~al.},\ }\bibfield  {title} {\bibinfo
  {title} {Tuning phase transitions in fese thin flakes by field-effect
  transistor with solid ion conductor as the gate dielectric},\ }\href
  {https://doi.org/https://doi.org/10.1103/PhysRevB.95.020503} {\bibfield
  {journal} {\bibinfo  {journal} {Phys. Rev. B}\ }\textbf {\bibinfo {volume}
  {95}},\ \bibinfo {pages} {020503(R)} (\bibinfo {year} {2017})}\BibitemShut
  {NoStop}%
\bibitem [{\citenamefont {Wu}\ \emph {et~al.}(2018)\citenamefont {Wu},
  \citenamefont {Lian}, \citenamefont {He}, \citenamefont {Liu}, \citenamefont
  {Wang}, \citenamefont {Xing}, \citenamefont {Mao},\ and\ \citenamefont
  {Liu}}]{WuAPL2018}%
  \BibitemOpen
  \bibfield  {author} {\bibinfo {author} {\bibfnamefont {Y.}~\bibnamefont
  {Wu}}, \bibinfo {author} {\bibfnamefont {H.}~\bibnamefont {Lian}}, \bibinfo
  {author} {\bibfnamefont {J.}~\bibnamefont {He}}, \bibinfo {author}
  {\bibfnamefont {J.}~\bibnamefont {Liu}}, \bibinfo {author} {\bibfnamefont
  {S.}~\bibnamefont {Wang}}, \bibinfo {author} {\bibfnamefont {H.}~\bibnamefont
  {Xing}}, \bibinfo {author} {\bibfnamefont {Z.}~\bibnamefont {Mao}},\ and\
  \bibinfo {author} {\bibfnamefont {Y.}~\bibnamefont {Liu}},\ }\bibfield
  {title} {\bibinfo {title} {Lithium ion intercalation in thin crystals of
  hexagonal {TaSe\ped{2}} gated by a polymer electrolyte},\ }\href
  {https://doi.org/https://doi.org/10.1063/1.5008623} {\bibfield  {journal}
  {\bibinfo  {journal} {Appl. Phys. Lett.}\ }\textbf {\bibinfo {volume}
  {112}},\ \bibinfo {pages} {023502} (\bibinfo {year} {2018})}\BibitemShut
  {NoStop}%
\bibitem [{\citenamefont {Kwabena~Bediako}\ \emph {et~al.}(2018)\citenamefont
  {Kwabena~Bediako}, \citenamefont {Rezaee}, \citenamefont {Yoo}, \citenamefont
  {Larson}, \citenamefont {Zhao}, \citenamefont {Taniguchi}, \citenamefont
  {Watanabe}, \citenamefont {Brower-Thomas}, \citenamefont {Kaxiras},\ and\
  \citenamefont {Kim}}]{KwabenaNature2018}%
  \BibitemOpen
  \bibfield  {author} {\bibinfo {author} {\bibfnamefont {D.}~\bibnamefont
  {Kwabena~Bediako}}, \bibinfo {author} {\bibfnamefont {M.}~\bibnamefont
  {Rezaee}}, \bibinfo {author} {\bibfnamefont {H.}~\bibnamefont {Yoo}},
  \bibinfo {author} {\bibfnamefont {D.~T.}\ \bibnamefont {Larson}}, \bibinfo
  {author} {\bibfnamefont {S.~F.}\ \bibnamefont {Zhao}}, \bibinfo {author}
  {\bibfnamefont {T.}~\bibnamefont {Taniguchi}}, \bibinfo {author}
  {\bibfnamefont {K.}~\bibnamefont {Watanabe}}, \bibinfo {author}
  {\bibfnamefont {T.~L.}\ \bibnamefont {Brower-Thomas}}, \bibinfo {author}
  {\bibfnamefont {E.}~\bibnamefont {Kaxiras}},\ and\ \bibinfo {author}
  {\bibfnamefont {P.}~\bibnamefont {Kim}},\ }\bibfield  {title} {\bibinfo
  {title} {Heterointerface effects in the electrointercalation of van der
  {Waals} heterostructures},\ }\href
  {https://doi.org/https://doi.org/10.1038/s41586-018-0205-0} {\bibfield
  {journal} {\bibinfo  {journal} {Nature}\ }\textbf {\bibinfo {volume} {558}},\
  \bibinfo {pages} {425} (\bibinfo {year} {2018})}\BibitemShut {NoStop}%
\bibitem [{\citenamefont {Che}\ \emph {et~al.}(2019)\citenamefont {Che},
  \citenamefont {Deng}, \citenamefont {Fang}, \citenamefont {Pan},
  \citenamefont {Yu},\ and\ \citenamefont {Huang}}]{CheAEM2019}%
  \BibitemOpen
  \bibfield  {author} {\bibinfo {author} {\bibfnamefont {X.}~\bibnamefont
  {Che}}, \bibinfo {author} {\bibfnamefont {Y.}~\bibnamefont {Deng}}, \bibinfo
  {author} {\bibfnamefont {Y.}~\bibnamefont {Fang}}, \bibinfo {author}
  {\bibfnamefont {J.}~\bibnamefont {Pan}}, \bibinfo {author} {\bibfnamefont
  {Y.}~\bibnamefont {Yu}},\ and\ \bibinfo {author} {\bibfnamefont
  {F.}~\bibnamefont {Huang}},\ }\bibfield  {title} {\bibinfo {title}
  {Gate-tunable electrical transport in thin {2M-WS\ped{2}} flakes},\ }\href
  {https://doi.org/https://doi.org/10.1002/aelm.201900462} {\bibfield
  {journal} {\bibinfo  {journal} {Adv. Electron. Mater.}\ }\textbf {\bibinfo
  {volume} {5}},\ \bibinfo {pages} {1900462} (\bibinfo {year}
  {2019})}\BibitemShut {NoStop}%
\bibitem [{\citenamefont {Shang}\ \emph {et~al.}(2019)\citenamefont {Shang},
  \citenamefont {Lei}, \citenamefont {Zhuo}, \citenamefont {Zhang},
  \citenamefont {Zhu}, \citenamefont {Cui}, \citenamefont {Luo}, \citenamefont
  {Wang}, \citenamefont {Meng}, \citenamefont {Ma} \emph
  {et~al.}}]{ShangPRB2019}%
  \BibitemOpen
  \bibfield  {author} {\bibinfo {author} {\bibfnamefont {C.}~\bibnamefont
  {Shang}}, \bibinfo {author} {\bibfnamefont {B.}~\bibnamefont {Lei}}, \bibinfo
  {author} {\bibfnamefont {W.}~\bibnamefont {Zhuo}}, \bibinfo {author}
  {\bibfnamefont {Q.}~\bibnamefont {Zhang}}, \bibinfo {author} {\bibfnamefont
  {C.}~\bibnamefont {Zhu}}, \bibinfo {author} {\bibfnamefont {J.}~\bibnamefont
  {Cui}}, \bibinfo {author} {\bibfnamefont {X.}~\bibnamefont {Luo}}, \bibinfo
  {author} {\bibfnamefont {N.}~\bibnamefont {Wang}}, \bibinfo {author}
  {\bibfnamefont {F.}~\bibnamefont {Meng}}, \bibinfo {author} {\bibfnamefont
  {L.}~\bibnamefont {Ma}}, \emph {et~al.},\ }\bibfield  {title} {\bibinfo
  {title} {Structural and electronic phase transitions driven by electric field
  in metastable {MoS\ped{2}} thin flakes},\ }\href
  {https://doi.org/https://doi.org/10.1103/PhysRevB.100.020508} {\bibfield
  {journal} {\bibinfo  {journal} {Phys. Rev. B}\ }\textbf {\bibinfo {volume}
  {100}},\ \bibinfo {pages} {020508(R)} (\bibinfo {year} {2019})}\BibitemShut
  {NoStop}%
\bibitem [{\citenamefont {Song}\ \emph {et~al.}(2019)\citenamefont {Song},
  \citenamefont {Liang}, \citenamefont {Guo}, \citenamefont {Deng},
  \citenamefont {Gao},\ and\ \citenamefont {Chen}}]{SongPRM2019}%
  \BibitemOpen
  \bibfield  {author} {\bibinfo {author} {\bibfnamefont {Y.}~\bibnamefont
  {Song}}, \bibinfo {author} {\bibfnamefont {X.}~\bibnamefont {Liang}},
  \bibinfo {author} {\bibfnamefont {J.}~\bibnamefont {Guo}}, \bibinfo {author}
  {\bibfnamefont {J.}~\bibnamefont {Deng}}, \bibinfo {author} {\bibfnamefont
  {G.}~\bibnamefont {Gao}},\ and\ \bibinfo {author} {\bibfnamefont
  {X.}~\bibnamefont {Chen}},\ }\bibfield  {title} {\bibinfo {title}
  {Superconductivity in {Li}-intercalated {1T-SnSe\ped{2}} driven by electric
  field gating},\ }\href
  {https://doi.org/https://doi.org/10.1103/PhysRevMaterials.3.054804}
  {\bibfield  {journal} {\bibinfo  {journal} {Phys. Rev. Materials}\ }\textbf
  {\bibinfo {volume} {3}},\ \bibinfo {pages} {054804} (\bibinfo {year}
  {2019})}\BibitemShut {NoStop}%
\bibitem [{\citenamefont {Lu}\ \emph {et~al.}(2017{\natexlab{b}})\citenamefont
  {Lu}, \citenamefont {Zhang}, \citenamefont {Zhang}, \citenamefont {Qiao},
  \citenamefont {He}, \citenamefont {Li}, \citenamefont {Wang}, \citenamefont
  {Guo}, \citenamefont {Zhang}, \citenamefont {Duan} \emph
  {et~al.}}]{LuNature2017}%
  \BibitemOpen
  \bibfield  {author} {\bibinfo {author} {\bibfnamefont {N.}~\bibnamefont
  {Lu}}, \bibinfo {author} {\bibfnamefont {P.}~\bibnamefont {Zhang}}, \bibinfo
  {author} {\bibfnamefont {Q.}~\bibnamefont {Zhang}}, \bibinfo {author}
  {\bibfnamefont {R.}~\bibnamefont {Qiao}}, \bibinfo {author} {\bibfnamefont
  {Q.}~\bibnamefont {He}}, \bibinfo {author} {\bibfnamefont {H.-B.}\
  \bibnamefont {Li}}, \bibinfo {author} {\bibfnamefont {Y.}~\bibnamefont
  {Wang}}, \bibinfo {author} {\bibfnamefont {J.}~\bibnamefont {Guo}}, \bibinfo
  {author} {\bibfnamefont {D.}~\bibnamefont {Zhang}}, \bibinfo {author}
  {\bibfnamefont {Z.}~\bibnamefont {Duan}}, \emph {et~al.},\ }\bibfield
  {title} {\bibinfo {title} {Electric-field control of tri-state phase
  transformation with a selective dual-ion switch},\ }\href
  {https://doi.org/https://doi.org/10.1038/nature22389} {\bibfield  {journal}
  {\bibinfo  {journal} {Nature}\ }\textbf {\bibinfo {volume} {546}},\ \bibinfo
  {pages} {124} (\bibinfo {year} {2017}{\natexlab{b}})}\BibitemShut {NoStop}%
\bibitem [{\citenamefont {Leng}\ \emph {et~al.}(2017)\citenamefont {Leng},
  \citenamefont {Pereiro}, \citenamefont {Strle}, \citenamefont {Dubuis},
  \citenamefont {Bollinger}, \citenamefont {Gozar}, \citenamefont {Wu},
  \citenamefont {Litombe}, \citenamefont {Panagopoulos}, \citenamefont {Pavuna}
  \emph {et~al.}}]{LengNPJQM2017}%
  \BibitemOpen
  \bibfield  {author} {\bibinfo {author} {\bibfnamefont {X.}~\bibnamefont
  {Leng}}, \bibinfo {author} {\bibfnamefont {J.}~\bibnamefont {Pereiro}},
  \bibinfo {author} {\bibfnamefont {J.}~\bibnamefont {Strle}}, \bibinfo
  {author} {\bibfnamefont {G.}~\bibnamefont {Dubuis}}, \bibinfo {author}
  {\bibfnamefont {A.}~\bibnamefont {Bollinger}}, \bibinfo {author}
  {\bibfnamefont {A.}~\bibnamefont {Gozar}}, \bibinfo {author} {\bibfnamefont
  {J.}~\bibnamefont {Wu}}, \bibinfo {author} {\bibfnamefont {N.}~\bibnamefont
  {Litombe}}, \bibinfo {author} {\bibfnamefont {C.}~\bibnamefont
  {Panagopoulos}}, \bibinfo {author} {\bibfnamefont {D.}~\bibnamefont
  {Pavuna}}, \emph {et~al.},\ }\bibfield  {title} {\bibinfo {title} {Insulator
  to metal transition in {WO\ped{3}} induced by electrolyte gating},\ }\href
  {https://doi.org/https://doi.org/10.1038/s41535-017-0039-2} {\bibfield
  {journal} {\bibinfo  {journal} {npj Quant. Mater.}\ }\textbf {\bibinfo
  {volume} {2}},\ \bibinfo {pages} {35} (\bibinfo {year} {2017})}\BibitemShut
  {NoStop}%
\bibitem [{\citenamefont {Cui}\ \emph {et~al.}(2018)\citenamefont {Cui},
  \citenamefont {Zhang}, \citenamefont {Li}, \citenamefont {Lin}, \citenamefont
  {Zhu}, \citenamefont {Wen}, \citenamefont {Wang}, \citenamefont {Sun},
  \citenamefont {Ma}, \citenamefont {Li} \emph {et~al.}}]{CuiSB2018}%
  \BibitemOpen
  \bibfield  {author} {\bibinfo {author} {\bibfnamefont {Y.}~\bibnamefont
  {Cui}}, \bibinfo {author} {\bibfnamefont {G.}~\bibnamefont {Zhang}}, \bibinfo
  {author} {\bibfnamefont {H.}~\bibnamefont {Li}}, \bibinfo {author}
  {\bibfnamefont {H.}~\bibnamefont {Lin}}, \bibinfo {author} {\bibfnamefont
  {X.}~\bibnamefont {Zhu}}, \bibinfo {author} {\bibfnamefont {H.-H.}\
  \bibnamefont {Wen}}, \bibinfo {author} {\bibfnamefont {G.}~\bibnamefont
  {Wang}}, \bibinfo {author} {\bibfnamefont {J.}~\bibnamefont {Sun}}, \bibinfo
  {author} {\bibfnamefont {M.}~\bibnamefont {Ma}}, \bibinfo {author}
  {\bibfnamefont {Y.}~\bibnamefont {Li}}, \emph {et~al.},\ }\bibfield  {title}
  {\bibinfo {title} {Protonation induced high-{T\ped{c}} phases in iron-based
  superconductors evidenced by nmr and magnetization measurements},\ }\href
  {https://doi.org/https://doi.org/10.1016/j.scib.2017.12.009} {\bibfield
  {journal} {\bibinfo  {journal} {Sci. Bull.}\ }\textbf {\bibinfo {volume}
  {63}},\ \bibinfo {pages} {11} (\bibinfo {year} {2018})}\BibitemShut {NoStop}%
\bibitem [{\citenamefont {Rafique}\ \emph {et~al.}(2019)\citenamefont
  {Rafique}, \citenamefont {Feng}, \citenamefont {Lin}, \citenamefont {Wei},
  \citenamefont {Liao}, \citenamefont {Zhang}, \citenamefont {Jin},\ and\
  \citenamefont {Xue}}]{RafiqueNL2019}%
  \BibitemOpen
  \bibfield  {author} {\bibinfo {author} {\bibfnamefont {M.}~\bibnamefont
  {Rafique}}, \bibinfo {author} {\bibfnamefont {Z.}~\bibnamefont {Feng}},
  \bibinfo {author} {\bibfnamefont {Z.}~\bibnamefont {Lin}}, \bibinfo {author}
  {\bibfnamefont {X.}~\bibnamefont {Wei}}, \bibinfo {author} {\bibfnamefont
  {M.}~\bibnamefont {Liao}}, \bibinfo {author} {\bibfnamefont {D.}~\bibnamefont
  {Zhang}}, \bibinfo {author} {\bibfnamefont {K.}~\bibnamefont {Jin}},\ and\
  \bibinfo {author} {\bibfnamefont {Q.-K.}\ \bibnamefont {Xue}},\ }\bibfield
  {title} {\bibinfo {title} {Ionic liquid gating induced protonation of
  electron-doped cuprate superconductors},\ }\href
  {https://doi.org/https://doi.org/10.1021/acs.nanolett.9b02776} {\bibfield
  {journal} {\bibinfo  {journal} {Nano Lett.}\ }\textbf {\bibinfo {volume}
  {19}},\ \bibinfo {pages} {7775} (\bibinfo {year} {2019})}\BibitemShut
  {NoStop}%
\bibitem [{\citenamefont {Cui}\ \emph {et~al.}(2019)\citenamefont {Cui},
  \citenamefont {Hu}, \citenamefont {Zhang}, \citenamefont {Ma}, \citenamefont
  {Ma}, \citenamefont {Ma}, \citenamefont {Wang}, \citenamefont {Yan},
  \citenamefont {Sun}, \citenamefont {Cheng} \emph {et~al.}}]{CuiCPL2019}%
  \BibitemOpen
  \bibfield  {author} {\bibinfo {author} {\bibfnamefont {Y.}~\bibnamefont
  {Cui}}, \bibinfo {author} {\bibfnamefont {Z.}~\bibnamefont {Hu}}, \bibinfo
  {author} {\bibfnamefont {J.-S.}\ \bibnamefont {Zhang}}, \bibinfo {author}
  {\bibfnamefont {W.-L.}\ \bibnamefont {Ma}}, \bibinfo {author} {\bibfnamefont
  {M.-W.}\ \bibnamefont {Ma}}, \bibinfo {author} {\bibfnamefont
  {Z.}~\bibnamefont {Ma}}, \bibinfo {author} {\bibfnamefont {C.}~\bibnamefont
  {Wang}}, \bibinfo {author} {\bibfnamefont {J.-Q.}\ \bibnamefont {Yan}},
  \bibinfo {author} {\bibfnamefont {J.-P.}\ \bibnamefont {Sun}}, \bibinfo
  {author} {\bibfnamefont {J.-G.}\ \bibnamefont {Cheng}}, \emph {et~al.},\
  }\bibfield  {title} {\bibinfo {title} {Ionic-liquid-gating induced
  protonation and superconductivity in {FeSe}, {FeSe\ped{0.93}S\ped{0.07}},
  {ZrNCl}, {1T-TaS\ped{2}} and {Bi\ped{2}Se\ped{3}}},\ }\href
  {https://doi.org/https://doi.org/10.1088/0256-307X/36/7/077401} {\bibfield
  {journal} {\bibinfo  {journal} {Chin. Phys. Lett.}\ }\textbf {\bibinfo
  {volume} {36}},\ \bibinfo {pages} {077401} (\bibinfo {year}
  {2019})}\BibitemShut {NoStop}%
\bibitem [{\citenamefont {Li}\ \emph {et~al.}(2020)\citenamefont {Li},
  \citenamefont {Shen}, \citenamefont {Tian}, \citenamefont {Hwangbo},
  \citenamefont {Wang}, \citenamefont {Wang}, \citenamefont {Bartram},
  \citenamefont {He}, \citenamefont {Lyu}, \citenamefont {Dong} \emph
  {et~al.}}]{LiarXiv}%
  \BibitemOpen
  \bibfield  {author} {\bibinfo {author} {\bibfnamefont {Z.}~\bibnamefont
  {Li}}, \bibinfo {author} {\bibfnamefont {S.}~\bibnamefont {Shen}}, \bibinfo
  {author} {\bibfnamefont {Z.}~\bibnamefont {Tian}}, \bibinfo {author}
  {\bibfnamefont {K.}~\bibnamefont {Hwangbo}}, \bibinfo {author} {\bibfnamefont
  {M.}~\bibnamefont {Wang}}, \bibinfo {author} {\bibfnamefont {Y.}~\bibnamefont
  {Wang}}, \bibinfo {author} {\bibfnamefont {F.~M.}\ \bibnamefont {Bartram}},
  \bibinfo {author} {\bibfnamefont {L.}~\bibnamefont {He}}, \bibinfo {author}
  {\bibfnamefont {Y.}~\bibnamefont {Lyu}}, \bibinfo {author} {\bibfnamefont
  {Y.}~\bibnamefont {Dong}}, \emph {et~al.},\ }\bibfield  {title} {\bibinfo
  {title} {Reversible manipulation of the magnetic state in {SrRuO\ped{3}}
  through electric-field controlled proton evolution},\ }\href
  {https://doi.org/https://doi.org/10.1038/s41467-019-13999-1} {\bibfield
  {journal} {\bibinfo  {journal} {Nat. Commun.}\ }\textbf {\bibinfo {volume}
  {11}},\ \bibinfo {pages} {184} (\bibinfo {year} {2020})}\BibitemShut
  {NoStop}%
\bibitem [{\citenamefont {Meng}\ \emph {et~al.}(2022)\citenamefont {Meng},
  \citenamefont {Xing}, \citenamefont {Yi}, \citenamefont {Li}, \citenamefont
  {Zhou}, \citenamefont {Li}, \citenamefont {Zhang}, \citenamefont {Wei},
  \citenamefont {Feng}, \citenamefont {Terashima} \emph
  {et~al.}}]{MengPRB2022}%
  \BibitemOpen
  \bibfield  {author} {\bibinfo {author} {\bibfnamefont {Y.}~\bibnamefont
  {Meng}}, \bibinfo {author} {\bibfnamefont {X.}~\bibnamefont {Xing}}, \bibinfo
  {author} {\bibfnamefont {X.}~\bibnamefont {Yi}}, \bibinfo {author}
  {\bibfnamefont {B.}~\bibnamefont {Li}}, \bibinfo {author} {\bibfnamefont
  {N.}~\bibnamefont {Zhou}}, \bibinfo {author} {\bibfnamefont {M.}~\bibnamefont
  {Li}}, \bibinfo {author} {\bibfnamefont {Y.}~\bibnamefont {Zhang}}, \bibinfo
  {author} {\bibfnamefont {W.}~\bibnamefont {Wei}}, \bibinfo {author}
  {\bibfnamefont {J.}~\bibnamefont {Feng}}, \bibinfo {author} {\bibfnamefont
  {K.}~\bibnamefont {Terashima}}, \emph {et~al.},\ }\bibfield  {title}
  {\bibinfo {title} {Protonation-induced discrete superconducting phases in
  bulk {FeSe} single crystals},\ }\href
  {https://doi.org/https://doi.org/10.1103/PhysRevB.105.134506} {\bibfield
  {journal} {\bibinfo  {journal} {Phys. Rev. B}\ }\textbf {\bibinfo {volume}
  {105}},\ \bibinfo {pages} {134506} (\bibinfo {year} {2022})}\BibitemShut
  {NoStop}%
\bibitem [{Pia()}]{PiattiArXiv2022}%
  \BibitemOpen
  \href@noop {} {\bibinfo {title} {{E. Piatti, G. Prando, M. Meinero, C.
  Tresca, M. Putti, S. Roddaro, G. Lamura, T. Shiroka, P. Carretta, G. Profeta,
  D. Daghero and R. S. Gonnelli. Coexisting superconductivity and
  charge-density wave in hydrogen-doped titanium diselenide via ionic liquid
  gating-induced protonation.
  \href{https://doi.org/10.48550/arXiv.2205.12951}{arXiv:2205.12951}
  }}}\BibitemShut {NoStop}%
\bibitem [{\citenamefont {Jeong}\ \emph {et~al.}(2013)\citenamefont {Jeong},
  \citenamefont {Aetukuri}, \citenamefont {Graf}, \citenamefont {Schladt},
  \citenamefont {Samant},\ and\ \citenamefont {Parkin}}]{JeongScience2013}%
  \BibitemOpen
  \bibfield  {author} {\bibinfo {author} {\bibfnamefont {J.}~\bibnamefont
  {Jeong}}, \bibinfo {author} {\bibfnamefont {N.}~\bibnamefont {Aetukuri}},
  \bibinfo {author} {\bibfnamefont {T.}~\bibnamefont {Graf}}, \bibinfo {author}
  {\bibfnamefont {T.~D.}\ \bibnamefont {Schladt}}, \bibinfo {author}
  {\bibfnamefont {M.~G.}\ \bibnamefont {Samant}},\ and\ \bibinfo {author}
  {\bibfnamefont {S.~S.}\ \bibnamefont {Parkin}},\ }\bibfield  {title}
  {\bibinfo {title} {Suppression of metal-insulator transition in {VO\ped{2}}
  by electric field--induced oxygen vacancy formation},\ }\href
  {https://doi.org/https://doi.org/10.1126/science.1230512} {\bibfield
  {journal} {\bibinfo  {journal} {Science}\ }\textbf {\bibinfo {volume}
  {339}},\ \bibinfo {pages} {1402} (\bibinfo {year} {2013})}\BibitemShut
  {NoStop}%
\bibitem [{\citenamefont {Schladt}\ \emph {et~al.}(2013)\citenamefont
  {Schladt}, \citenamefont {Graf}, \citenamefont {Aetukuri}, \citenamefont
  {Li}, \citenamefont {Fantini}, \citenamefont {Jiang}, \citenamefont
  {Samant},\ and\ \citenamefont {Parkin}}]{SchladtACSNano2013}%
  \BibitemOpen
  \bibfield  {author} {\bibinfo {author} {\bibfnamefont {T.~D.}\ \bibnamefont
  {Schladt}}, \bibinfo {author} {\bibfnamefont {T.}~\bibnamefont {Graf}},
  \bibinfo {author} {\bibfnamefont {N.~B.}\ \bibnamefont {Aetukuri}}, \bibinfo
  {author} {\bibfnamefont {M.}~\bibnamefont {Li}}, \bibinfo {author}
  {\bibfnamefont {A.}~\bibnamefont {Fantini}}, \bibinfo {author} {\bibfnamefont
  {X.}~\bibnamefont {Jiang}}, \bibinfo {author} {\bibfnamefont {M.~G.}\
  \bibnamefont {Samant}},\ and\ \bibinfo {author} {\bibfnamefont {S.~S.}\
  \bibnamefont {Parkin}},\ }\bibfield  {title} {\bibinfo {title}
  {Crystal-facet-dependent metallization in electrolyte-gated rutile tio\ped{2}
  single crystals},\ }\href {https://doi.org/https://doi.org/10.1021/nn403340d}
  {\bibfield  {journal} {\bibinfo  {journal} {ACS Nano}\ }\textbf {\bibinfo
  {volume} {7}},\ \bibinfo {pages} {8074} (\bibinfo {year} {2013})}\BibitemShut
  {NoStop}%
\bibitem [{\citenamefont {Petach}\ \emph {et~al.}(2014)\citenamefont {Petach},
  \citenamefont {Lee}, \citenamefont {Davis}, \citenamefont {Mehta},\ and\
  \citenamefont {Goldhaber-Gordon}}]{PetachPRB2014}%
  \BibitemOpen
  \bibfield  {author} {\bibinfo {author} {\bibfnamefont {T.~A.}\ \bibnamefont
  {Petach}}, \bibinfo {author} {\bibfnamefont {M.}~\bibnamefont {Lee}},
  \bibinfo {author} {\bibfnamefont {R.~C.}\ \bibnamefont {Davis}}, \bibinfo
  {author} {\bibfnamefont {A.}~\bibnamefont {Mehta}},\ and\ \bibinfo {author}
  {\bibfnamefont {D.}~\bibnamefont {Goldhaber-Gordon}},\ }\bibfield  {title}
  {\bibinfo {title} {Mechanism for the large conductance modulation in
  electrolyte-gated thin gold films},\ }\href
  {https://doi.org/https://doi.org/10.1103/PhysRevB.90.081108} {\bibfield
  {journal} {\bibinfo  {journal} {Phys. Rev. B}\ }\textbf {\bibinfo {volume}
  {90}},\ \bibinfo {pages} {081108(R)} (\bibinfo {year} {2014})}\BibitemShut
  {NoStop}%
\bibitem [{\citenamefont {Maruyama}\ \emph {et~al.}(2015)\citenamefont
  {Maruyama}, \citenamefont {Shin}, \citenamefont {Zhang}, \citenamefont
  {Suchoski}, \citenamefont {Yasui}, \citenamefont {Jin}, \citenamefont
  {Greene},\ and\ \citenamefont {Takeuchi}}]{MaruyamaAPL2015}%
  \BibitemOpen
  \bibfield  {author} {\bibinfo {author} {\bibfnamefont {S.}~\bibnamefont
  {Maruyama}}, \bibinfo {author} {\bibfnamefont {J.}~\bibnamefont {Shin}},
  \bibinfo {author} {\bibfnamefont {X.}~\bibnamefont {Zhang}}, \bibinfo
  {author} {\bibfnamefont {R.}~\bibnamefont {Suchoski}}, \bibinfo {author}
  {\bibfnamefont {S.}~\bibnamefont {Yasui}}, \bibinfo {author} {\bibfnamefont
  {K.}~\bibnamefont {Jin}}, \bibinfo {author} {\bibfnamefont {R.}~\bibnamefont
  {Greene}},\ and\ \bibinfo {author} {\bibfnamefont {I.}~\bibnamefont
  {Takeuchi}},\ }\bibfield  {title} {\bibinfo {title} {Reversible
  electrochemical modulation of the superconducting transition temperature of
  {LiTi\ped{2}O\ped{4}} ultrathin films by ionic liquid gating},\ }\href
  {https://doi.org/https://doi.org/10.1063/1.4932551} {\bibfield  {journal}
  {\bibinfo  {journal} {Appl. Phys. Lett.}\ }\textbf {\bibinfo {volume}
  {107}},\ \bibinfo {pages} {142602} (\bibinfo {year} {2015})}\BibitemShut
  {NoStop}%
\bibitem [{\citenamefont {Walter}\ \emph {et~al.}(2016)\citenamefont {Walter},
  \citenamefont {Wang}, \citenamefont {Luo}, \citenamefont {Frisbie},\ and\
  \citenamefont {Leighton}}]{WalterACSNano2016}%
  \BibitemOpen
  \bibfield  {author} {\bibinfo {author} {\bibfnamefont {J.}~\bibnamefont
  {Walter}}, \bibinfo {author} {\bibfnamefont {H.}~\bibnamefont {Wang}},
  \bibinfo {author} {\bibfnamefont {B.}~\bibnamefont {Luo}}, \bibinfo {author}
  {\bibfnamefont {C.~D.}\ \bibnamefont {Frisbie}},\ and\ \bibinfo {author}
  {\bibfnamefont {C.}~\bibnamefont {Leighton}},\ }\bibfield  {title} {\bibinfo
  {title} {Electrostatic versus electrochemical doping and control of
  ferromagnetism in ion-gel-gated ultrathin
  {La\ped{0.5}Sr\ped{0.5}CoO\ped{3-\delta}}},\ }\href
  {https://doi.org/https://doi.org/10.1021/acsnano.6b03403} {\bibfield
  {journal} {\bibinfo  {journal} {ACS Nano}\ }\textbf {\bibinfo {volume}
  {10}},\ \bibinfo {pages} {7799} (\bibinfo {year} {2016})}\BibitemShut
  {NoStop}%
\bibitem [{\citenamefont {Zhang}\ \emph {et~al.}(2017)\citenamefont {Zhang},
  \citenamefont {Zeng}, \citenamefont {Yin}, \citenamefont {Asmara},
  \citenamefont {Yang}, \citenamefont {Han}, \citenamefont {Cao}, \citenamefont
  {Zhou}, \citenamefont {Wan}, \citenamefont {Tang} \emph
  {et~al.}}]{ZhangACSNAno2017}%
  \BibitemOpen
  \bibfield  {author} {\bibinfo {author} {\bibfnamefont {L.}~\bibnamefont
  {Zhang}}, \bibinfo {author} {\bibfnamefont {S.}~\bibnamefont {Zeng}},
  \bibinfo {author} {\bibfnamefont {X.}~\bibnamefont {Yin}}, \bibinfo {author}
  {\bibfnamefont {T.~C.}\ \bibnamefont {Asmara}}, \bibinfo {author}
  {\bibfnamefont {P.}~\bibnamefont {Yang}}, \bibinfo {author} {\bibfnamefont
  {K.}~\bibnamefont {Han}}, \bibinfo {author} {\bibfnamefont {Y.}~\bibnamefont
  {Cao}}, \bibinfo {author} {\bibfnamefont {W.}~\bibnamefont {Zhou}}, \bibinfo
  {author} {\bibfnamefont {D.}~\bibnamefont {Wan}}, \bibinfo {author}
  {\bibfnamefont {C.~S.}\ \bibnamefont {Tang}}, \emph {et~al.},\ }\bibfield
  {title} {\bibinfo {title} {The mechanism of electrolyte gating on
  high-t\ped{c} cuprates: the role of oxygen migration and electrostatics},\
  }\href {https://doi.org/https://doi.org/10.1021/acsnano.7b03978} {\bibfield
  {journal} {\bibinfo  {journal} {ACS Nano}\ }\textbf {\bibinfo {volume}
  {11}},\ \bibinfo {pages} {9950} (\bibinfo {year} {2017})}\BibitemShut
  {NoStop}%
\bibitem [{\citenamefont {Zeng}\ \emph {et~al.}(2018)\citenamefont {Zeng},
  \citenamefont {Yin}, \citenamefont {Herng}, \citenamefont {Han},
  \citenamefont {Huang}, \citenamefont {Zhang}, \citenamefont {Li},
  \citenamefont {Zhou}, \citenamefont {Wan}, \citenamefont {Yang} \emph
  {et~al.}}]{ZengPRL2018}%
  \BibitemOpen
  \bibfield  {author} {\bibinfo {author} {\bibfnamefont {S.}~\bibnamefont
  {Zeng}}, \bibinfo {author} {\bibfnamefont {X.}~\bibnamefont {Yin}}, \bibinfo
  {author} {\bibfnamefont {T.}~\bibnamefont {Herng}}, \bibinfo {author}
  {\bibfnamefont {K.}~\bibnamefont {Han}}, \bibinfo {author} {\bibfnamefont
  {Z.}~\bibnamefont {Huang}}, \bibinfo {author} {\bibfnamefont
  {L.}~\bibnamefont {Zhang}}, \bibinfo {author} {\bibfnamefont
  {C.}~\bibnamefont {Li}}, \bibinfo {author} {\bibfnamefont {W.}~\bibnamefont
  {Zhou}}, \bibinfo {author} {\bibfnamefont {D.}~\bibnamefont {Wan}}, \bibinfo
  {author} {\bibfnamefont {P.}~\bibnamefont {Yang}}, \emph {et~al.},\
  }\bibfield  {title} {\bibinfo {title} {Oxygen electromigration and energy
  band reconstruction induced by electrolyte field effect at oxide
  interfaces},\ }\href
  {https://doi.org/https://doi.org/10.1103/PhysRevLett.121.146802} {\bibfield
  {journal} {\bibinfo  {journal} {Phys. Rev. Lett.}\ }\textbf {\bibinfo
  {volume} {121}},\ \bibinfo {pages} {146802} (\bibinfo {year}
  {2018})}\BibitemShut {NoStop}%
\bibitem [{\citenamefont {Saleem}\ \emph {et~al.}(2019)\citenamefont {Saleem},
  \citenamefont {Cui}, \citenamefont {Song}, \citenamefont {Sun}, \citenamefont
  {Gu}, \citenamefont {Zhang}, \citenamefont {Fayaz}, \citenamefont {Zhou},
  \citenamefont {Werner}, \citenamefont {Parkin} \emph
  {et~al.}}]{SaleemAMI2019}%
  \BibitemOpen
  \bibfield  {author} {\bibinfo {author} {\bibfnamefont {M.~S.}\ \bibnamefont
  {Saleem}}, \bibinfo {author} {\bibfnamefont {B.}~\bibnamefont {Cui}},
  \bibinfo {author} {\bibfnamefont {C.}~\bibnamefont {Song}}, \bibinfo {author}
  {\bibfnamefont {Y.}~\bibnamefont {Sun}}, \bibinfo {author} {\bibfnamefont
  {Y.}~\bibnamefont {Gu}}, \bibinfo {author} {\bibfnamefont {R.}~\bibnamefont
  {Zhang}}, \bibinfo {author} {\bibfnamefont {M.~U.}\ \bibnamefont {Fayaz}},
  \bibinfo {author} {\bibfnamefont {X.}~\bibnamefont {Zhou}}, \bibinfo {author}
  {\bibfnamefont {P.}~\bibnamefont {Werner}}, \bibinfo {author} {\bibfnamefont
  {S.~S.}\ \bibnamefont {Parkin}}, \emph {et~al.},\ }\bibfield  {title}
  {\bibinfo {title} {Electric field control of phase transition and tunable
  resistive switching in {SrFeO\ped{2.5}}},\ }\href
  {https://doi.org/https://doi.org/10.1021/acsami.8b18251} {\bibfield
  {journal} {\bibinfo  {journal} {ACS Appl. Mater. Interfaces}\ }\textbf
  {\bibinfo {volume} {11}},\ \bibinfo {pages} {6581} (\bibinfo {year}
  {2019})}\BibitemShut {NoStop}%
\bibitem [{\citenamefont {Zhang}\ \emph {et~al.}(2014)\citenamefont {Zhang},
  \citenamefont {Oka}, \citenamefont {Suzuki}, \citenamefont {Ye},\ and\
  \citenamefont {Iwasa}}]{ZhangScience2014}%
  \BibitemOpen
  \bibfield  {author} {\bibinfo {author} {\bibfnamefont {Y.}~\bibnamefont
  {Zhang}}, \bibinfo {author} {\bibfnamefont {T.}~\bibnamefont {Oka}}, \bibinfo
  {author} {\bibfnamefont {R.}~\bibnamefont {Suzuki}}, \bibinfo {author}
  {\bibfnamefont {J.}~\bibnamefont {Ye}},\ and\ \bibinfo {author}
  {\bibfnamefont {Y.}~\bibnamefont {Iwasa}},\ }\bibfield  {title} {\bibinfo
  {title} {Electrically switchable chiral light-emitting transistor},\ }\href
  {https://doi.org/https://doi.org/10.1126/science.1251329} {\bibfield
  {journal} {\bibinfo  {journal} {Science}\ }\textbf {\bibinfo {volume}
  {344}},\ \bibinfo {pages} {725} (\bibinfo {year} {2014})}\BibitemShut
  {NoStop}%
\bibitem [{\citenamefont {Chen}\ \emph {et~al.}(2018)\citenamefont {Chen},
  \citenamefont {Lu}, \citenamefont {Liang}, \citenamefont {Zheliuk},
  \citenamefont {Ali El~Yumin},\ and\ \citenamefont {Ye}}]{ChenAdvMater2018}%
  \BibitemOpen
  \bibfield  {author} {\bibinfo {author} {\bibfnamefont {Q.}~\bibnamefont
  {Chen}}, \bibinfo {author} {\bibfnamefont {J.}~\bibnamefont {Lu}}, \bibinfo
  {author} {\bibfnamefont {L.}~\bibnamefont {Liang}}, \bibinfo {author}
  {\bibfnamefont {O.}~\bibnamefont {Zheliuk}}, \bibinfo {author} {\bibfnamefont
  {A.}~\bibnamefont {Ali El~Yumin}},\ and\ \bibinfo {author} {\bibfnamefont
  {J.}~\bibnamefont {Ye}},\ }\bibfield  {title} {\bibinfo {title} {Continuous
  low-bias switching of superconductivity in a {MoS\ped{2}} transistor},\
  }\href {https://doi.org/https://doi.org/10.1002/adma.201800399} {\bibfield
  {journal} {\bibinfo  {journal} {Adv. Mater.}\ }\textbf {\bibinfo {volume}
  {30}},\ \bibinfo {pages} {1800399} (\bibinfo {year} {2018})}\BibitemShut
  {NoStop}%
\bibitem [{\citenamefont {De~Simoni}\ \emph {et~al.}(2018)\citenamefont
  {De~Simoni}, \citenamefont {Paolucci}, \citenamefont {Solinas}, \citenamefont
  {Strambini},\ and\ \citenamefont {Giazotto}}]{DeSimoniNatNano2018}%
  \BibitemOpen
  \bibfield  {author} {\bibinfo {author} {\bibfnamefont {G.}~\bibnamefont
  {De~Simoni}}, \bibinfo {author} {\bibfnamefont {F.}~\bibnamefont {Paolucci}},
  \bibinfo {author} {\bibfnamefont {P.}~\bibnamefont {Solinas}}, \bibinfo
  {author} {\bibfnamefont {E.}~\bibnamefont {Strambini}},\ and\ \bibinfo
  {author} {\bibfnamefont {F.}~\bibnamefont {Giazotto}},\ }\bibfield  {title}
  {\bibinfo {title} {Metallic supercurrent field-effect transistor},\ }\href
  {https://doi.org/https://doi.org/10.1038/s41565-018-0190-3} {\bibfield
  {journal} {\bibinfo  {journal} {Nat. Nanotechnol.}\ }\textbf {\bibinfo
  {volume} {13}},\ \bibinfo {pages} {802} (\bibinfo {year} {2018})}\BibitemShut
  {NoStop}%
\bibitem [{\citenamefont {Paolucci}\ \emph {et~al.}(2018)\citenamefont
  {Paolucci}, \citenamefont {De~Simoni}, \citenamefont {Strambini},
  \citenamefont {Solinas},\ and\ \citenamefont {Giazotto}}]{PaolucciNL2018}%
  \BibitemOpen
  \bibfield  {author} {\bibinfo {author} {\bibfnamefont {F.}~\bibnamefont
  {Paolucci}}, \bibinfo {author} {\bibfnamefont {G.}~\bibnamefont {De~Simoni}},
  \bibinfo {author} {\bibfnamefont {E.}~\bibnamefont {Strambini}}, \bibinfo
  {author} {\bibfnamefont {P.}~\bibnamefont {Solinas}},\ and\ \bibinfo {author}
  {\bibfnamefont {F.}~\bibnamefont {Giazotto}},\ }\bibfield  {title} {\bibinfo
  {title} {Ultra-efficient superconducting {Dayem} bridge field-effect
  transistor},\ }\href
  {https://doi.org/https://doi.org/10.1021/acs.nanolett.8b01010} {\bibfield
  {journal} {\bibinfo  {journal} {Nano Lett.}\ }\textbf {\bibinfo {volume}
  {18}},\ \bibinfo {pages} {4195} (\bibinfo {year} {2018})}\BibitemShut
  {NoStop}%
\bibitem [{\citenamefont {Paolucci}\ \emph
  {et~al.}(2019{\natexlab{a}})\citenamefont {Paolucci}, \citenamefont
  {De~Simoni}, \citenamefont {Solinas}, \citenamefont {Strambini},
  \citenamefont {Ligato}, \citenamefont {Virtanen}, \citenamefont {Braggio},\
  and\ \citenamefont {Giazotto}}]{PaolucciPRApp2019}%
  \BibitemOpen
  \bibfield  {author} {\bibinfo {author} {\bibfnamefont {F.}~\bibnamefont
  {Paolucci}}, \bibinfo {author} {\bibfnamefont {G.}~\bibnamefont {De~Simoni}},
  \bibinfo {author} {\bibfnamefont {P.}~\bibnamefont {Solinas}}, \bibinfo
  {author} {\bibfnamefont {E.}~\bibnamefont {Strambini}}, \bibinfo {author}
  {\bibfnamefont {N.}~\bibnamefont {Ligato}}, \bibinfo {author} {\bibfnamefont
  {P.}~\bibnamefont {Virtanen}}, \bibinfo {author} {\bibfnamefont
  {A.}~\bibnamefont {Braggio}},\ and\ \bibinfo {author} {\bibfnamefont
  {F.}~\bibnamefont {Giazotto}},\ }\bibfield  {title} {\bibinfo {title}
  {Magnetotransport experiments on fully metallic superconducting
  {Dayem}-bridge field-effect transistors},\ }\href
  {https://doi.org/https://doi.org/10.1103/PhysRevApplied.11.024061} {\bibfield
   {journal} {\bibinfo  {journal} {Phys. Rev. Applied}\ }\textbf {\bibinfo
  {volume} {11}},\ \bibinfo {pages} {024061} (\bibinfo {year}
  {2019}{\natexlab{a}})}\BibitemShut {NoStop}%
\bibitem [{\citenamefont {Paolucci}\ \emph
  {et~al.}(2019{\natexlab{b}})\citenamefont {Paolucci}, \citenamefont {Vischi},
  \citenamefont {De~Simoni}, \citenamefont {Guarcello}, \citenamefont
  {Solinas},\ and\ \citenamefont {Giazotto}}]{PaolucciNL2019}%
  \BibitemOpen
  \bibfield  {author} {\bibinfo {author} {\bibfnamefont {F.}~\bibnamefont
  {Paolucci}}, \bibinfo {author} {\bibfnamefont {F.}~\bibnamefont {Vischi}},
  \bibinfo {author} {\bibfnamefont {G.}~\bibnamefont {De~Simoni}}, \bibinfo
  {author} {\bibfnamefont {C.}~\bibnamefont {Guarcello}}, \bibinfo {author}
  {\bibfnamefont {P.}~\bibnamefont {Solinas}},\ and\ \bibinfo {author}
  {\bibfnamefont {F.}~\bibnamefont {Giazotto}},\ }\bibfield  {title} {\bibinfo
  {title} {Field-effect controllable metallic {Josephson} interferometer},\
  }\href {https://doi.org/https://doi.org/10.1021/acs.nanolett.9b02369}
  {\bibfield  {journal} {\bibinfo  {journal} {Nano Lett.}\ }\textbf {\bibinfo
  {volume} {19}},\ \bibinfo {pages} {6263} (\bibinfo {year}
  {2019}{\natexlab{b}})}\BibitemShut {NoStop}%
\bibitem [{\citenamefont {Cho}\ \emph {et~al.}(2008)\citenamefont {Cho},
  \citenamefont {Lee}, \citenamefont {Xia}, \citenamefont {Kim}, \citenamefont
  {He}, \citenamefont {Renn}, \citenamefont {Lodge},\ and\ \citenamefont
  {Frisbie}}]{ChoNatMater2008}%
  \BibitemOpen
  \bibfield  {author} {\bibinfo {author} {\bibfnamefont {J.~H.}\ \bibnamefont
  {Cho}}, \bibinfo {author} {\bibfnamefont {J.}~\bibnamefont {Lee}}, \bibinfo
  {author} {\bibfnamefont {Y.}~\bibnamefont {Xia}}, \bibinfo {author}
  {\bibfnamefont {B.}~\bibnamefont {Kim}}, \bibinfo {author} {\bibfnamefont
  {Y.}~\bibnamefont {He}}, \bibinfo {author} {\bibfnamefont {M.~J.}\
  \bibnamefont {Renn}}, \bibinfo {author} {\bibfnamefont {T.~P.}\ \bibnamefont
  {Lodge}},\ and\ \bibinfo {author} {\bibfnamefont {C.~D.}\ \bibnamefont
  {Frisbie}},\ }\bibfield  {title} {\bibinfo {title} {Printable ion-gel gate
  dielectrics for low-voltage polymer thin-film transistors on plastic},\
  }\href {https://doi.org/https://doi.org/10.1038/nmat2291} {\bibfield
  {journal} {\bibinfo  {journal} {Nat. Mater.}\ }\textbf {\bibinfo {volume}
  {7}},\ \bibinfo {pages} {900} (\bibinfo {year} {2008})}\BibitemShut {NoStop}%
\bibitem [{\citenamefont {Lee}\ \emph {et~al.}(2014)\citenamefont {Lee},
  \citenamefont {Kabir}, \citenamefont {Sharma}, \citenamefont {Kim},
  \citenamefont {Cho},\ and\ \citenamefont {Ahn}}]{LeeNanotech2014}%
  \BibitemOpen
  \bibfield  {author} {\bibinfo {author} {\bibfnamefont {S.-K.}\ \bibnamefont
  {Lee}}, \bibinfo {author} {\bibfnamefont {S.~H.}\ \bibnamefont {Kabir}},
  \bibinfo {author} {\bibfnamefont {B.~K.}\ \bibnamefont {Sharma}}, \bibinfo
  {author} {\bibfnamefont {B.~J.}\ \bibnamefont {Kim}}, \bibinfo {author}
  {\bibfnamefont {J.~H.}\ \bibnamefont {Cho}},\ and\ \bibinfo {author}
  {\bibfnamefont {J.-H.}\ \bibnamefont {Ahn}},\ }\bibfield  {title} {\bibinfo
  {title} {Photo-patternable ion gel-gated graphene transistors and inverters
  on plastic},\ }\href
  {https://doi.org/https://doi.org/10.1088/0957-4484/25/1/014002} {\bibfield
  {journal} {\bibinfo  {journal} {Nanotechnology}\ }\textbf {\bibinfo {volume}
  {25}},\ \bibinfo {pages} {014002} (\bibinfo {year} {2014})}\BibitemShut
  {NoStop}%
\bibitem [{\citenamefont {Piatti}\ \emph {et~al.}(2021)\citenamefont {Piatti},
  \citenamefont {Arbab}, \citenamefont {Galanti}, \citenamefont {Carey},
  \citenamefont {Anzi}, \citenamefont {Spurling}, \citenamefont {Roy},
  \citenamefont {Zhussupbekova}, \citenamefont {Patel}, \citenamefont {Kim}
  \emph {et~al.}}]{PiattiNatElectron2021}%
  \BibitemOpen
  \bibfield  {author} {\bibinfo {author} {\bibfnamefont {E.}~\bibnamefont
  {Piatti}}, \bibinfo {author} {\bibfnamefont {A.}~\bibnamefont {Arbab}},
  \bibinfo {author} {\bibfnamefont {F.}~\bibnamefont {Galanti}}, \bibinfo
  {author} {\bibfnamefont {T.}~\bibnamefont {Carey}}, \bibinfo {author}
  {\bibfnamefont {L.}~\bibnamefont {Anzi}}, \bibinfo {author} {\bibfnamefont
  {D.}~\bibnamefont {Spurling}}, \bibinfo {author} {\bibfnamefont
  {A.}~\bibnamefont {Roy}}, \bibinfo {author} {\bibfnamefont {A.}~\bibnamefont
  {Zhussupbekova}}, \bibinfo {author} {\bibfnamefont {K.~A.}\ \bibnamefont
  {Patel}}, \bibinfo {author} {\bibfnamefont {J.~M.}\ \bibnamefont {Kim}},
  \emph {et~al.},\ }\bibfield  {title} {\bibinfo {title} {Charge transport
  mechanisms in inkjet-printed thin-film transistors based on two-dimensional
  materials},\ }\href
  {https://doi.org/https://doi.org/10.1038/s41928-021-00684-9} {\bibfield
  {journal} {\bibinfo  {journal} {Nat. Electron.}\ }\textbf {\bibinfo {volume}
  {4}},\ \bibinfo {pages} {893} (\bibinfo {year} {2021})}\BibitemShut {NoStop}%
\bibitem [{\citenamefont {Shimizu}\ \emph {et~al.}(2019)\citenamefont
  {Shimizu}, \citenamefont {Shiogai}, \citenamefont {Takemori}, \citenamefont
  {Sakai}, \citenamefont {Ikeda}, \citenamefont {Arita}, \citenamefont
  {Nojima}, \citenamefont {Tsukazaki},\ and\ \citenamefont
  {Iwasa}}]{ShimizuNatCommun2019}%
  \BibitemOpen
  \bibfield  {author} {\bibinfo {author} {\bibfnamefont {S.}~\bibnamefont
  {Shimizu}}, \bibinfo {author} {\bibfnamefont {J.}~\bibnamefont {Shiogai}},
  \bibinfo {author} {\bibfnamefont {N.}~\bibnamefont {Takemori}}, \bibinfo
  {author} {\bibfnamefont {S.}~\bibnamefont {Sakai}}, \bibinfo {author}
  {\bibfnamefont {H.}~\bibnamefont {Ikeda}}, \bibinfo {author} {\bibfnamefont
  {R.}~\bibnamefont {Arita}}, \bibinfo {author} {\bibfnamefont
  {T.}~\bibnamefont {Nojima}}, \bibinfo {author} {\bibfnamefont
  {A.}~\bibnamefont {Tsukazaki}},\ and\ \bibinfo {author} {\bibfnamefont
  {Y.}~\bibnamefont {Iwasa}},\ }\bibfield  {title} {\bibinfo {title} {Giant
  thermoelectric power factor in ultrathin {FeSe} superconductor},\ }\href
  {https://doi.org/https://doi.org/10.1038/s41467-019-08784-z} {\bibfield
  {journal} {\bibinfo  {journal} {Nat. Commun.}\ }\textbf {\bibinfo {volume}
  {10}},\ \bibinfo {pages} {825} (\bibinfo {year} {2019})}\BibitemShut
  {NoStop}%
\bibitem [{\citenamefont {Huang}\ \emph {et~al.}(2021)\citenamefont {Huang},
  \citenamefont {Zhang}, \citenamefont {Song}, \citenamefont {Wang},
  \citenamefont {Hou}, \citenamefont {Hu}, \citenamefont {Chen},\ and\
  \citenamefont {Zhai}}]{Huang2021}%
  \BibitemOpen
  \bibfield  {author} {\bibinfo {author} {\bibfnamefont {W.}~\bibnamefont
  {Huang}}, \bibinfo {author} {\bibfnamefont {Y.}~\bibnamefont {Zhang}},
  \bibinfo {author} {\bibfnamefont {M.}~\bibnamefont {Song}}, \bibinfo {author}
  {\bibfnamefont {B.}~\bibnamefont {Wang}}, \bibinfo {author} {\bibfnamefont
  {H.}~\bibnamefont {Hou}}, \bibinfo {author} {\bibfnamefont {X.}~\bibnamefont
  {Hu}}, \bibinfo {author} {\bibfnamefont {X.}~\bibnamefont {Chen}},\ and\
  \bibinfo {author} {\bibfnamefont {T.}~\bibnamefont {Zhai}},\ }\bibfield
  {title} {\bibinfo {title} {Encapsulation strategies on {2D} materials for
  field effect transistors and photodetectors},\ }\href
  {https://doi.org/https://doi.org/10.1016/j.cclet.2021.08.086} {\bibfield
  {journal} {\bibinfo  {journal} {Chinese Chemical Letters}\ ,\ \bibinfo
  {pages} {In press}} (\bibinfo {year} {2021})}\BibitemShut {NoStop}%
\bibitem [{\citenamefont {Jo}\ \emph {et~al.}(2017)\citenamefont {Jo},
  \citenamefont {Choi}, \citenamefont {Hyun}, \citenamefont {Seo},
  \citenamefont {Kim}, \citenamefont {Lee}, \citenamefont {Kwon}, \citenamefont
  {Moon}, \citenamefont {Kwon}, \citenamefont {Ahn} \emph
  {et~al.}}]{JoSciRep2017}%
  \BibitemOpen
  \bibfield  {author} {\bibinfo {author} {\bibfnamefont {H.}~\bibnamefont
  {Jo}}, \bibinfo {author} {\bibfnamefont {J.-H.}\ \bibnamefont {Choi}},
  \bibinfo {author} {\bibfnamefont {C.-M.}\ \bibnamefont {Hyun}}, \bibinfo
  {author} {\bibfnamefont {S.-Y.}\ \bibnamefont {Seo}}, \bibinfo {author}
  {\bibfnamefont {C.-M.}\ \bibnamefont {Kim}}, \bibinfo {author} {\bibfnamefont
  {M.-J.}\ \bibnamefont {Lee}}, \bibinfo {author} {\bibfnamefont {J.-D.}\
  \bibnamefont {Kwon}}, \bibinfo {author} {\bibfnamefont {H.-S.}\ \bibnamefont
  {Moon}}, \bibinfo {author} {\bibfnamefont {S.-H.}\ \bibnamefont {Kwon}},
  \bibinfo {author} {\bibfnamefont {J.-H.}\ \bibnamefont {Ahn}}, \emph
  {et~al.},\ }\bibfield  {title} {\bibinfo {title} {A hybrid gate dielectrics
  of ion gel with ultra-thin passivation layer for high-performance transistors
  based on two-dimensional semiconductor channels},\ }\href
  {https://doi.org/https://doi.org/10.1038/s41598-017-14649-6} {\bibfield
  {journal} {\bibinfo  {journal} {Sci. Rep.}\ }\textbf {\bibinfo {volume}
  {7}},\ \bibinfo {pages} {14194} (\bibinfo {year} {2017})}\BibitemShut
  {NoStop}%
\bibitem [{\citenamefont {Dane}\ \emph {et~al.}(2017)\citenamefont {Dane},
  \citenamefont {McCaughan}, \citenamefont {Zhu}, \citenamefont {Zhao},
  \citenamefont {Kim}, \citenamefont {Calandri}, \citenamefont {Agarwal},
  \citenamefont {Bellei},\ and\ \citenamefont {Berggren}}]{dane2017bias}%
  \BibitemOpen
  \bibfield  {author} {\bibinfo {author} {\bibfnamefont {A.~E.}\ \bibnamefont
  {Dane}}, \bibinfo {author} {\bibfnamefont {A.~N.}\ \bibnamefont {McCaughan}},
  \bibinfo {author} {\bibfnamefont {D.}~\bibnamefont {Zhu}}, \bibinfo {author}
  {\bibfnamefont {Q.}~\bibnamefont {Zhao}}, \bibinfo {author} {\bibfnamefont
  {C.-S.}\ \bibnamefont {Kim}}, \bibinfo {author} {\bibfnamefont
  {N.}~\bibnamefont {Calandri}}, \bibinfo {author} {\bibfnamefont
  {A.}~\bibnamefont {Agarwal}}, \bibinfo {author} {\bibfnamefont
  {F.}~\bibnamefont {Bellei}},\ and\ \bibinfo {author} {\bibfnamefont {K.~K.}\
  \bibnamefont {Berggren}},\ }\bibfield  {title} {\bibinfo {title} {Bias
  sputtered {NbN} and superconducting nanowire devices},\ }\href
  {https://doi.org/https://doi.org/10.1063/1.4990066} {\bibfield  {journal}
  {\bibinfo  {journal} {Appl. Phys. Lett.}\ }\textbf {\bibinfo {volume}
  {111}},\ \bibinfo {pages} {122601} (\bibinfo {year} {2017})}\BibitemShut
  {NoStop}%
\bibitem [{\citenamefont {Medeiros}\ \emph {et~al.}(2019)\citenamefont
  {Medeiros}, \citenamefont {Colangelo}, \citenamefont {Charaev},\ and\
  \citenamefont {Berggren}}]{medeiros2019measuring}%
  \BibitemOpen
  \bibfield  {author} {\bibinfo {author} {\bibfnamefont {O.}~\bibnamefont
  {Medeiros}}, \bibinfo {author} {\bibfnamefont {M.}~\bibnamefont {Colangelo}},
  \bibinfo {author} {\bibfnamefont {I.}~\bibnamefont {Charaev}},\ and\ \bibinfo
  {author} {\bibfnamefont {K.~K.}\ \bibnamefont {Berggren}},\ }\bibfield
  {title} {\bibinfo {title} {Measuring thickness in thin {NbN} films for
  superconducting devices},\ }\href
  {https://doi.org/https://doi.org/10.1116/1.5088061} {\bibfield  {journal}
  {\bibinfo  {journal} {J. Vac. Sci. Technol. A}\ }\textbf {\bibinfo {volume}
  {37}},\ \bibinfo {pages} {041501} (\bibinfo {year} {2019})}\BibitemShut
  {NoStop}%
\bibitem [{SM()}]{SM}%
  \BibitemOpen
  \href@noop {} {\bibinfo {title} {{See the Supplemental Material at [insert
  link here] for a detailed description of the samples and the fabrication
  process, the full response of an encapsulated ultrathin film to a triangular
  gate voltage wave, the sheet resistance modulation at low temperatures, the
  control experiments in non-encapsulated ultrathin films, the
  parallel-resistor model for the resistance modulation, and additional details
  on the X-ray photoelectron spectroscopy of encapsulated ultrathin films and
  on the measurement and fitting of the tunnelling V(I) curves.}}}\BibitemShut
  {Stop}%
\bibitem [{\citenamefont {Zhou}\ and\ \citenamefont
  {Ramanathan}(2012)}]{ZhouJAP2012}%
  \BibitemOpen
  \bibfield  {author} {\bibinfo {author} {\bibfnamefont {Y.}~\bibnamefont
  {Zhou}}\ and\ \bibinfo {author} {\bibfnamefont {S.}~\bibnamefont
  {Ramanathan}},\ }\bibfield  {title} {\bibinfo {title} {Relaxation dynamics of
  ionic liquid-{VO\ped{2}} interfaces and influence in electric double-layer
  transistors},\ }\href {https://doi.org/https://doi.org/10.1063/1.4704689}
  {\bibfield  {journal} {\bibinfo  {journal} {J. Appl. Phys.}\ }\textbf
  {\bibinfo {volume} {111}},\ \bibinfo {pages} {084508} (\bibinfo {year}
  {2012})}\BibitemShut {NoStop}%
\bibitem [{\citenamefont {Scholtz}(2010)}]{ScholtzBook}%
  \BibitemOpen
  \bibfield  {author} {\bibinfo {author} {\bibfnamefont {F.}~\bibnamefont
  {Scholtz}},\ }\href
  {https://doi.org/https://doi.org/10.1007/978-3-642-02915-8} {\emph {\bibinfo
  {title} {Electroanalytical Methods}}}\ (\bibinfo  {publisher}
  {Springer-Verlag},\ \bibinfo {address} {Berlin Heidelberg},\ \bibinfo {year}
  {2010})\BibitemShut {NoStop}%
\bibitem [{\citenamefont {Pignolet}\ \emph {et~al.}(1995)\citenamefont
  {Pignolet}, \citenamefont {Rao},\ and\ \citenamefont
  {Krupanidhi}}]{PignoletTF1995}%
  \BibitemOpen
  \bibfield  {author} {\bibinfo {author} {\bibfnamefont {A.}~\bibnamefont
  {Pignolet}}, \bibinfo {author} {\bibfnamefont {G.~M.}\ \bibnamefont {Rao}},\
  and\ \bibinfo {author} {\bibfnamefont {S.~B.}\ \bibnamefont {Krupanidhi}},\
  }\bibfield  {title} {\bibinfo {title} {Rapid thermal processed thin films of
  niobium pentoxide ({Nb\ped{2}O\ped{5}}) deposited by reactive magnetron
  sputtering},\ }\href
  {https://doi.org/https://doi.org/10.1016/S0040-6090(94)06470-9} {\bibfield
  {journal} {\bibinfo  {journal} {Thin Solid Films}\ }\textbf {\bibinfo
  {volume} {261}},\ \bibinfo {pages} {18} (\bibinfo {year} {1995})}\BibitemShut
  {NoStop}%
\bibitem [{\citenamefont {Fuchs}(1938)}]{Fuchs1938}%
  \BibitemOpen
  \bibfield  {author} {\bibinfo {author} {\bibfnamefont {K.}~\bibnamefont
  {Fuchs}},\ }\bibfield  {title} {\bibinfo {title} {The conductivity of thin
  metallic films according to the electron theory of metals},\ }\href
  {https://doi.org/https://doi:10.1017/S0305004100019952} {\bibfield  {journal}
  {\bibinfo  {journal} {Mathematical Proceedings of the Cambridge Philosophical
  Society}\ }\textbf {\bibinfo {volume} {34}},\ \bibinfo {pages} {100 108}
  (\bibinfo {year} {1938})}\BibitemShut {NoStop}%
\bibitem [{\citenamefont {Chockalingam}\ \emph {et~al.}(2008)\citenamefont
  {Chockalingam}, \citenamefont {Chand}, \citenamefont {Jesudasan},
  \citenamefont {Tripathi},\ and\ \citenamefont
  {Raychaudhuri}}]{ChockalingamPRB2008}%
  \BibitemOpen
  \bibfield  {author} {\bibinfo {author} {\bibfnamefont {S.~P.}\ \bibnamefont
  {Chockalingam}}, \bibinfo {author} {\bibfnamefont {M.}~\bibnamefont {Chand}},
  \bibinfo {author} {\bibfnamefont {J.}~\bibnamefont {Jesudasan}}, \bibinfo
  {author} {\bibfnamefont {V.}~\bibnamefont {Tripathi}},\ and\ \bibinfo
  {author} {\bibfnamefont {P.}~\bibnamefont {Raychaudhuri}},\ }\bibfield
  {title} {\bibinfo {title} {Superconducting properties and {Hall} effect of
  epitaxial {NbN} thin films},\ }\href
  {https://doi.org/https://doi.org/10.1103/PhysRevB.77.214503} {\bibfield
  {journal} {\bibinfo  {journal} {Phys. Rev. B}\ }\textbf {\bibinfo {volume}
  {77}},\ \bibinfo {pages} {214503} (\bibinfo {year} {2008})}\BibitemShut
  {NoStop}%
\bibitem [{\citenamefont {El-Shazly}\ \emph {et~al.}(2018)\citenamefont
  {El-Shazly}, \citenamefont {Hassan}, \citenamefont {Abd-el Rehim},\ and\
  \citenamefont {Allam}}]{TamerJPE2018}%
  \BibitemOpen
  \bibfield  {author} {\bibinfo {author} {\bibfnamefont {T.~S.}\ \bibnamefont
  {El-Shazly}}, \bibinfo {author} {\bibfnamefont {W.~M.}\ \bibnamefont
  {Hassan}}, \bibinfo {author} {\bibfnamefont {S.~S.}\ \bibnamefont {Abd-el
  Rehim}},\ and\ \bibinfo {author} {\bibfnamefont {N.~K.}\ \bibnamefont
  {Allam}},\ }\bibfield  {title} {\bibinfo {title} {Optical and electronic
  properties of niobium oxynitrides with various {N/O} ratios: insights from
  first-principles calculations},\ }\href
  {https://doi.org/https://doi.org/10.1117/1.JPE.8.026501} {\bibfield
  {journal} {\bibinfo  {journal} {J. Photonics Energy}\ }\textbf {\bibinfo
  {volume} {8}},\ \bibinfo {pages} {026501} (\bibinfo {year}
  {2018})}\BibitemShut {NoStop}%
\bibitem [{\citenamefont {Ummarino}\ \emph {et~al.}(2017)\citenamefont
  {Ummarino}, \citenamefont {Piatti}, \citenamefont {Daghero}, \citenamefont
  {Gonnelli}, \citenamefont {Sklyadneva}, \citenamefont {Chulkov},\ and\
  \citenamefont {Heid}}]{UmmarinoPRB2017}%
  \BibitemOpen
  \bibfield  {author} {\bibinfo {author} {\bibfnamefont {G.~A.}\ \bibnamefont
  {Ummarino}}, \bibinfo {author} {\bibfnamefont {E.}~\bibnamefont {Piatti}},
  \bibinfo {author} {\bibfnamefont {D.}~\bibnamefont {Daghero}}, \bibinfo
  {author} {\bibfnamefont {R.~S.}\ \bibnamefont {Gonnelli}}, \bibinfo {author}
  {\bibfnamefont {I.~Y.}\ \bibnamefont {Sklyadneva}}, \bibinfo {author}
  {\bibfnamefont {E.~V.}\ \bibnamefont {Chulkov}},\ and\ \bibinfo {author}
  {\bibfnamefont {R.}~\bibnamefont {Heid}},\ }\bibfield  {title} {\bibinfo
  {title} {Proximity {Eliashberg} theory of electrostatic field-effect doping
  in superconducting films},\ }\href
  {https://doi.org/https://doi.org/10.1103/PhysRevB.96.064509} {\bibfield
  {journal} {\bibinfo  {journal} {Phys. Rev. B}\ }\textbf {\bibinfo {volume}
  {96}},\ \bibinfo {pages} {064509} (\bibinfo {year} {2017})}\BibitemShut
  {NoStop}%
\bibitem [{\citenamefont {Ummarino}\ and\ \citenamefont
  {Romanin}(2020)}]{UmmarinoPSSB2020}%
  \BibitemOpen
  \bibfield  {author} {\bibinfo {author} {\bibfnamefont {G.~A.}\ \bibnamefont
  {Ummarino}}\ and\ \bibinfo {author} {\bibfnamefont {D.}~\bibnamefont
  {Romanin}},\ }\bibfield  {title} {\bibinfo {title} {Theoretical explanation
  of electric field-induced superconductive critical temperature shifts in
  indium thin films},\ }\href
  {https://doi.org/https://doi.org/10.1002/pssb.201900651} {\bibfield
  {journal} {\bibinfo  {journal} {Phys. Status Solidi B}\ }\textbf {\bibinfo
  {volume} {257}},\ \bibinfo {pages} {1900651} (\bibinfo {year}
  {2020})}\BibitemShut {NoStop}%
\bibitem [{\citenamefont {Piatti}\ \emph
  {et~al.}(2018{\natexlab{c}})\citenamefont {Piatti}, \citenamefont {Romanin},
  \citenamefont {Gonnelli},\ and\ \citenamefont
  {Daghero}}]{PiattiApSuSc2018nbn}%
  \BibitemOpen
  \bibfield  {author} {\bibinfo {author} {\bibfnamefont {E.}~\bibnamefont
  {Piatti}}, \bibinfo {author} {\bibfnamefont {D.}~\bibnamefont {Romanin}},
  \bibinfo {author} {\bibfnamefont {R.~S.}\ \bibnamefont {Gonnelli}},\ and\
  \bibinfo {author} {\bibfnamefont {D.}~\bibnamefont {Daghero}},\ }\bibfield
  {title} {\bibinfo {title} {Anomalous screening of an electrostatic field at
  the surface of niobium nitride},\ }\href
  {https://doi.org/https://doi.org/10.1016/j.apsusc.2018.05.181} {\bibfield
  {journal} {\bibinfo  {journal} {Appl. Surf. Sci.}\ }\textbf {\bibinfo
  {volume} {461}},\ \bibinfo {pages} {17} (\bibinfo {year}
  {2018}{\natexlab{c}})}\BibitemShut {NoStop}%
\bibitem [{\citenamefont {Charaev}\ \emph {et~al.}(2017)\citenamefont
  {Charaev}, \citenamefont {Silbernagel}, \citenamefont {Bachowsky},
  \citenamefont {Kuzmin}, \citenamefont {Doerner}, \citenamefont {Ilin},
  \citenamefont {Semenov}, \citenamefont {Roditchev}, \citenamefont
  {Vodolazov},\ and\ \citenamefont {Siegel}}]{CharaevJAP2017}%
  \BibitemOpen
  \bibfield  {author} {\bibinfo {author} {\bibfnamefont {I.}~\bibnamefont
  {Charaev}}, \bibinfo {author} {\bibfnamefont {T.}~\bibnamefont
  {Silbernagel}}, \bibinfo {author} {\bibfnamefont {B.}~\bibnamefont
  {Bachowsky}}, \bibinfo {author} {\bibfnamefont {A.}~\bibnamefont {Kuzmin}},
  \bibinfo {author} {\bibfnamefont {S.}~\bibnamefont {Doerner}}, \bibinfo
  {author} {\bibfnamefont {K.}~\bibnamefont {Ilin}}, \bibinfo {author}
  {\bibfnamefont {A.}~\bibnamefont {Semenov}}, \bibinfo {author} {\bibfnamefont
  {D.}~\bibnamefont {Roditchev}}, \bibinfo {author} {\bibfnamefont {D.~Y.}\
  \bibnamefont {Vodolazov}},\ and\ \bibinfo {author} {\bibfnamefont
  {M.}~\bibnamefont {Siegel}},\ }\bibfield  {title} {\bibinfo {title}
  {Enhancement of superconductivity in {NbN} nanowires by negative
  electron-beam lithography with positive resist},\ }\href
  {https://doi.org/https://doi.org/10.1063/1.4986416} {\bibfield  {journal}
  {\bibinfo  {journal} {J. Appl. Phys.}\ }\textbf {\bibinfo {volume} {122}},\
  \bibinfo {pages} {083901} (\bibinfo {year} {2017})}\BibitemShut {NoStop}%
\bibitem [{\citenamefont {Neylon}\ \emph {et~al.}(2002)\citenamefont {Neylon},
  \citenamefont {Bej}, \citenamefont {Bennett},\ and\ \citenamefont
  {Thompson}}]{NeylonACA2002}%
  \BibitemOpen
  \bibfield  {author} {\bibinfo {author} {\bibfnamefont {M.~K.}\ \bibnamefont
  {Neylon}}, \bibinfo {author} {\bibfnamefont {S.~K.}\ \bibnamefont {Bej}},
  \bibinfo {author} {\bibfnamefont {C.~A.}\ \bibnamefont {Bennett}},\ and\
  \bibinfo {author} {\bibfnamefont {L.~T.}\ \bibnamefont {Thompson}},\
  }\bibfield  {title} {\bibinfo {title} {Ethanol amination catalysis over early
  transition metal nitrides},\ }\href
  {https://doi.org/https://doi.org/10.1016/S0926-860X(02)00054-6} {\bibfield
  {journal} {\bibinfo  {journal} {Appl. Catal. A - Gen.}\ }\textbf {\bibinfo
  {volume} {232}},\ \bibinfo {pages} {13} (\bibinfo {year} {2002})}\BibitemShut
  {NoStop}%
\bibitem [{\citenamefont {Ermolieff}\ \emph {et~al.}(1985)\citenamefont
  {Ermolieff}, \citenamefont {Girard}, \citenamefont {Raoul}, \citenamefont
  {Bertrand},\ and\ \citenamefont {Duc}}]{ErmolieffApSuSc1985}%
  \BibitemOpen
  \bibfield  {author} {\bibinfo {author} {\bibfnamefont {A.}~\bibnamefont
  {Ermolieff}}, \bibinfo {author} {\bibfnamefont {M.}~\bibnamefont {Girard}},
  \bibinfo {author} {\bibfnamefont {C.}~\bibnamefont {Raoul}}, \bibinfo
  {author} {\bibfnamefont {C.}~\bibnamefont {Bertrand}},\ and\ \bibinfo
  {author} {\bibfnamefont {T.~M.}\ \bibnamefont {Duc}},\ }\bibfield  {title}
  {\bibinfo {title} {An {XPS} comparative study on thermal oxide barrier
  formation on {Nb} and {NbN} thin films},\ }\href
  {https://doi.org/https://doi.org/10.1016/0378-5963(85)90008-X} {\bibfield
  {journal} {\bibinfo  {journal} {Appl. Surf. Sci.}\ }\textbf {\bibinfo
  {volume} {21}},\ \bibinfo {pages} {65} (\bibinfo {year} {1985})}\BibitemShut
  {NoStop}%
\bibitem [{\citenamefont {Lan}\ \emph {et~al.}(2022)\citenamefont {Lan},
  \citenamefont {Fu}, \citenamefont {Zhao}, \citenamefont {Liu}, \citenamefont
  {Zhou}, \citenamefont {Ning},\ and\ \citenamefont {Guo}}]{LanCEJ2022}%
  \BibitemOpen
  \bibfield  {author} {\bibinfo {author} {\bibfnamefont {Z.}~\bibnamefont
  {Lan}}, \bibinfo {author} {\bibfnamefont {H.}~\bibnamefont {Fu}}, \bibinfo
  {author} {\bibfnamefont {R.}~\bibnamefont {Zhao}}, \bibinfo {author}
  {\bibfnamefont {H.}~\bibnamefont {Liu}}, \bibinfo {author} {\bibfnamefont
  {W.}~\bibnamefont {Zhou}}, \bibinfo {author} {\bibfnamefont {H.}~\bibnamefont
  {Ning}},\ and\ \bibinfo {author} {\bibfnamefont {J.}~\bibnamefont {Guo}},\
  }\bibfield  {title} {\bibinfo {title} {Roles of in situ-formed {NbN} and
  {Nb\ped{2}O\ped{5}} from {N}-doped {Nb\ped{2}C} {MXene} in regulating the
  re/hydrogenation and cycling performance of magnesium hydride},\ }\href
  {https://doi.org/https://doi.org/10.1016/j.cej.2021.133985} {\bibfield
  {journal} {\bibinfo  {journal} {Chem. Eng. J.}\ }\textbf {\bibinfo {volume}
  {431}},\ \bibinfo {pages} {133985} (\bibinfo {year} {2022})}\BibitemShut
  {NoStop}%
\bibitem [{\citenamefont {Darlinski}\ and\ \citenamefont
  {Halbritter}(1987)}]{DarlinskiSIA1987}%
  \BibitemOpen
  \bibfield  {author} {\bibinfo {author} {\bibfnamefont {A.}~\bibnamefont
  {Darlinski}}\ and\ \bibinfo {author} {\bibfnamefont {J.}~\bibnamefont
  {Halbritter}},\ }\bibfield  {title} {\bibinfo {title} {Angle-resolved {XPS}
  studies of oxides at {NbN}, {NbC}, and {Nb} surfaces},\ }\href
  {https://doi.org/https://doi.org/10.1002/sia.740100502} {\bibfield  {journal}
  {\bibinfo  {journal} {Surf. Interface Anal.}\ }\textbf {\bibinfo {volume}
  {10}},\ \bibinfo {pages} {223} (\bibinfo {year} {1987})}\BibitemShut
  {NoStop}%
\bibitem [{\citenamefont {Simmons}(1963)}]{Simmons1963}%
  \BibitemOpen
  \bibfield  {author} {\bibinfo {author} {\bibfnamefont {J.~G.}\ \bibnamefont
  {Simmons}},\ }\bibfield  {title} {\bibinfo {title} {Generalized formula for
  the electric tunnel effect between similar electrodes separated by a thin
  insulating film},\ }\href {https://doi.org/https://doi.org/10.1063/1.1702682}
  {\bibfield  {journal} {\bibinfo  {journal} {J. Appl. Phys.}\ }\textbf
  {\bibinfo {volume} {34}},\ \bibinfo {pages} {1793} (\bibinfo {year}
  {1963})}\BibitemShut {NoStop}%
\end{thebibliography}%

\end{document}


\title{Reversible Tuning of Superconductivity in Ion-Gated NbN Ultrathin Films by Self-Encapsulation with a High-$\kappa$ Dielectric Layer\\ \medskip
Supplemental Material}
\author{Erik Piatti}
\thanks{These authors contributed equally.}
\affiliation{
	Department of Applied Science and Technology, Politecnico di Torino, I-10129 Torino, Italy
}
\author{Marco Colangelo}
\thanks{These authors contributed equally.}
\affiliation{
	Department of Electrical Engineering and Computer Science, Massachusetts Institute of Technology, Cambridge, MA 02139, USA
}
\author{Mattia Bartoli}
\affiliation{
	Center for Sustainable Future Technologies-CSFT@POLITO, I-10144 Torino, Italy
}
\affiliation{
	Consorzio Interuniversitario Nazionale per la Scienza e Tecnologia dei Materiali (INSTM), I-850121 Firenze, Italy
}
\author{Owen Medeiros}
\affiliation{
	Department of Electrical Engineering and Computer Science, Massachusetts Institute of Technology, Cambridge, MA 02139, USA
}
\author{Renato S. Gonnelli}
\affiliation{
	Department of Applied Science and Technology, Politecnico di Torino, I-10129 Torino, Italy
}
\author{Karl K. Berggren}
\affiliation{
	Department of Electrical Engineering and Computer Science, Massachusetts Institute of Technology, Cambridge, MA 02139, USA
}
\author{Dario Daghero}
\email{dario.daghero@polito.it}
\affiliation{
	Department of Applied Science and Technology, Politecnico di Torino, I-10129 Torino, Italy
}

\maketitle
\renewcommand{\thefigure}{S\arabic{figure}}

\tableofcontents

\clearpage

\section{Samples and fabrication process}\label{sec:fab}
The Hall bar structure was fabricated starting from a $\approx 5\,\mathrm{nm}$-thick NbN layer deposited on a substrate with a $300\,\mathrm{nm}$-thick thermal oxide layer on silicon. Details on the deposition process (room temperature reactive sputtering) can be found in Ref.\,\onlinecite{dane2017bias}. The Hall bar was patterned with a direct writing lithography process. The sample was spin-coated with a  $1.2\,\mathrm{\upmu m}$-thick layer of Microposit S1813 resist. The resist was patterned using Heidelberg $\upmu$PG 101. The pattern was developed in Microposit MF-CD-26 and transferred into the NbN with reactive ion etching in CF$_4$ plasma. Gold pads were fabricated on the Hall bar leads to simplify the electrical contact. The sample was spin-coated with a $500\,\mathrm{nm}$-thick layer of NANO PMGI SF9 and a $1.2\,\mathrm{\upmu m}$-thick layer of Microposit S1813. The patterns for the pads were defined with aligned direct writing lithography using Heidelberg $\upmu$PG 101 and developed in Microposit MF-CD-26. A $5\,\mathrm{nm}$-thick titanium adhesion layer and a $20\,\mathrm{nm}$-thick gold layer were deposited with electron-beam evaporation and lifted off in 1-Methyl-2-pyrrolidinone. To grow the encapsulation layers, the samples were annealed with a rapid thermal annealing (RTA) process. The samples used for electrical characterization and measurements were annealed for a total of 20 minutes at 200 degrees Celsius with 5 SPLM oxygen.  The samples used for XPS were annealed for a total of 20 minutes at 200 degrees Celsius with 1 SPLM oxygen. This difference in annealing methods was caused by the dismissal of the first RTA tool during the experiments.
In Table\,\ref{tab:samples}, we show the thickness of NbN and Nb$_2$O$_5$ measured with ellipsometry\,\cite{medeiros2019measuring} after the annealing process. For the Hall bar sample, the measurement was performed on a monitor chip, with a NbN film nominally identical (from the same deposition batch) to the Hall bar device.

\begin{table}[h]
	\begin{tabular}{cccccc}
		\toprule
		\textbf{Sample} & & \textbf{NbN} & &\textbf{Nb$_2$O$_5$} \\ \midrule
		Hall bar (monitor sample) & & $5.73\,\mathrm{nm}$ & & $2.06\,\mathrm{nm}$\\ \midrule
		XPS & & $5.21\,\mathrm{nm}$ & & $2.66\,\mathrm{nm}$  \\ \midrule
		
		\bottomrule
	\end{tabular}
	\caption
	{Thickness of samples used for the experiments.}
	\label{tab:samples}
\end{table} 

\clearpage

\section{Full response of an encapsulated ultrathin film to a triangular gate voltage wave}
\begin{figure}[h]
	\centering
	\includegraphics[width=\mywidth]{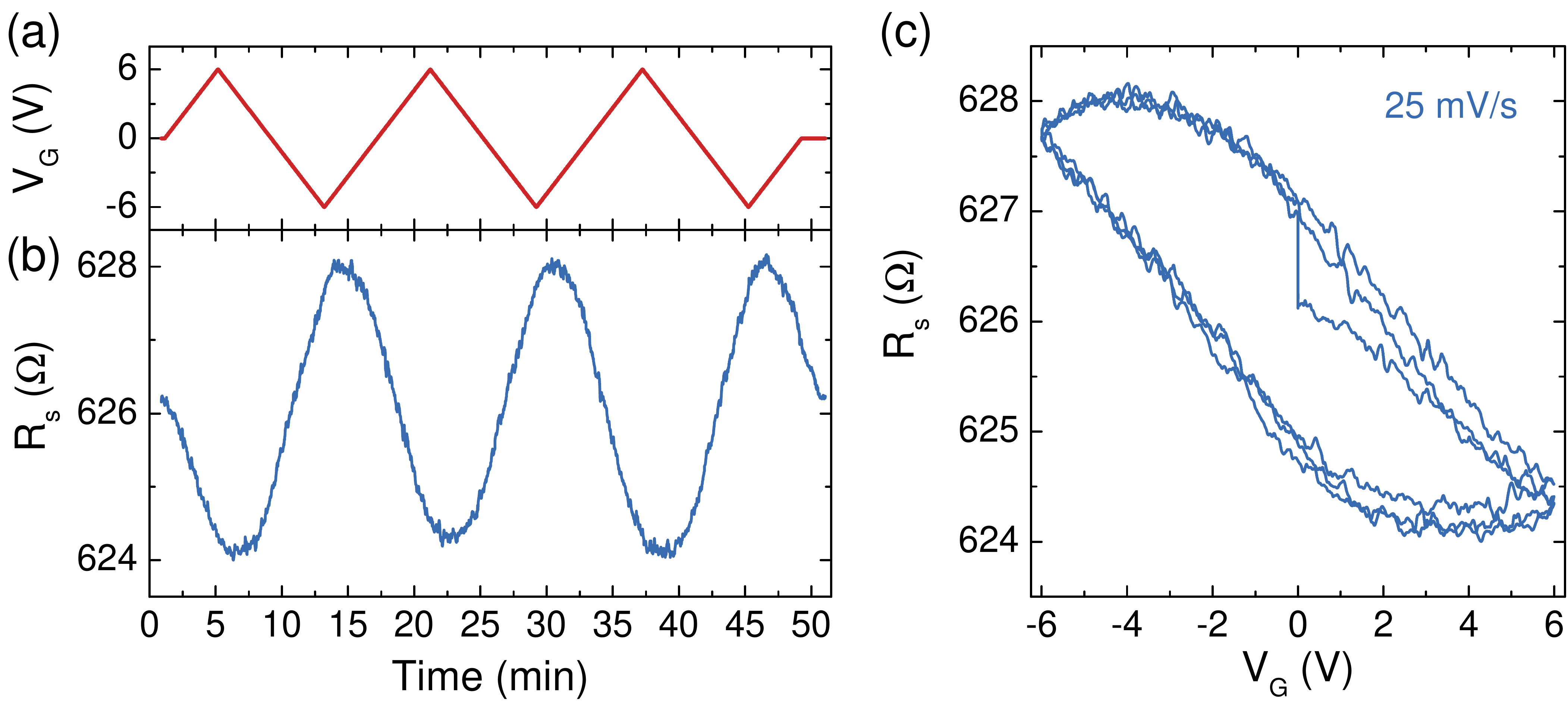}
	\caption
	{(a) Application of a triangular gate voltage $V\ped{G}$ wave with a sweep rate of $25$\,mV\,s\apex{-1}, and 
	(b) typical response of $R\ped{s}$ in an encapsulated NbN film at $T=220$\,K.
	(c) Corresponding $R\ped{s}$ vs. $V\ped{G}$ curves.}
	\label{fig:gate_sweep}
\end{figure}

\section{Sheet resistance modulation at low temperatures}
\begin{figure}[h]
	\centering
	\includegraphics[width=0.5\textwidth]{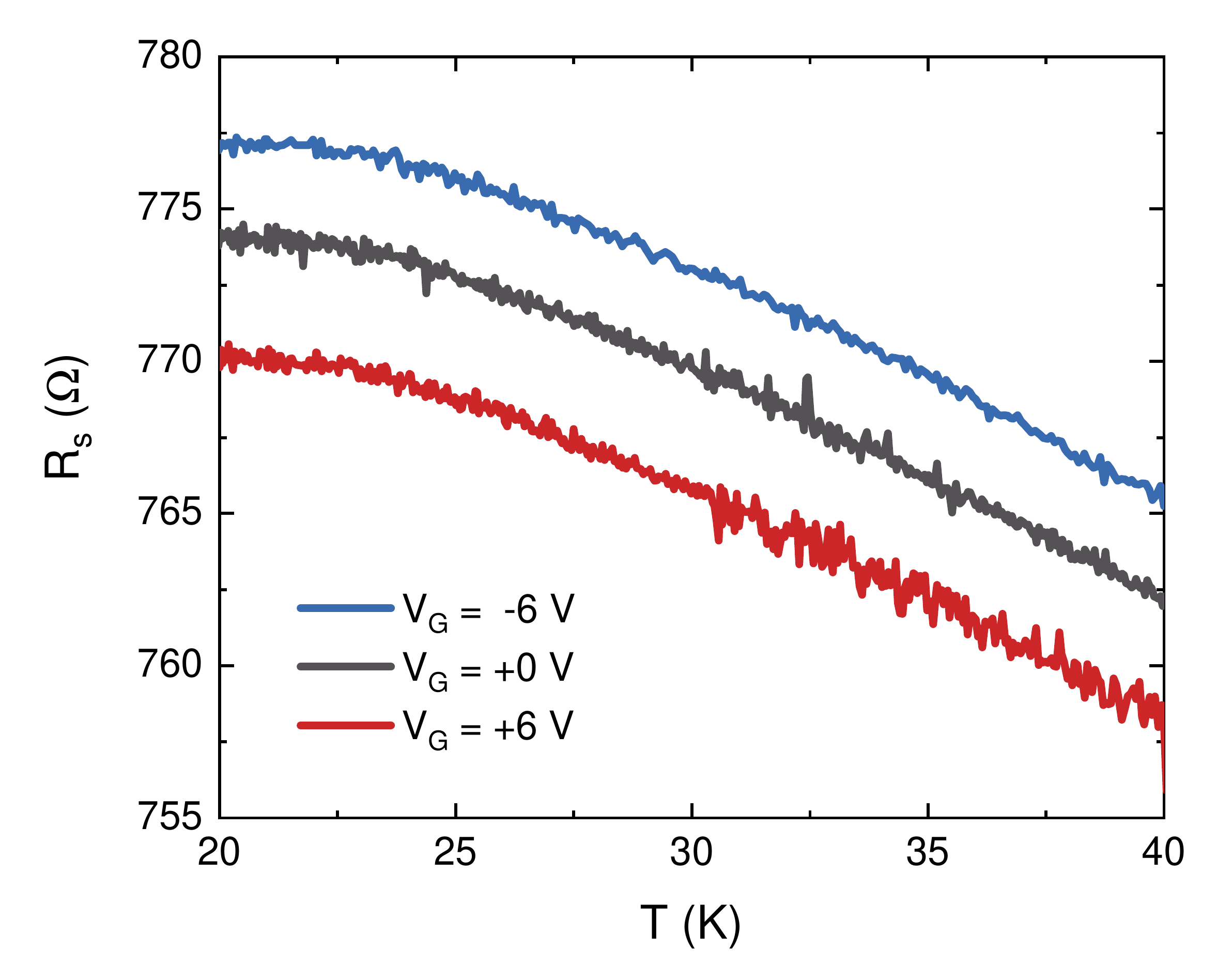}
	\caption
	{Sheet resistance $R\ped{s}$ of the gated channel as a function of temperature $T$ between 20 and 40\,K, for three different values of $V\ped{G}$. The clear gate-induced modulation of $R\ped{s}$ can be observed even in this $T$ range way below the freezing point of the ionic liquid, where the gate leakage current is zero due to the complete halt of any ionic motion.}
	\label{fig:Rs_modulation_lowT}
\end{figure}

\clearpage

\section{Control experiments in non-encapsulated ultrathin films}

\begin{figure}[h]
\centering
\includegraphics[width=\mywidth]{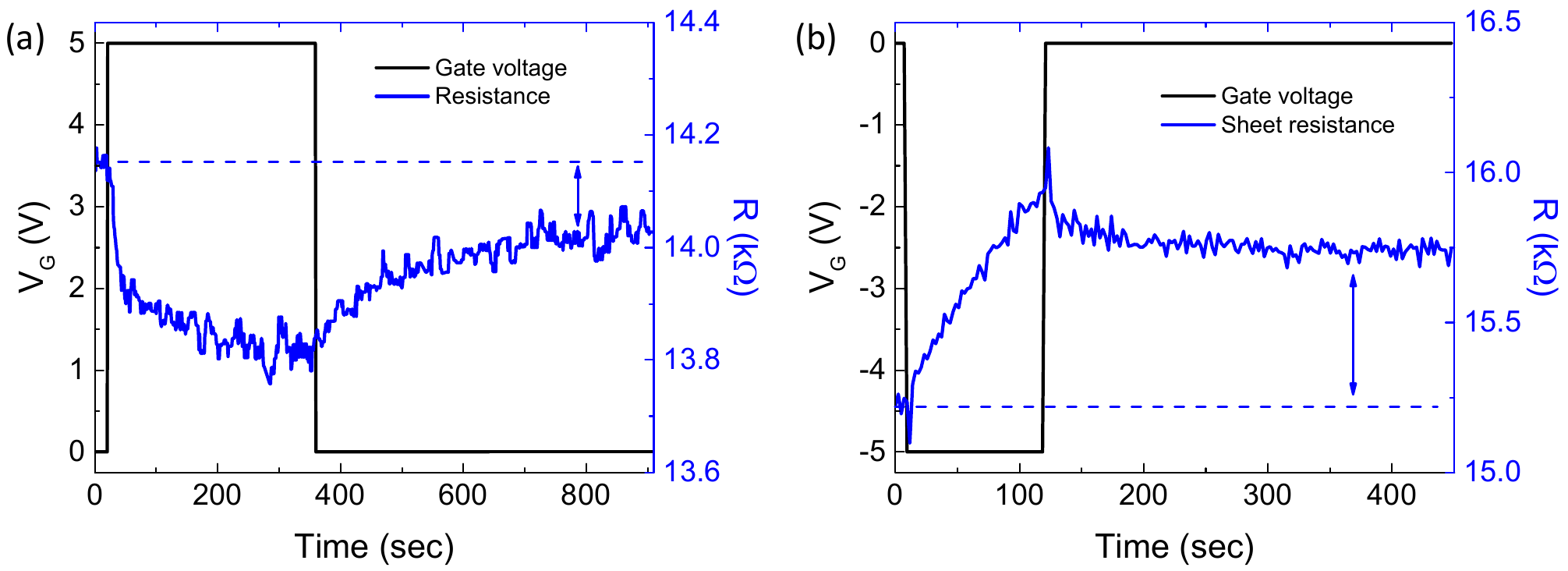}
\caption
{Typical response of $R\ped{s}$ (blue line, right scale) to the step-like application and removal of positive (a) and negative (b) values of $V\ped{G}$ (black line, left scale) at $T=220$~K in a non-encapsulated ultrathin NbN film. The blue arrows indicate the amount of non-reversible $R\ped{s}$ modulation upon removal of the applied $V\ped{G}$.}
\label{fig:gate_op}
\end{figure}
\begin{figure}[h]
\centering
\includegraphics[width=\mywidth]{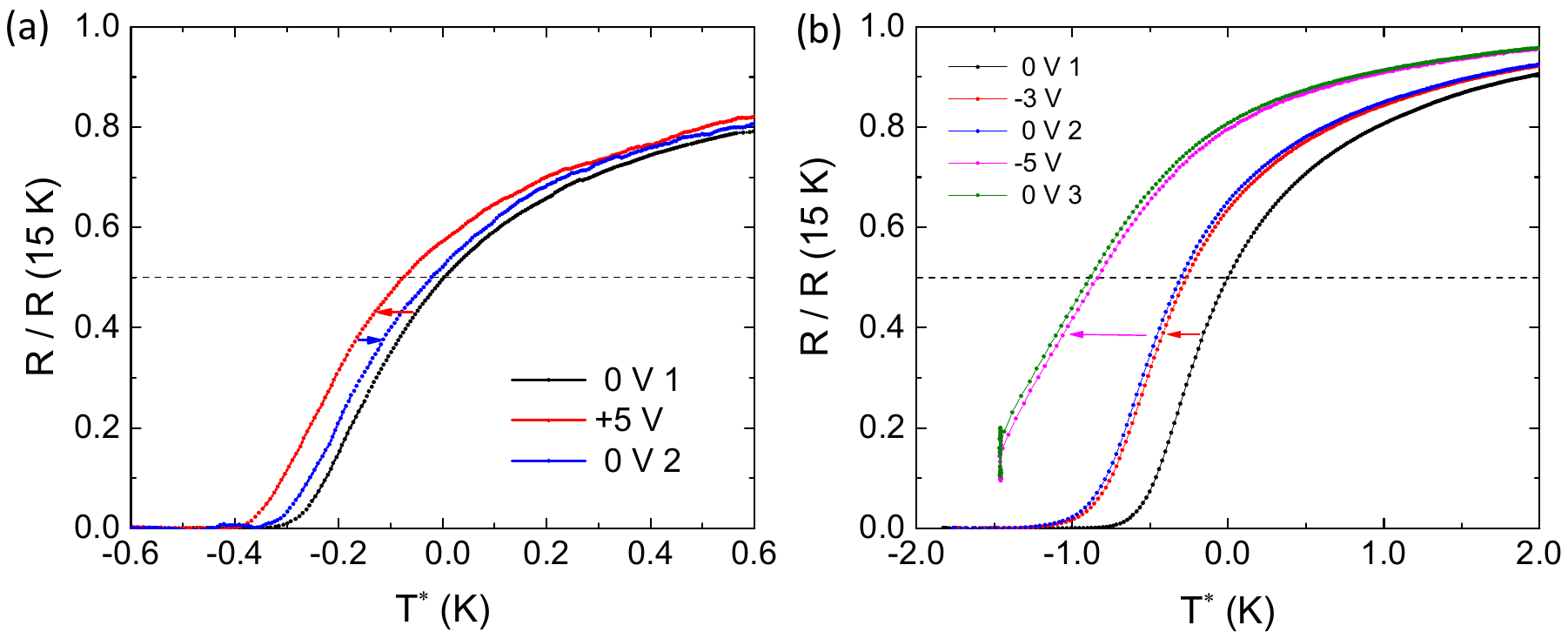}
\caption
{Normalized resistance $R\ped{s} / R\ped{s} (15 \mathrm{K})$ vs. referenced temperature $T^* = \left[T - T\ped{c}\apex{ref}\right]_{V\ped{G}} - [T\ped{c}\apex{act} - T\ped{c}\apex{ref}]_{0}$ for different values of the applied gate voltage $V\ped{G}$ in a non-encapsulated ultrathin NbN film. Dashed lines highlight the criteria used to obtain $T\ped{c}^{50}$ from the resistive transitions. Arrows highlight the order in which the measurements were performed.}
\label{fig:SC_transition}
\end{figure}

\clearpage
\section{The parallel-resistor model for the resistance modulation}\label{sec:deltaR}
In order to properly interpret the trend of the normalized resistance modulation  $\displaystyle \Delta R/R'$ as a function of $\Delta n\ped{2D}$, we must generalize the free-electron model we used for homogeneous films \cite{DagheroPRL2012,TortelloApSuSc2013} to account for the presence of the oxynitride layer on top of the NbN film.  
Let us assume that the film has length $\ell$ and width $w$ and is made of two layers of thickness $d_1$ (the oxynitride layer) and $d_2$ (the underlying NbN layer), respectively. Let $n_i$, $m_i$ and $\tau_i$ be the unperturbed \textit{volume} density of charge carriers, the effective mass and the scattering lifetime of the $i$-th layer. If $z$ is the axis normal to the interface, the additional charge induced by the electric field clearly follows a density profile $\Delta n\ped{3D}(z)$, which decays on a length scale defined by the screening length $\xi$. In materials with high intrinsic carrier density, $\xi$ is usually thought to be of the order of the Thomas-Fermi screening length $\lambda\ped{TF}$. However, as experimentally observed, and as demonstrated by first-principle calculations in the FET configuration\,\cite{PiattiApSuSc2018nbn}, in the presence of the very intense fields reached in ionic-gating experiments the linear Thomas-Fermi approximation breaks down and the screening length increases (well beyond $\lambda\ped{TF}$) as a function of the effective surface density of induced charges, $\Delta n\ped{2D}$. Indeed, as shown in Ref.\,\onlinecite{PiattiPRB2017}, everything happens as if there were a saturation in the volume charge density and, consequently, an extension of the depth of the layer of charge accumulation when the total induced charge increases. In the case of NbN, this depth can become as large as a few nanometers for $\Delta n\ped{2D} > 10^{15}\,\mathrm{cm}^{-2}$. 

\begin{figure}[h]
    \centering
    \includegraphics[width=0.9\textwidth]{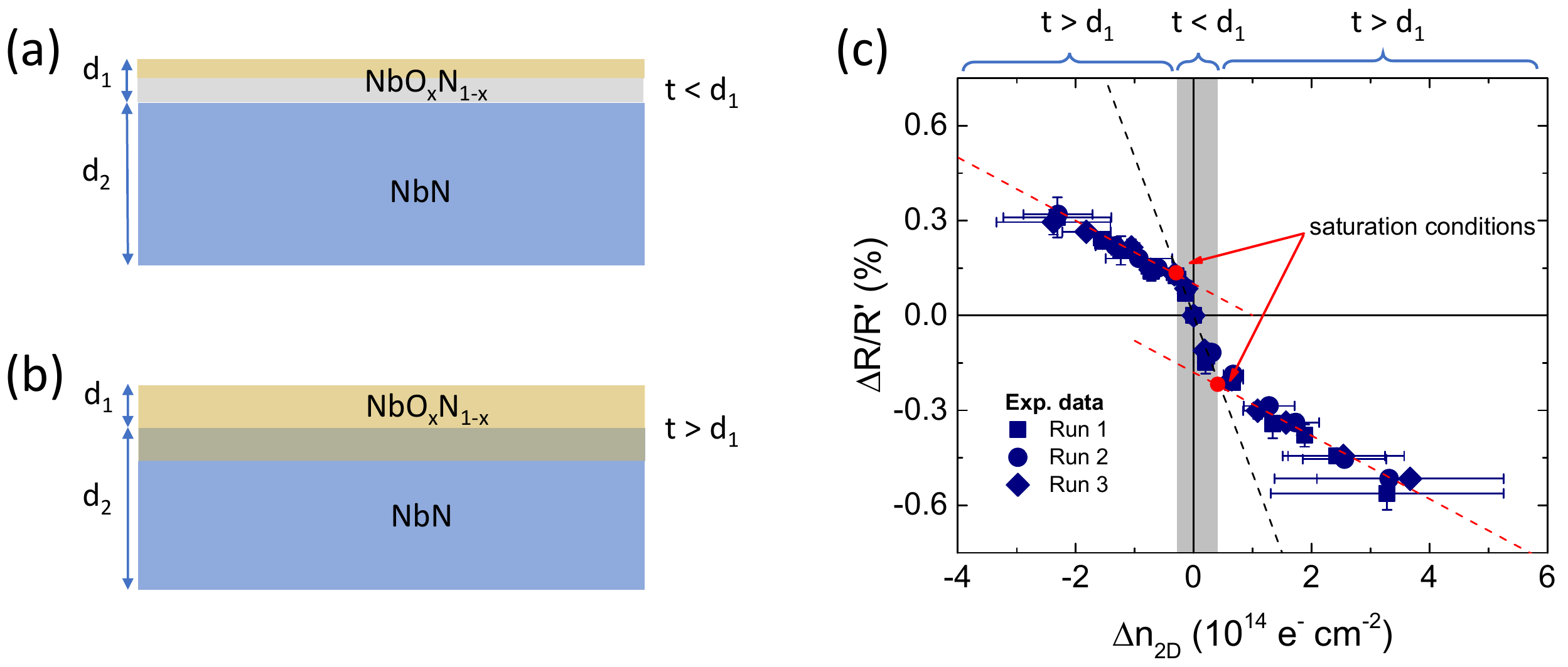}
    \caption{(a, b) Scheme of the charge accumulation in the NbN films (blue), taking into account the presence of the oxynitride layer (grey). The charge accumulation layer is depicted in yellow. In the low-field regime it is limited to the oxynitride layer (a) while in the high-field regime it extends down to the NbN film (b). The two regimes are separated by the saturation point in which $t=d_1$. (c) Experimental behaviour of $\Delta R/R'$ as a function of $\Delta n\ped{2D}$ (blue symbols) together with the linear trends that approximate it in the low-field (black dashed line) and in the high-field (red dashed lines) regime. The red point indicates the saturation point.}
    \label{fig:DeltaR}
\end{figure}

Here, as we did in the case of homogeneous films\,\cite{DagheroPRL2012} we will choose for $\Delta n\ped{3D}(z)$ a step profile, thus assuming for simplicity that the induced charge is uniformly distributed within a depth $t$ that acts as an effective screening length. However, the presence of two layers of different materials, with very different intrinsic properties, forces us to consider them separately. In the absence of field (no gate voltage) the conductance of the compound slab is
\begin{equation}
G_0 = \frac{e^2 w}{\ell}\left(\frac{n_1 \tau_1}{m_1}d_1 +\frac{n_2 \tau_2}{m_2}d_2\right) = \frac{e^2 w}{\ell}\left(n_1  A_1 d_1 + n_2 A_2 d_2\right)
\label{eq:G0}
\end{equation}
having defined the parameters \[A_i = \tau_i/m_i\] that are material-dependent and account for the bandstructure and the scattering lifetime.
When a small gate voltage is applied, induced charges accumulate only in the topmost layer, i.e. $t<d_1$ (see Fig.\,\ref{fig:DeltaR}a). Let us call $n'_1$ the perturbed volume charge density, which is uniform in the accumulation layer, and is such that $\Delta n\ped{2D} = (n'_1 - n_1) t$. In these conditions, the conductance of the compound slab is:
\begin{equation}
G_1 = \frac{e^2 w}{\ell}\left[n'_1 A_1 t + n_1 A_1 (d_1 -t) + n_2 A_2 d_2\right]. 
\label{eq:G1}
\end{equation}
The normalized resistance variation is thus
\begin{equation}
\frac{\Delta R}{R'} = \frac{G_0-G_1}{G_0} =  -(n'_1-n_1) t \frac{A_1}{n_1  A_1 d_1 + n_2 A_2 d_2}=-\Delta n\ped{2D}\frac{A_1}{n_1  A_1 d_1 + n_2 A_2 d_2}.
\label{eq:deltaR1}
\end{equation}
This equation corresponds to the low-field region of the experimental curves, i.e. it describes the black dashed line in Fig.\,\ref{fig:DeltaR}c. 
On increasing the electric field and the total induced charge, the thickness of the accumulation layer $t$ increases. When it becomes equal to $d_1$ (that means that the whole oxynitride layer has been perturbed) the normalized resistance variation reaches the saturation value (red point in Fig.\,\ref{fig:DeltaR}c):
\begin{equation}
\left.\frac{\Delta R}{R'}\right|\ped{sat} = -(n'_1-n_1) d_1 \frac{A_1}{n_1  A_1 d_1 + n_2 A_2 d_2}=-\Delta n\ped{2D,sat}\frac{A_1}{n_1  A_1 d_1 + n_2 A_2 d_2}.
\label{eq:deltaRlim}
\end{equation}

On further increasing the field, the contribution of the oxynitride layer to the slab conductance remains constant, but the accumulation layer extends beyond $d_1$ and the charge induction interests also the underlying NbN layer (Fig.\,\ref{fig:DeltaR}b). In the conditions where $t>d_1$, and calling $t' = t-d_1$ the thickness of the NbN layer interested by charge accumulation, one obtains that the conductance of the compound slab is
\begin{equation}
G_2 = \frac{e^2 w}{\ell}\left[n'_1 A_1 d_1 + n'_2 A_2 t' + n_2 A_2 (d_2-t')\right]. 
\label{eq:G2}
\end{equation}
and the normalized resistance variation is
\begin{eqnarray}
\frac{\Delta R}{R'} & =& \frac{G_0-G_2}{G_0} =  -(n'_1-n_1) d_1 \frac{A_1}{n_1  A_1 d_1 + n_2 A_2 d_2} - (n'_2-n_2) t \frac{A_2}{n_1  A_1 d_1 + n_2 A_2 d_2} \\
& = & \left.\frac{\Delta R}{R'}\right|\ped{sat}-(\Delta n\ped{2D}-\Delta n\ped{2D,sat})\frac{A_2}{n_1  A_1 d_1 + n_2 A_2 d_2} \\
& = & -\Delta n\ped{2D}\frac{A_2}{n_1  A_1 d_1 + n_2 A_2 d_2} + \mathrm{const}.
\label{eq:deltaR2}
\end{eqnarray}
This equation corresponds to the red dashed line in Fig.\,\ref{fig:DeltaR}c that approximates the high-field linear behaviour of the experimental points.  
By comparing Eqs.\,\ref{eq:deltaR1} and \ref{eq:deltaR2}, it turns out that the  slopes of the $\Delta R/R'$ vs $\Delta n\ped{2D}$ curves in the two regimes are different because of the different properties of the two materials. Here we have assumed that neither $\tau_i$ nor $m_i$ are affected by the charge perturbation, so that the $A_i$ are always equal to their intrinsic values -- an approximation which is reasonable for materials with a high intrinsic carrier density.
The ratio between the two slopes turns out to be equal to the ratio $A_2/A_1$. Experimentally, this ratio is equal to 0.2, which means that $A_1 = 5A_2$. This result is reasonable, at least if the contribution of the effective mass is considered, since niobium oxynitrides are predicted to exhibit smaller effective masses with respect to stoichiometric NbN\,\cite{TamerJPE2018, ChockalingamPRB2008}. 

Up to now, we have not used the experimental information about the thickness of the layers, but only about the slope of the experimental points in Fig.\,\ref{fig:DeltaR}. However, starting from Eq.\,\ref{eq:deltaRlim} and using the coordinates of the saturation point (for electron doping) $\displaystyle \left.\frac{\Delta R}{R'}\right|\ped{sat} = - 0.217 \times 10^{-2}$ and $\Delta n\ped{2D,sat}=0.407\times 10^{14}\, \mathrm{cm^{-2}}$, together with the fact that $A_1=5A_2$, one can obtain
\[5 n_1 d_1 + n_2 d_2 = 9.38\, 10^{16}\, \mathrm{cm^{-2}}.\] 
Approximating the experimental values of the thicknesses obtained by tunnel spectroscopy, XPS and ellipsometry to $d_1 \simeq 1\, \mathrm{nm}$ and $d_2 \simeq 5 \,\mathrm{nm}$, this equation provides
\[ n_1 + n_2 \simeq 1.88 \, 10^{23}\, \mathrm{cm^{-3}}\]
which is of the proper order of magnitude. Although we cannot obtain $n_1$ and $n_2$ separately, this result somehow refines the rough estimation of the intrinsic 3D charge density provided in the main text, which was based solely on the slopes of $\Delta R/R'$ as a function of $\Delta n\ped{2D}$ and was extracted by using the parallel-resistor model in the case of a \textit{homogeneous} material (thus neglecting the fact that the slab is actually made of two layers). In that case, the slope is simply $-1/n\ped{3D} d$ and the estimated $n_1 = 2 \times 10^{23}\, \mathrm{cm^{-3}}$, $n_2 = 0.4 \times 10^{23}\, \mathrm{cm^{-3}}$ were obtained by taking $d$ as the (approximated) total thickness of the slab.
\clearpage

\section{X-ray photoelectron spectroscopy of encapsulated ultrathin films}

\begin{figure}[h]
	\centering
	\includegraphics[width=0.7\textwidth]{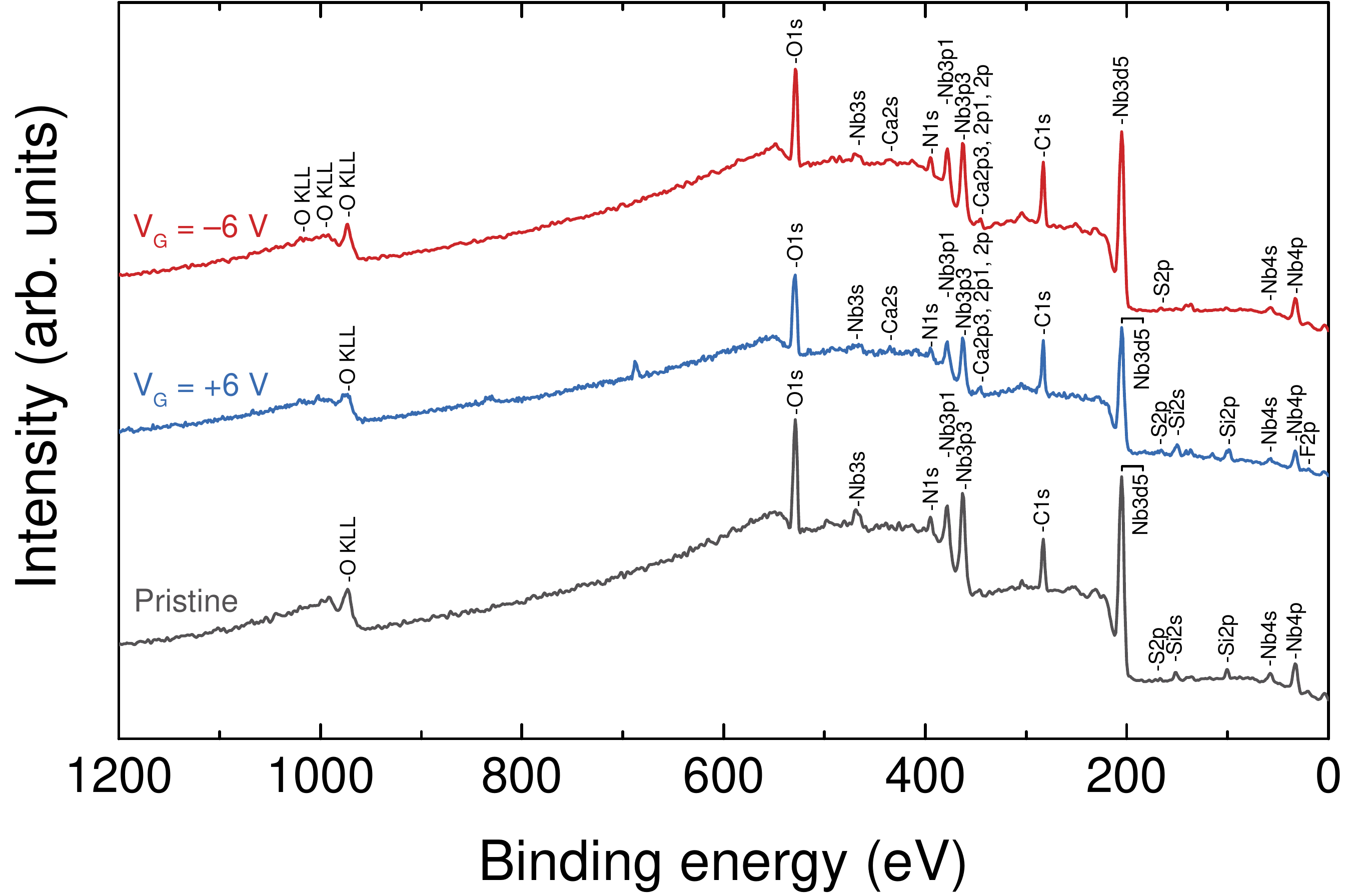}
	\caption
	{X-ray photoelectron spectroscopy survey scans of a pristine NbN film (black line), a NbN film gated at $V\ped{G}=+6$\,V (blue line), and a NbN film gated at $V\ped{G}=-6$\,V (red line), together with peak assignments obtained using the Multi Pak Software 9.7 for peak recognition.}
	\label{fig:XPS_survey}
\end{figure}

X-ray photoelectron spectroscopy measurements were performed on three samples cut from the same large-area, unpatterned film: a pristine film, a film gated at $V\ped{G}=+6$\,V, and a film gated at $V\ped{G}\,-6$\,V, as discussed in the Main Text. The survey spectra of all three samples (shown in Fig.\,\ref{fig:XPS_survey}) show a massive presence of carbon contamination. Considering that all films were exposed to the DEME-TFSI ionic liquid and were coated by a protective polymethyl methacrylate (PMMA) layer during storage and shipping, it is reasonable to assume that the carbon contamination is due to residues arising from both, which could not be fully removed by the cleaning process prior to the XPS measurements. Within this surface contamination layer, the amount of ionic liquid is reasonably inferior compared with that of PMMA, as suggested by the small sulphur signal as reported in Table\,\ref{tab:XPS_survey}, from the oxygen spectra, and from the ionic liquid composition (DEME-TFSI Elemental Analysis: C 28.17\%; H 4.73\%; F 26.73\%; N 6.57\%; O 18.76\%; S 15.04\%). Different silicon percentages arising from the underlying SiO\ped{2} substrate were detected in the three samples, which were reasonably due to a slighlty different thickness of the Nb\ped{2}O\ped{5}/NbN stack in each analysis spot.

\begin{table}[h]
	\begin{tabular}{ccccccc}
		\toprule
		\textbf{Element} & & \textbf{Pristine} & &\textbf{ Gated +6\,V} & & \textbf{Gated --6\,V} \\
		& & (atom \%) & & (atom \%) & & (atom \%) \\
		\midrule
		C & & 34 & & 36 & & 43 \\
		\midrule
		O & & 38 & & 31 & & 33 \\
		 \midrule
		N & & 7 & & 6 & & 7 \\
		 \midrule
		Nb & & 17 & & 11 & & 15 \\
		 \midrule
		S & & 4 & & 2 & & 1 \\
		 \midrule
		Si & & $<$1 & & 9 & & 1 \\
		 \midrule
		Ca & & Not detected & & 1 & & Not detected \\
		 \midrule
		F & & Not detected & & 4 & & Not detected \\
		\bottomrule
	\end{tabular}
	\caption
	{Atomic compositions of a pristine NbN film, a NbN film gated at $V\ped{G}=+6$\,V, and a NbN film gated at $V\ped{G}=-6$\,V, according to XPS survey scan analysis.}
	\label{tab:XPS_survey}
\end{table} 

\clearpage

\section{Measurement and fitting of the tunnelling $V(I)$ curves}\label{sec:tunnel}

Due to the small thickness and width of the NbN strip, the resistance $r\ped{s}$ of the portion of film between the point contact and the $V^-$ contact can be comparable to the resistance of the junction itself. It thus heavily affects the voltage drop measured between the $V^+$ and the $V^-$ contacts, that turns out to be equal to $V\ped{exp}(I)= V(I) + r\ped{s} I$. In order to determine the true $V(I)$ we need to eliminate the contribution of $r\ped{s}$. In principle this can be done by carrying out the measurements in the superconducting state; however, the features associated to the energy gap in NbN can disturb the fitting of the $V(I)$ with simple models thought for N/I/N junctions. Therefore, we took the $V\ped{exp}(I)$ in the normal state and rescaled the values in such a way that $V(I) = V\ped{exp}(I) - r\ped{s} I$ is identical to the $V\ped{exp}(I)$ curve measured in the superconducting state, apart from the superconducting gap features. This procedure allows determining the value of $r\ped{s}$ but, more important, allows constructing the intrinsic current/voltage characteristic of the junction. This curve can be directly compared to the Simmons model in the intermediate voltage range\,\cite{Simmons1963}, according to which the current flowing through the junction can be expressed as:
\begin{equation}
I = A \frac{e^2}{2\pi h s^2}\left[\left(\phi-\frac{V}{2}\right) e^{-\frac{4\pi s}{h}\sqrt{2me\left(\phi - \frac{V}{2}\right)}} -\left(\phi+\frac{V}{2}\right) e^{-\frac{4\pi s}{h}\sqrt{2me\left(\phi + \frac{V}{2}\right)}}\right]
\end{equation}
where $e$ is the elementary charge, $h$ is the Planck constant, $m$ is the electron mass, $A$ is the junction area, $\phi$ is the height of the potential barrier and $s$ is the barrier thickness. Thanks to the geometry of our strip, which is 100 $\upmu$m wide, we were able to estimate that the upper limit for the junction area is $A\ped{max} \simeq 1.2 \cdot 10^{-8}\, \mathrm{m^2}$ in all our junctions. In order to facilitate the convergence of the fitting procedure, we chose to fix the area at different values (generally from $A\ped{max} $ down to $0.1 A\ped{max}$). For each fit, the relative uncertainties of the parameters $\phi$ and $s$ are very small (in the worst cases,  3\% and 1.5\% ). We then took $\phi$ and $s$ as the midpoint of the relevant interval of values determined by the fits with different $A$. The points in Fig.\,6c and d were obtained in this way, and the error bars correspond to 1/2 of the interval width. In some cases, such uncertainty is smaller than the size of the symbols.  
Note that the results provide values of $\phi$ that are larger than the maximum voltage applied through the junction, which justifies the use of the model for the intermediate regime.

%